\documentclass[
aip,jcp,superscriptaddress,
reprint, floatfix, longbibliography
]{revtex4-1}

\usepackage[version=3]{mhchem}

\usepackage{dcolumn}%
\usepackage{bm}%
\usepackage{amsfonts}
\usepackage{bbm}
\usepackage{todonotes}
\usepackage{tikz}
\usepackage{cool}
\usepackage[fleqn]{mathtools}
\usepackage{isomath}
\usepackage{lmodern}
\usepackage{multirow}
\usepackage[normalem]{ulem}
\usepackage{subcaption}
\usepackage{float}
\usepackage{hyperref}
\begin{document}

\title{A Fully Quantum-Mechanical Treatment for Kaolinite}

\author{Sam Shepherd}
\author{Gareth A. Tribello}
\author{David M. Wilkins}
\email{d.wilkins@qub.ac.uk}
\affiliation{Centre for Quantum Materials and Technologies, School of Mathematics and Physics, Queen’s University Belfast, Belfast BT7 1NN, Northern Ireland, United Kingdom}
\date{}

\begin{abstract}
Neural network potentials for kaolinite minerals have been fitted to data extracted from density functional theory calculation that were performed using the revPBE + D3 and revPBE + vdW functionals.  These potentials have then been used to calculate static and dynamic properties of the mineral.  We show that revPBE + vdW is better at reproducing the static properties.  However, revPBE + D3 does a better job of reproducing the experimental IR spectrum.  We also consider what happens to these properties when a fully-quantum treatment of the nuclei is employed.  We find that nuclear quantum effects (NQEs) do not make a substantial difference to the static properties.  However, when NQEs are included the dynamic properties of the material change substantially.

\end{abstract}

\maketitle

\section{Introduction}
\label{sec:introduction}

Clay minerals represent some of the most abundant, yet experimentally frustrating minerals on the planet. They are employed in a large number of applications\cite{murray_overview_1991} ranging from catalysis\cite{adams_clay_2006, ferris_mineral_2005, johns_clay_1979, cecilia_catalytic_2021} and aqueous contaminant removal\cite{bhattacharyya_adsorption_2008, struijk_novel_2017, crasto_de_lima_retention_2017, zhang_adsorption_2021}, to those involving the modification of the clay minerals via intercalation between the clay layers\cite{liu_polymer_2007, varma_clay_2002}. The presence of clay mineral particles also hinders industrial processes such as oil well drilling. When drilling through clay, it is important to understand the damage that is done to the rock network~\cite{wilson_influence_2014} and to prevent the migration of fine particles within the oil~\cite{prempeh_effects_2020}. Due to the diversity of these applications and the range of clay minerals used within them, interest in these processes is ongoing and expanding. The experimental understanding of these processes is however, incomplete. Numerous methods have been used to investigate clay minerals\cite{madejova_ir_2017, prost_infared_1989, lanson_decomposition_1997, pan_review_2020, zhou_xrd-based_2018, martins_neutron_2014}, but unearthing the properties and structures of individual minerals and surface motifs remains a difficult task\cite{pan_review_2020, aharvey_distinguishing_2019, lagaly_characterization_1981}.

One such clay mineral, kaolinite, finds practical use in a large number of fields, ranging from applications within medicine\cite{awad_kaolinite_2017, gianni_kaolinite_2020} and catalysis \cite{cao_recent_2021} to the creation of nano-hybrid devices\cite{dedzo_functional_2016}. Kaolinite is also an illustrative example of the incomplete picture currently available experimentally\cite{bish_rietveld_1993, white_what_2009}. The small (sub-$\mu$m) size and roughness of the grain surface makes it difficult to determine the mineral structure using standard experimental methods\cite{yin_surface_2012}.   

To aid experimental understanding, theoretical approaches offer us a method to reconcile the difficulties encountered when using experimental techniques. Computational techniques have been used to investigate the formation of kaolinite particles \cite{volkova_molecular_2021}, ice formation on layer surfaces of kaolinite \cite{sosso_ice_2016} and halloysite nanotube formation\cite{aprishchenko_molecular_2018}. These studies employed classical force fields \cite{cygan_molecular_2004, cygan_advances_2021, pouvreau_structure_2017} that allow one to study processes on long length and time scales. These timescales are necessary to enable thorough investigation of the vibrational dynamics\cite{greathouse_implementation_2009} of the system and interactions with aqueous interlayer species\cite{cygan_advances_2021}. Classical force fields are typically fitted to reproduce experimental results or to quantum mechanical calculations and hence, may not work well when the system moves to parts of phase space that lie outside of the region where they are fitted.  Other research has thus used electronic structure theory to study the mechanisms via which transition metals and organic molecules\cite{liu_adsorption_2019, wang_dft_2013, ren_adsorption_2020} adsorb on the surface and enter the interlayer regions\cite{chen_mechanism_2019, chen_mechanism_2019-1, wang_theoretical_2015, wang_adsorption_2017, zhao_theoretical_2014, he_adsorption_2013}. In these more-accurate simulations, small system sizes are studied over short length scales, which limits available insight into processes which occur over longer timescales. 

When studying other materials, researchers have leveraged the power of machine learning \cite{schran_machine_2021, schran_transferability_2021, deringer_machine_2019, bartok_gaussian_2010, bartok_gaussian_2015, chmiela_towards_2018, zhang_efficient_2020, deringer_gaussian_2021, musil_machine_2019}. In one such application, an ab-initio method is used to build a database of structures and their associated energies and atomic forces. A less computationally-expensive potential is then fitted from this database.  One can thus run long length and time scale simulations with ab-initio accuracy without incurring the additional computational expense associated with the reference method.  In this paper we therefore extend our previous work\cite{kurapothula_hydrogen-bonding_2022} in which we investigated how nuclear quantum effects (NQEs) impact the structural and dynamical properties of the clay mineral kaolinite and pyrophyllite.  However, instead of employing the classical force field, CLAYFF\cite{cygan_molecular_2004, cygan_advances_2021, greathouse_implementation_2009} we use neural network potentials\cite{behler_representing_2014, behler_constructing_2015} that were trained from ab-initio data.  

We use these potentials to perform path integral \textit{ab initio} molecular dynamics\cite{parrinello_study_1984} simulations at the \textit{ab-initio} level and thus study the impact nuclear quantum effects have on the system when the electrons are modelled \textit{ab-initio}.

As we observed in our previous simulations\cite{kurapothula_hydrogen-bonding_2022}, nuclear quantum effects do not change the static properties of the material.  However, they have a substantial effect on the dynamical properties.  The fact that we observe similar differences when the electrons are modelled classically and when they are modelled using \textit{ab initio} techniques tells us that nuclear quantum effects are important and must be included.  The discrepancies that we observed between the experimental spectra and the spectra that was obtained from the classical simulation in our previous work is not simply a consequence of an inaccurate forcefield.  There is important physics missing from the model when the nuclei are treated classically.

The remainder of this paper is laid out as follows.  In section \ref{sec:background} we provide some background information on the structure of kaolinite and the previous work that has been done on this material using ab initio simulation.  We then describe how the machine learning potentials were trained in section \ref{subsec:NNP-Creation}. We detail how molecular dynamics simulations were run and analysed in section \ref{subsec:MD}. A comparison of the structural and dynamic properties we obtain from our simulations using our machine learned potentials when nuclear quantum effects are both included and not included is then presented in sections \ref{subsec:Structure} and \ref{subsec:vibrations}.

\section{Background}
\label{sec:background}
The complexity involved with the computational modelling of kaolinite is predominantly due to the layered structure of the mineral. As shown in figure \ref{fig:KaoliniteImages}, Kaolinite is a 1:1 type kaolin clay, with each individual layer comprised of an octahedral aluminol sheet covalently bonded to a tetrahedral silica sheet\cite{bish_rietveld_1989, bish_rietveld_1993}. These layers are held together through hydrogen bonds and dispersion interactions between the partially positive aluminol surface and the partially negative silica surface. In the case of 1:1 type clay minerals like kaolinite, these interactions are dominated by hydrogen bonding between the terminal O-H groups of the aluminol layer and the interfacial oxygen atoms of the silica layer. Ensuring these interactions are accounted for when studying clay materials is extremely important for the accurate modelling of the mineral\cite{tunega_assessment_2012, zen_toward_2016, voora_density_2011, crasto_de_lima_retention_2017}. The combination of hydrogen bonding and dispersion interactions ensures that any theoretical study into kaolinite using \textit{ab initio} methods is a computationally expensive undertaking.

In spite of this expense, traditional electronic structure methods like density functional theory (DFT) have been used extensively in the study of kaolinite\cite{chen_mechanism_2019-1, wang_theoretical_2015, zhao_theoretical_2014} and other clay minerals\cite{chatterjee_dft_1999, wungu_absorption_2011}. These have allowed researchers to gain insight into numerous properties of kaolinite, mainly regarding the absorption mechanisms\cite{zhao_theoretical_2014} and vibrational properties\cite{balan_first-principles_2001, balan_first-principles_2005, balan_spectroscopic_2011, madejova_ir_2017} of the mineral. These studies have been limited in both time and length scale, but have shown that the accuracy of the calculations depend quite sensitively on the level of theory used when modelling the system. Both dispersion corrections\cite{tunega_assessment_2012, zen_toward_2016, voora_density_2011} and a full quantum mechanical treatment of the nuclei\cite{kurapothula_hydrogen-bonding_2022} are necessary to reproduce the relevant physics. The necessity of including all of the aforementioned physics has hitherto limited the ability of researchers to study these minerals in greater detail. 

\section{Computational Methods}
\label{sec:computational-methods}

\subsection{Neural Network Potential Creation}
\label{subsec:NNP-Creation}

\begin{figure}[b!]
    \centering
    \begin{subfigure}[b]{0.5\linewidth}
         \centering
         \includegraphics[width=\linewidth]{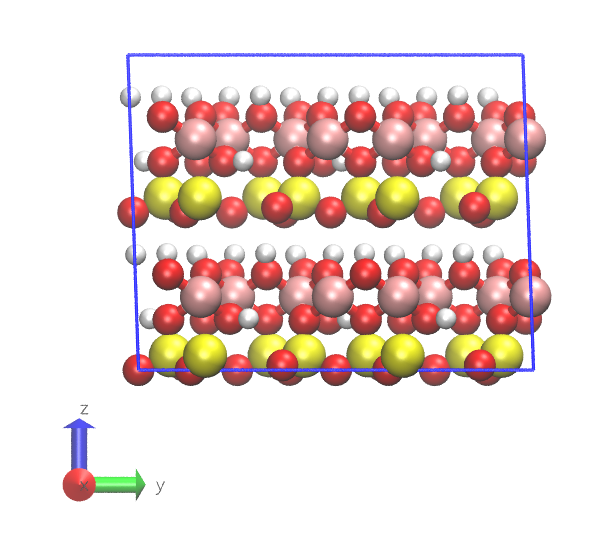}
         \label{FrontView}
     \end{subfigure}
     \hfill
     \begin{subfigure}[b]{0.44\linewidth}
         \centering
         \includegraphics[width=\linewidth]{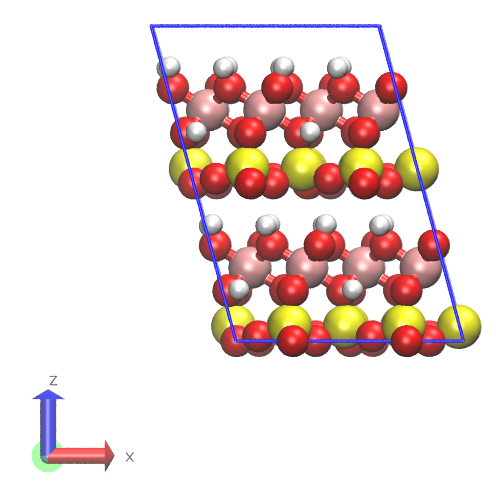}
         \label{SideView}
     \end{subfigure}
     \hfill
    \caption{Experimental structure of kaolinite used within this work. Initial structure was obtained from Bish\cite{bish_rietveld_1993} and replicated in all  directions to create a $2 \times 2 \time 2$ supercell to serve as a suitable starting point for simulations.}
    \label{fig:KaoliniteImages}
\end{figure}

We used the kaolinite structure from neutron scattering experiments\cite{bish_rietveld_1993} as a starting point. This structure was replicated twice in each direction to create the $2$ x $2$ x $2$ supercell that is pictured in Fig. \ref{fig:KaoliniteImages}.  

To create an initial dataset for the training of a machine learning potential, we performed a number of both canonical (NVT) and constant-stress (NST) ensemble molecular dynamics simulations at the density functional tight binding\cite{elstner_self-consistent-charge_1998, porezag_construction_1995, seifert_calculations_1996} (DFTB) level of theory with the parameters of Frenzel \textit{et al}\cite{frenzel_semirelativistic_nodate}. These simulations were performed using CP2K\cite{kuhne_cp2k_2020}.  These simulations were all started from the structure shown in figure \ref{fig:KaoliniteImages} and run a range of temperatures (200 K - 400 K). The final dataset that was used for training was then assembled using farthest point sampling (FPS)\cite{imbalzano_automatic_2018} to select the most diverse configurations from all available data.

The total energy and the forces were calculated for a subset of the configurations within the database using density functional theory (DFT). Two potentials were trained, which we will refer to as ClayNN-D3 and ClayNN-vdW. The revPBE functional \cite{perdew_generalized_1996, zhang_comment_1998}was used when calculating reference energies for both potentials.  The difference was in the way dispersion interactions are incorporated.  In particular, Grimme's D3 corrections\cite{grimme_consistent_2010, grimme_effect_2011} was used for the ClayNN-D3 potential, while Dion et al's dispersion interactions\cite{dion_van_2004} were used for the ClayNN-vdW potential. These two different dispersion recipes were employed because previous work by Tunega et al\cite{tunega_assessment_2012} using the revPBE functional argued that dispersion corrections must be included if the non-bonded interactions within clay minerals are to be reproduced accurately. Grimme's D3 corrections were selected based on the ability of revPBE + D3 to accurately model dispersion interactions within water\cite{ohto_accessing_2019}. Dion's corrections were selected due to previous work by Crasto de Lima\cite{crasto_de_lima_retention_2017} who utilized the revPBE + vdW functional to study the adsorption mechanisms of contaminants onto clay mineral surfaces. More recently, Kobayashi \textit{et al.}\cite{kobayashi_machine_2022} were able to achieve excellent agreement for a wide array of structural, dynamical and mechanical properties of kaolinite by using the meta-GGA functional, SCAN without the explicit addition of dispersion corrections.  These findings suggest that the SCAN 
functional incorporates dispersion interactions correctly. It is still important to use different treatments of dispersion in this work, however, as we are using a lower level of theory.
  
To train the machine learning potentials we used the well-established method based on high-dimensional neural networks that was introduced by Behler and Parinello \cite{behler_constructing_2015, behler_four_2021}.  An extensive description of the construction, architecture and symmetry functions we used is included in the SI. Initial neural network potentials (NNP) were trained using N2P2\cite{singraber_library-based_2019, singraber_parallel_2019} with a training dataset consisting of 6000 frames. Expansion of these potentials was completed using the `query by committee'\cite{schran_machine_2021} procedure, which evaluated the total energy and atomic forces predicted by the driving NNP using a 10 membered committee of similar (but importantly, non-identical) potentials. If the standard deviation of the committees average atomic force prediction was below a set range (20 - 50 meV/ \r{A}) the configuration was well represented within the training dataset and would contribute nothing upon its addition to said training dataset. In the event of a configuration producing a relatively high error (with a preset upper limit, to avoid the addition of non-physical systems), this configuration was added to the training data of the potentials.  The energies and forces for this configuration were then recomputed using the reference method for the potential. To prevent overfitting during training, we employed the `early stopping' method, and selected the weights that gave the lowest testing error.
In all cases, excellent agreement between the testing data and the predictions of the network indicate that the short-ranged HDNNPs used are sufficient to capture the important physics.

The final potentials presented within this work (ClayNN-D3 and ClayNN-vdW) consist of a total of 7748 kaolinite configurations sampled from initial DFTB simulations and NNP driven NVT and NST simulations. All DFT calculations used throughout this work were carried out using CP2K interfaced with AiiDA\cite{huber_aiida_2020, uhrin_workflows_2021}. Example input files for theses computations can be found within our \href{https://github.com/sshepherd637/KnnP}{github repository}.

ClayNN-D3 and ClayNN-vdW were validated by comparing the predicted system energy and atomic forces to reference values that were computed using revPBE + D3 and revPBE + vdW calculations in CP2K. Both potentials were validated using a similar strategy to that of Schran et al.\cite{schran_transferability_2021, schran_automated_2020}, with structural and dynamical properties of the system used to validate the potentials as well as the predicted and calculated energies and forces. The structural properties used included the computed and predicted radial distribution functions and the phonon density of states (PDOS). The way in which both the energies and the forces were used for validation are presented within the SI.

\subsection{Molecular Dynamic Simulations}
\label{subsec:MD}

To equilibrate the system initial MD simulations in the NVT ensemble were run for 500 ps at 300 K prior to a further 500 ps equilibration in the NST ensemble at 300 K.  The timestep in these simulations was set equal to 0.25~fs.  Furthermore, in these NST simulations the diagonal elements of the stress tensor were set equal to 1 bar and all off-diagonal components were set equal to zero. Temperature was controlled using a Langevin thermostat\cite{bussi_accurate_2007} with a friction of 1 ps.  To obtain sufficient statistics for the structural properties of the clay system, a subsequent 1 ns of NST simulation was performed. 

To study the vibrational dynamics of the system, we ran 500 ps simulations at temperatures ranging from 10 K to 300 K in the NST ensmble. Ten configurations from each of these runs were then used as starting points for microcanonical (NVE) simulations. Separate estimates of the vibrational density of states (VDOS) were calculated from each of these simulations by finding the Fourier transform of the velocity-velocity autocorrelation function. The final VDOS results at each temperature that we show in the results section are thus averages over 10 independent estimates.

Path integral molecular dynamics (PIMD) was used to account for nuclear quantum effects. All simulations using PIMD were run using 16 replicas at 300 K.  This number of replicas is justified as we observed a change of less than 0.5 \% of the total system energy when the number of replicas was increased from 16 to 32. 

When the system was modelled with PIMD it was allowed to equilibrate for 200 ps in the NVT ensemble.  It was then subsequently equilibrated for a further 100 ps in the NST ensemble.  Static properties for the kaolinite system were then obtained from a 100 ps production simulation. The global path integral Langevin equation (PILE)\cite{ceriotti_efficient_2010} thermostat with global velocity rescaling was employed in all these simulations. When considering dynamical properties for the system, we employed both thermostatted ring polymer molecular dynamics (TRPMD)\cite{habershon_ring-polymer_2013, rossi_how_2014} and partially adiabatic centroid molecular dynamics (PACMD)\cite{hone_comparative_2006}. We provide input files for both of these simulations. We present the results from the PACMD simulations in the manuscript as they were superior.  However, information on the results we obtained with TRPMD are provided in the SI for completeness. 
 
All molecular dynamics simulations (both classical and path-integral) were performed using LAMMPS\cite{thompson_lammps_2022} interfaced with N2P2 and driven by the i-PI\cite{kapil_i-pi_2019} code. Example input files for all simulations are available in our \href{https://github.com/sshepherd637/KnnP}{github repository}.

\section{Results and Discussion}
\label{sec:Results}

\subsection{Structural Properties}
\label{subsec:Structure}

\begin{table*}
\centering
\caption{
Unit cell parameters and volumes for kaolinite obtained using various means.  The left part of the table contains results obtained from geometry optimization calculations using the reference DFT functionals (revPBE + D3 and revPBE + vdW) and the neural network potentials (ClayNN-D3 and ClayNN-vdW). The right part of the table contains results from  ClayNN-D3 and ClayNN-vdW driven classical (CL) and PI (NQE) MD calculations in the constant-stress (NST) ensemble. The $\delta_\text{NQE}$ column in this right part contains the differences between the results obtained with and without nuclear quantum effects. The $\delta_{\text{D3-vdW}}$ column contains the differences in the cell parameters for the two models of dispersion. This difference was calculated for simulations in which the nuclei were treated classically. The experimental values are given in the columns labelled EXP and are taken from Bish\cite{bish_rietveld_1993} and Bish and von Dreele\cite{bish_rietveld_1989}.
\label{tab:kaolinite_cellParams}}
\begin{tabular}{c|c|cccccc|c|ccccccc}
\hline\hline
&  EXP & \multicolumn{6}{c|}{Geometry Optimization} & EXP & \multicolumn{7}{c}{Molecular Dynamics at 300K} \\
\hline 
&  1.5 K & \multicolumn{3}{c|}{revPBE + D3} & \multicolumn{3}{c|}{revPBE + vdW} & 300 K & \multicolumn{3}{c|}{revPBE + D3} & \multicolumn{3}{c|}{revPBE + vdW} & \\ 
\hline
Property & Ref~\citenum{bish_rietveld_1993} & DFT & NNP & $\delta_{\text{NNP}}$ & DFT & NNP & $\delta_{\text{NNP}}$ & Ref~\citenum{bish_rietveld_1989} & CL & NQE & $\delta_{\text{NQE-D3}}$ & CL & NQE & $\delta_{\text{NQE-vdW}}$ & $\delta_{\text{D3-vdW}}$\\
\hline
$A (\text{\r{A}})$ & 5.153 & 5.095 & 5.115 & 0.02 & 5.194 & 5.235 & 0.039 & 5.154 & 5.128 & 5.137 & 0.011 & 5.243 & 5.252 & 0.009 & -0.115 \\
$B (\text{\r{A}})$ & 8.942 & 8.828 & 8.818 & -0.01 & 9.020 & 9.059 & 0.039 & 8.945 & 8.879 & 8.887 & 0.008 & 9.073 & 9.088 & 0.005 & -0.194 \\
$C (\text{\r{A}})$ & 7.391 & 7.513 & 7.523 & 0.01 & 7.439 & 7.525 & 0.086 & 7.405 & 7.773 & 7.876 & 0.103 & 7.586 & 7.605 & 0.019 & 0.187 \\
$\alpha (^{\circ})$  & 91.93 & 94.57 & 95.10 & 0.53 & 92.61 & 91.42 & -1.19 & 91.70 & 103.34 & 106.24 & 2.90 & 91.15 & 91.23 & 0.08 & 12.19 \\
$\beta (^{\circ})$ & 105.05 & 108.21 & 108.97 & 0.76 & 104.93 & 104.96 & 0.03 & 104.86 & 108.31 & 107.94 & -0.37 & 104.53 & 104.49 & -0.04 & 3.78 \\
$\gamma (^{\circ})$ & 89.80 & 90.05 & 90.37 & 0.32 & 89.89 & 90.03 & 0.04 & 89.82 & 90.61 & 90.61 & 0.00 & 90.09 & 90.09 & 0.00 & 0.52 \\
$ V (\text{\r{A}}^3)$ & 328.69 & 319.86 & 319.39 & -0.47 & 336.38 & 344.65 & 8.27 & 329.82 & 325.61 & 326.58 & 0.97 & 349.24 & 351.35 & 2.11 & -24.77    \\
\hline\hline
\end{tabular}
\end{table*}

\begin{figure*}
    \centering
    \includegraphics[width=\linewidth]{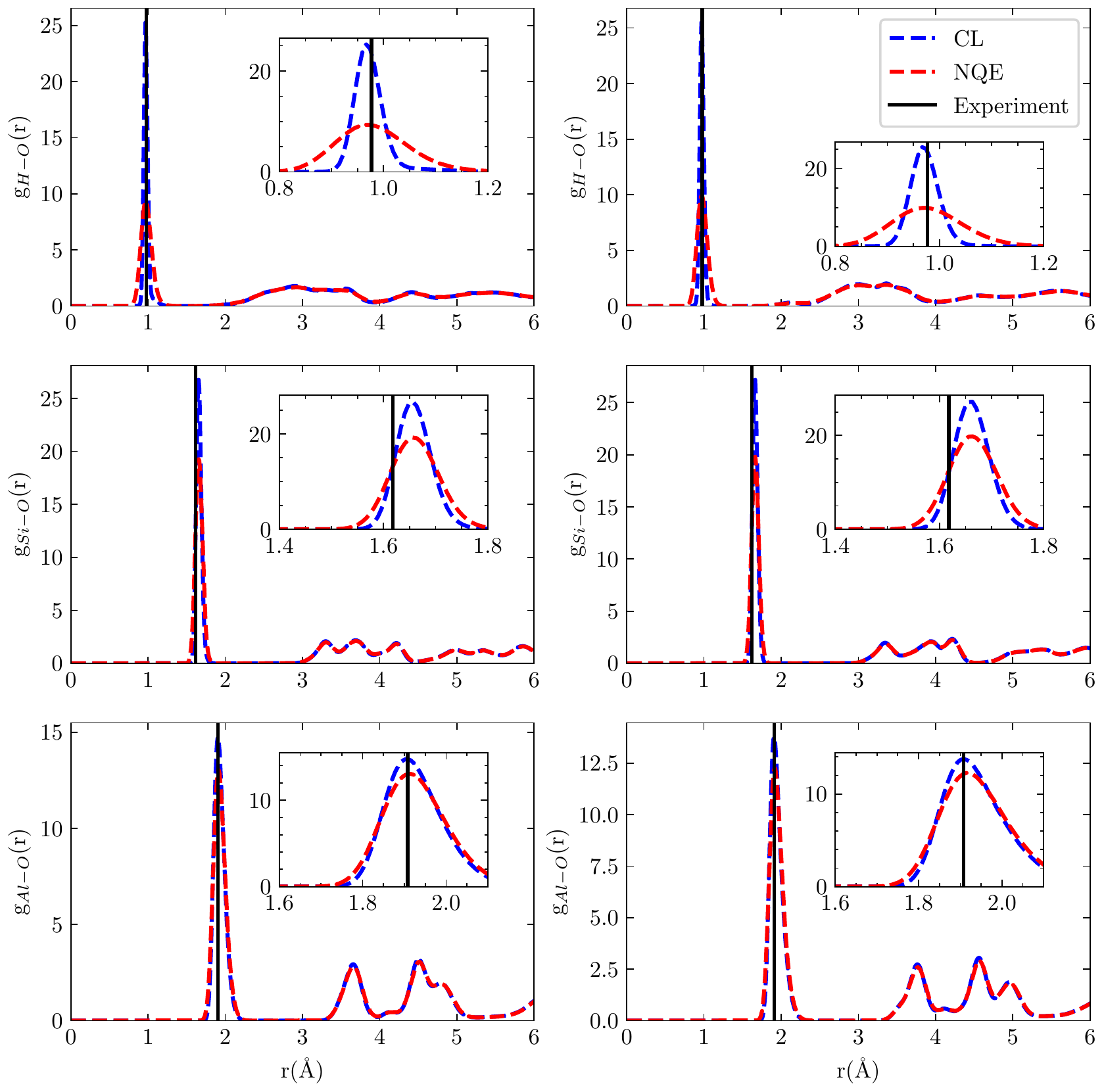}
    \caption{Radial distribution functions (RDFs) for O--H (top), Si--O (middle) and Al--O (bottom) elemental pairs. The left and right plots correspond to results obtained using ClayNN-D3 and ClayNN-vdW respectively. The dashed blue lines indicate the result obtained using classical molecular dynamics simulations while dashed red lines shows the result obtained when  NQEs are included by using PIMD. The vertical black lines indicate the interatomic distance for the corresponding atom pair that is obtained from experiment.  The inserts show a zoomed in view of the first peak in the RDF.}
    \label{fig:CharacteristicRDFS}
\end{figure*}

Table. \ref{tab:kaolinite_cellParams} shows the unit cell parameters for kaolinite that were obtained from experiments performed at 1.5~K\cite{bish_rietveld_1993}. The first part of the table shows the cell parameters that are obtained from geometry and cell optimization calculations. Geometry optimisations were performed using the reference DFT functionals (revPBE + D3 and revPBE + vdW) directly and using the fitted neural network potentials (ClayNN-D3 and ClayNN-vdW).  The $\delta_{NNP}$ column within Table \ref{tab:kaolinite_cellParams} shows the differences in the cell parameters that are obtained when the optimisation is performed using the functional directly and when the optimisation is performed using the neural network potential. The maximum difference between the cell parameters is 0.7\% for the ClayNN-D3 (this difference is in the $\beta$ parameter) and 1.14\% for ClayNN-vdW (this difference is in the $\alpha$ parameter) respectively. There is thus an overall excellent agreement between the reference DFT methods and the NNPs.  

Table \ref{tab:kaolinite_cellParams} show that the optimisations that were performed using revPBE + vdW have cell parameters that more closely match those in experiment.  For revPBE + D3 the percentage difference from the experiment are -0.058 \r{A} (1.12$\%$), -0.114 \r{A} (1.27$\%$) and 0.25$^{\circ}$ (0.28$\%$) for $A$, $B$ and $\gamma$ respectively.  For revPBE + vdW these differences take lower values of 0.041 \r{A} (0.79\%), 0.078 \r{A} (0.87\%) and 0.09$^{\circ}$ (0.1\%).  On these parameters, however, the agreement is excellent for both functionals. More marked differences in the parameters are observed for the other cell parameters.  The $C$, $\alpha$ and $\beta$ parameters for revPBE + D3 differ from experiment by 0.122 \r{A} (1.62$\%$), 2.64$^{\circ}$ (2.79$\%$) and 3.16$\circ$ (2.92$\%$) respectively.  For revPBE + vdW these differences are 0.048 \r{A} (0.65\%), 0.67$^{\circ}$ (0.74\%) and -0.120$^{\circ}$ (0.11\%). The differences between the two functionals is more marked for these parameters because these are the cell directions that are more strongly affected by the treatment of dispersion. There is a slight underbinding between the hydrogen bonded layers when the D3 dispersion correction is used.  Meanwhile, the vdW correction predicts cell parameters that are in line with those obtained using the dispersion corrected functionals that were considered by Tunega \textit{et al}.\cite{tunega_assessment_2012} However, the strength of the dispersion interaction is still underestimated slightly with this correction as the layers sit further apart than they would in experiment. 

The second part of Table \ref{tab:kaolinite_cellParams} shows the values of the cell parameters that are obtained from MD and PIMD simulations that were run using the fitted, neural network potentials ClayNN-D3 and ClayNN-vdW. These estimates were obtained from NST simulations of a $2 \times 2 \time 2$ supercell and are in good agreement with experiments performed at 300~ K\cite{bish_rietveld_1989}.

Comparing the parameters in the right half of the table with those in the left tells us that the unit cell size increases with temperature, as would be expected. The extent of this increase with temperature is, however, markedly larger than what is observed experimentally. Parameters which depend on the description of interlayer interactions are particularly affected. 
The $C$ parameter increases more strongly in our simulations because both functionals underestimate the strength of the dispersion-based interactions in the interlayer region. This underestimation affects the results at 300 K more strongly than the results at 0 K because bond lengths in the interlayer region undergo larger fluctuations. 

We have performed simulations with and without NQEs using two different treatments for dispersion. The molecular dynamics results in table \ref{tab:kaolinite_cellParams} show that changing the description of dispersion interactions has a much larger effect on the structure than including an explicit treatment of NQEs. We summarise this by calculating the percentage cell volume change between classical and path integral simulations. The cell volume increased by 7.26\% when the vdW correction is used in place of the D3 in a classical model of the nuclei.  By contrast, the cell volume increased by 0.60\% when NQEs are included in the model with the vdW correction. The percentage volume change when the D3 correction is used was 0.29\%, slightly less than the vdW value. This result is surprising as individual cell parameters when described using ClayNN-D3 are more keenly affected by NQEs than ClayNN-vdW. The largest increase due to NQEs was calculated as 0.103 \r{A}, observed for the $C$ cell parameter when using ClayNN-D3. This value is an order of magnitude larger than that observed for the $C$ cell parameter when using vdW corrections, which remains the most affected parameter. This result is hardly surprising for either potential, as the $C$ parameter is the one most acutely affected by dispersion interactions. The discrepancy between potentials we attribute to the less accurate reproduction of experimental results by ClayNN-D3. The comparatively small effect incurred by including NQEs is perhaps surprising given how much hydrogen bonding occurs within the structure of kaolinite. This observation is however, in agreement with our previous simulations with classical potentials\cite{kurapothula_hydrogen-bonding_2022}, where the percentage cell volume increase due to NQEs was 0.63\%.   

To investigate whether the discrepancies between the experimental and simulated cell parameters can be attributed to a poor description of the interaction between a particular pair of atom types, we calculated radial distribution functions (RDFs). Fig .\ref{fig:CharacteristicRDFS} shows three of these RDFs: $g_{\text{O-H}}(r)$, $g_{\text{Si-O}}(r)$, and $g_{\text{Al-O}}(r)$. RDFs for all other atom pairs can be found within the SI.  

Figure \ref{fig:CharacteristicRDFS} also shows average interatomic distances for the corresponding pairs of atoms that are obtained from experiments\cite{bish_rietveld_1993}. We note the good agreement between these experimentally obtained values and the positions of the first peak in $g_{\text{O-H}}(r)$ and $g_{\text{Al-O}}(r)$. By contrast, the first peak in the $g_{\text{Si-O}}(r)$ sits at a value of $r$ that is 2.47\% larger than the experimental average bond length. The Si-O interaction is thus weaker than it should be in our model. This under-binding of Si-O bonds by the studied dispersion corrected functionals is consistent with findings from Tunega \textit{et al}\cite{tunega_assessment_2012}.

Lastly, figure \ref{fig:CharacteristicRDFS} shows a comparison of the RDF that is obtained when the simulation includes nuclear quantum effects and when it does not. It is clear from the figure that including NQEs does not affect the positions of the peaks.  However, the peaks in $g_{\text{O-H}}(r)$ are broadened when nuclear quantum effects are included. One would expect this RDF to be most impacted by the inclusion of nuclear quantum effects because the atoms involved are the lightest. When nuclear quantum effects are included the O--H bond weakens and the width of the distribution for the lengths of the O--H bonds increases. This result explains our observation of cell parameter $C$ being substantially affected by NQEs in \ref{tab:kaolinite_cellParams}, as it is heavily dependent on interactions involving O--H groups. By comparison, the other RDFs pictured within figure \ref{fig:CharacteristicRDFS} are largely unaffected. Ultimately NQEs have a very small effect on the overall structure and mainly affect the value of the $C$ parameter. 

\subsection{Vibrational Properties}
\label{subsec:vibrations}
We have shown in the previous section that explicit treatment of the nuclear quantum has little effect on the structural dynamics of these clay systems. Nuclear quantum effects are much more likely to affect the vibrational properties that are acutely affected by subtle changes in the strength of the interatomic bonds.  In this next section we thus investigate the vibrational spectra of the material.

\begin{figure*}[ht!]
    \centering
    \includegraphics[width=\textwidth]{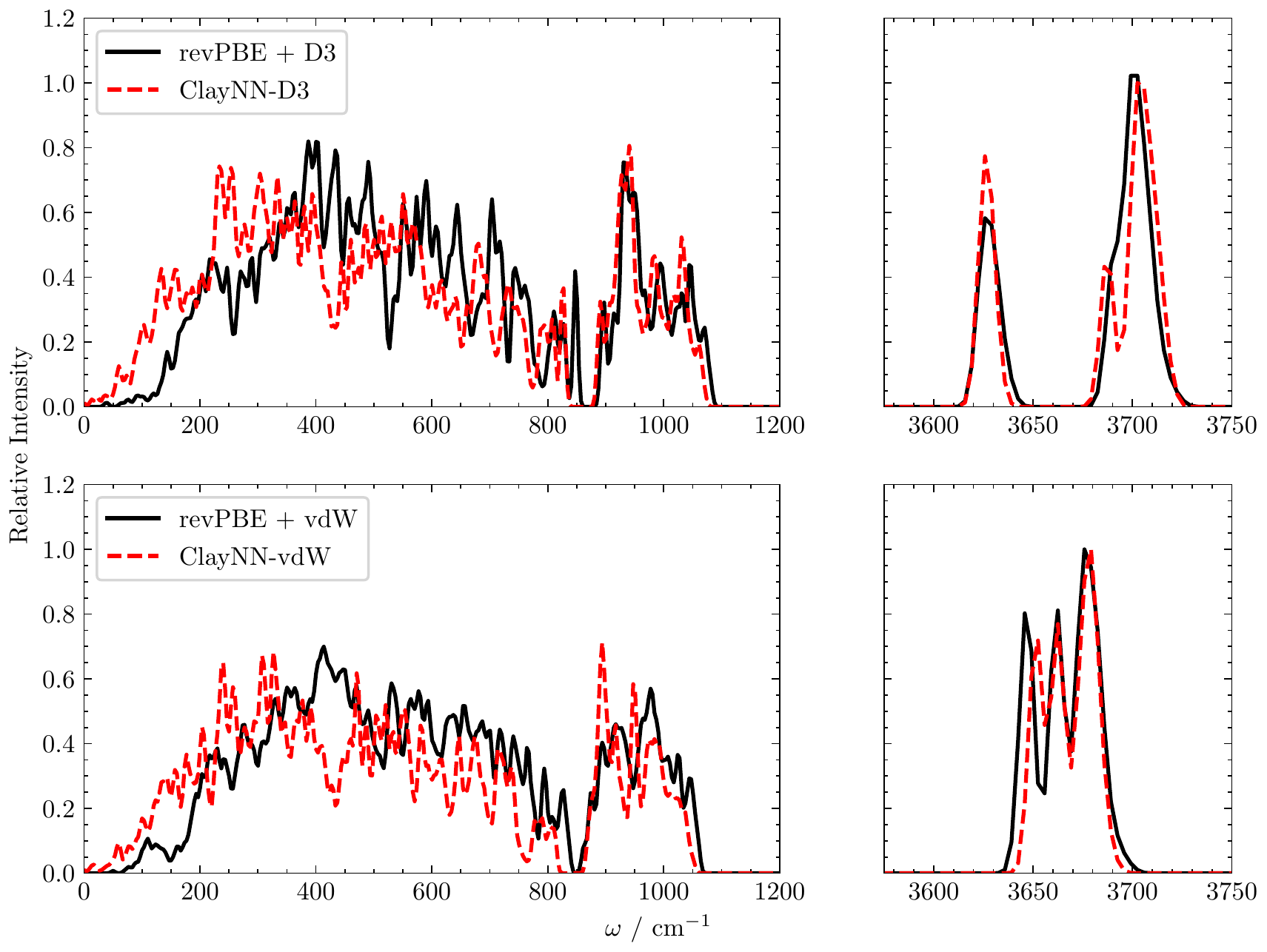}
    \caption{Computed phonon densities of states for kaolinite showing the low frequency region (left) and the high frequency region (right) for both the reference DFT methods and the trained NNPs that were used within this work. The upper panels show revPBE + D3 (solid black line) and ClayNN-D3 (dashed red line) whilst the lower panels show revPBE + vdW (solid black line) and ClayNN-vdW (dashed red line). All of the presented spectra have been normalised to have unit area.}
    \label{fig:PDOSvalidate}
\end{figure*}

Classical phonon densities of states (PDOS) were calculated using revPBE + D3, revPBE + vdW, ClayNN-D3 and ClayNN-vdW. PHONOPY\cite{togo_first_2015} was used in all these calculations with CP2K and the respective NNPs serving as external force calculators.  Fig. \ref{fig:PDOSvalidate} shows the PDOS that were obtained from this procedure. Both NN potentials produce PDOS that almost exactly match the \textit{ab initio} reference methods within the higher frequency region between 3600 to 3800 cm$^{-1}$. The NNPs thus successfully replicate the number and the positions of the vibrational O--H modes that are observed with the corresponding DFT functional. In the low frequency region (i.e. below 1200 cm$^{-1}$) there are larger discrepancies between the potentials and the reference DFT methods. This is most evident below 800 cm$^{-1}$. In this region it becomes more difficult to distinguish different lattice modes and the spectra predicted using the NNPs `shift' away from the reference.

\begin{figure}[bt!]
    \centering
    \includegraphics[width=\linewidth]{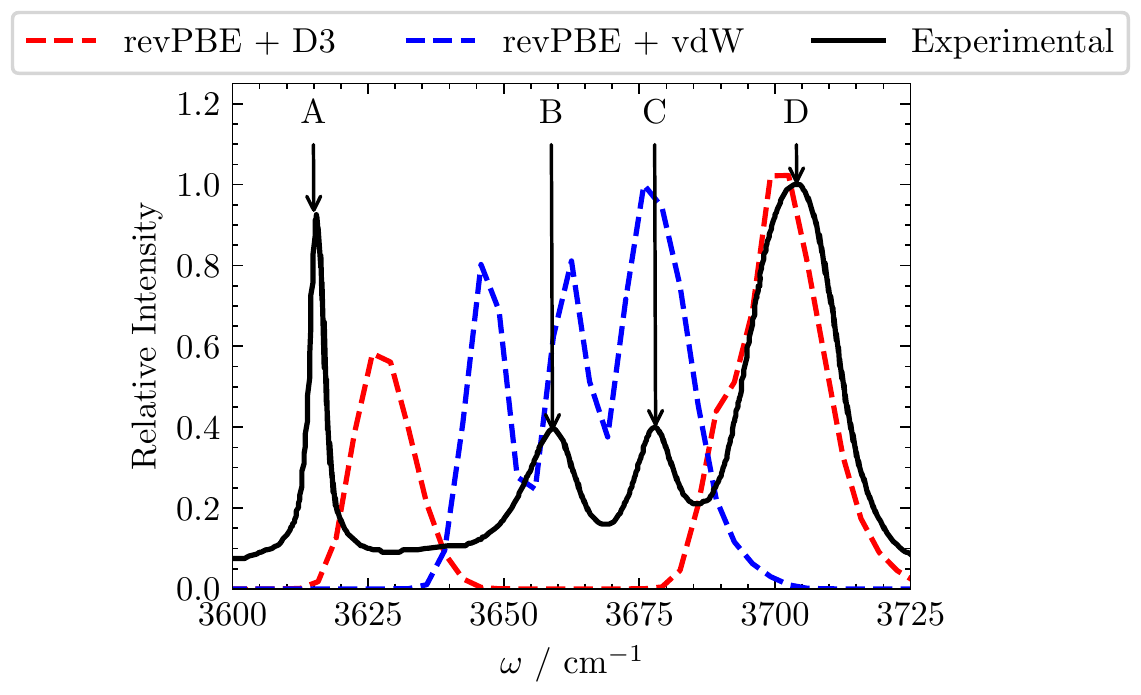}
    \caption{Computed phonon densities of states for kaolinite showing the high frequency region for both reference DFT methods. The dashed red line shows the result for revPBE + D3, the dashed blue line shows the result for revPBE + vdW and the black line shows the IR spectra that was obtained from experiments at 16~K\cite{t_johnston_low-temperature_2008}. All three spectra have been normalised to have unit area.}
    \label{fig:PDOSExperimentHigh}
\end{figure}

Figure \ref{fig:PDOSExperimentHigh} shows the high-frequency region of the classical PDOS computed using revPBE + D3 and revPBE + vdW in dashed red and dashed blue lines respectively. These are presented alongside experimental IR spectra obtained at 1.5 K\cite{balan_first-principles_2001}.  The complete computed PDOS, including the low frequency modes, is available within the SI. Neither functional reproduces the experimentally-observed O--H stretching vibrations.  In experiment there are four distinct vibrational modes.  The revPBE + D3 PDOS has only two distinct modes, while the PDOS for revPBE + vdW has three.

To aid in the assignment of the ClayNN computed modes, the peaks in the experimental IR spectrum shown in figure \ref{fig:PDOSExperimentHigh} are labelled using the assignment of Farmer\cite{farmer_transverse_2000}. Label A points to the inner O--H stretching mode around 3615 cm$^{-1}$. Label B points to an anti-phase O--H stretching mode of the interlayer O--H groups at 3659 cm$^{-1}$. Label C points to a higher frequency interlayer O--H anti-phase stretching mode at 3678 cm$^{-1}$. Finally, label D points to the in-phase interlayer O--H stretching mode located at 3704 cm$^{-1}$. 

We consider revPBE + D3 first.  Mode D agrees almost exactly with experiment. We believe the lower frequency peak around 3530 cm$^{-1}$ corresponds to mode A but it is much broader and is blue shifted. We also observe a slight shoulder toward the lower frequency side of mode D that may correspond to mode C. It is difficult to assign mode C, however, as there is nothing in the simulated spectra that corresponds to mode B. 

The reproduction of the high frequency vibrational modes of the experimental IR using revPBE + vdW is very different. We find that revPBE + vdW has features that correspond to modes B, C and D.  However, all of these features are red-shifted ~ 30 cm$^{-1}$ from those observed experimentally. There is no feature that corresponds to mode A in the simulated spectra but it may have merged with another peak.

The two functionals do not reproduce the four distinct vibrational modes within the O-H stretching region that are pictured in figure \ref{fig:PDOSExperimentHigh}. However, revPBE + D3 (and therefore, ClayNN-D3) provides a better description of the O-H vibrational dynamics and the in-phase interlayer O-H stretching mode at 3704 cm$^{-1}$ in particular. This result is annoying, as the previous section showed that revPBE + vdW better reproduces the structural properties. However, the fact that the vdW correction reproduces structural properties while the D3 correction reproduces dynamic properties agrees with previous work on aqueous systems \cite{marsalek_quantum_2017, ohto_accessing_2019}.

Figure \ref{fig:PDOSExperimentHigh} indicates that revPBE + D3 does a better job of reproducing the experimental IR structure.  We therefore used the NN potential that was fitted to the energies and forces that were calculated using this functional to computed the vibrational density of states (VDOS) at temperatures between 10 K and 300 K.  Averaged VDOS were obtained by performing Fourier transforms of the velocity autocorrelation function (VAF). These VAF were calculated from 10 individual canonical (NVE) ensemble simulations that were started from randomly sampled frames that were taken from an equilibrated region of the corresponding NST trajectories. 

\begin{figure}[ht!]
    \centering
    \includegraphics[width=\linewidth]{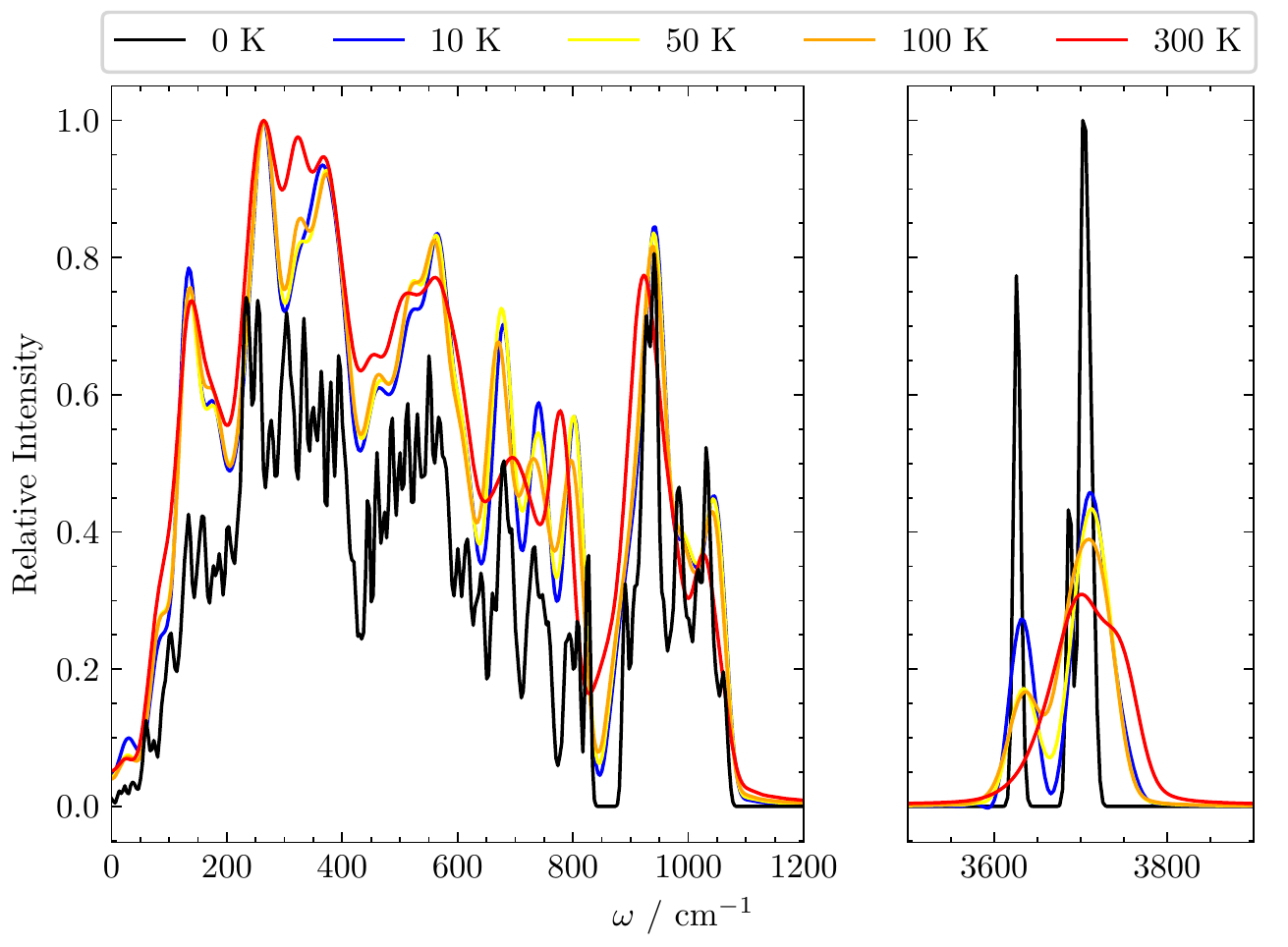}
    \caption{Vibrational densities of states for kaolinite showing the low frequency region (left) and the high frequency region (right) at a range of different temperatures. These densities of states were computed using ClayNN-D3. The black line shows the 0 K result, which was calculating by finding the PDOS. The blue, yellow, orange and red lines are then the results obtained at 10 K, 50 K, 100 K and 300 K respectively.}
    \label{fig:TempVDOSClassicalD3}
\end{figure}

Figure \ref{fig:TempVDOSClassicalD3} shows some of the VDOS that were obtained from our simulations. The full set of results are available within the SI. Increasing the temperature broadens the peaks in the spectrum.  This makes it more difficult to assign each peak to a specific vibrational mode. The same problem exists for experiments that are performed at finite temperature.  However, figure \ref{fig:TempVDOSClassicalD3} shows that there are still three clear peaks in the high-temperature experimental spectrum.    
The loss of all fine structure that is observed in our simulations thus suggests that there is some non-physical artifact in the simulations.  

\begin{figure*}[ht!]
    \centering
    \includegraphics{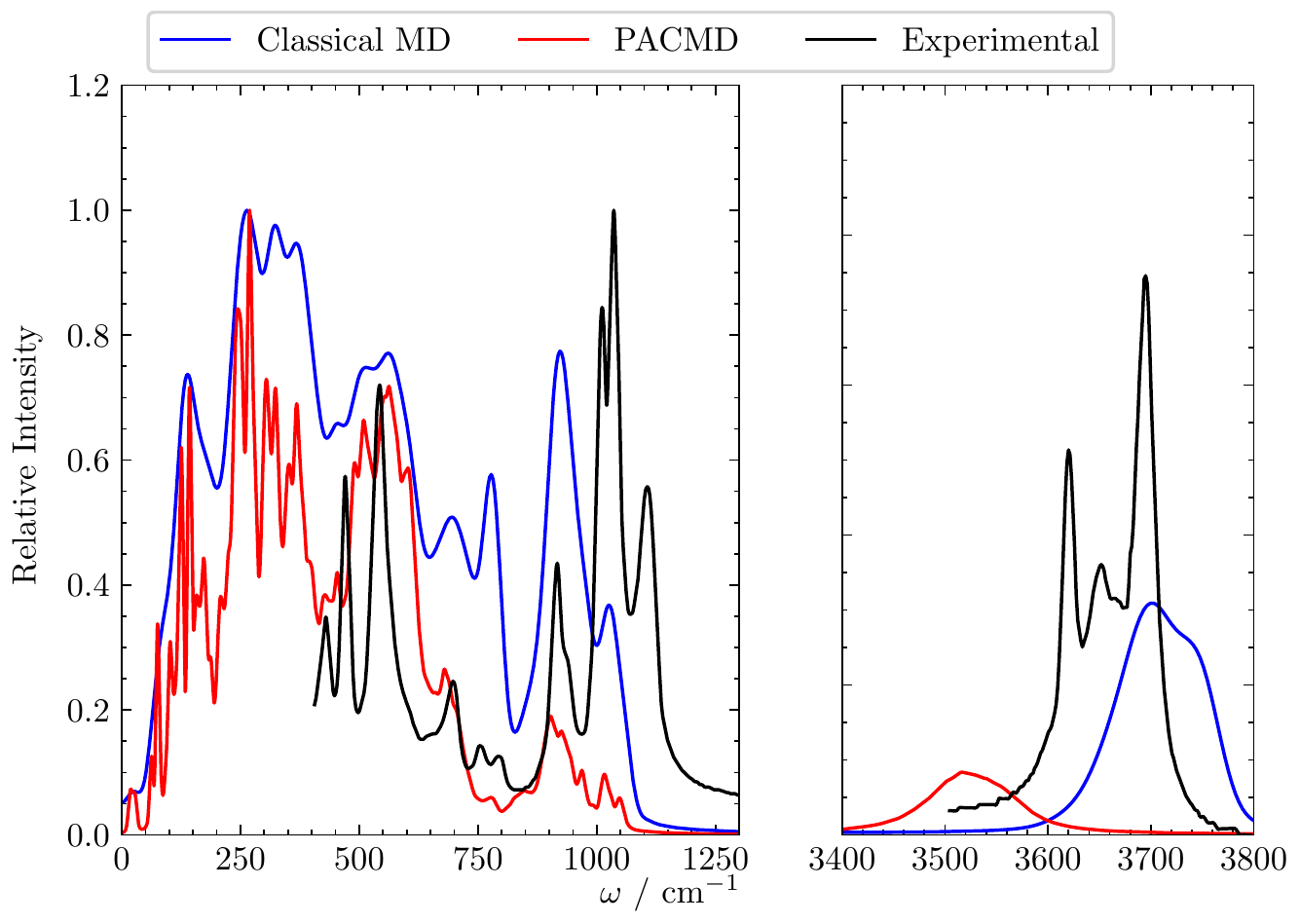}
    \caption{Vibrational densities of states for kaolinite showing the low frequency region (left) and the high frequency region (right). These densities of states were computed at 300 K using ClayNN-D3 and both classical nuclei (blue line) and path integrals (red line). Experimental data (black line) is taken from Ref.~\citenum{madejova_ir_2017}. We highlight the large shift away from the experimentally obtained O--H peak positions within the high frequency domain when we include NQEs.}
    \label{fig:PACMD}
\end{figure*}

As discussed in section \ref{subsec:Structure} kaolinite contains many light atoms. It is therefore important to consider the role of nuclear quantum effects. This holds doubly true for the dynamic properties, which are more sensitive to factors affecting interatomic interactions.  Figure \ref{fig:PACMD} shows the effects of NQEs on the computed IR spectrum. The blue line in this figure is the classical result for revPBE + D3 at 300 K that was shown in red in figure \ref{fig:TempVDOSClassicalD3}.  Meanwhile, the red line shows the spectrum that was obtained from PACMD simulations at 300 K.  You can clearly see that the peak in the high frequency spectrum is red shifted when nuclear quantum effects are included in the model. This result reflects the weakening of O--H bonds within the system when NQEs are included, and agrees with previous work which employed the revPBE + D3 functional for aqueous systems\cite{marsalek_quantum_2017}. In both that work and ours, there was good agreement between the computed and experimental peak positions within the IR spectra when NQEs were not included. However, the O--H frequencies were red shifted away from the experimental values when NQE were modelled explicitly. As such, the good fit between low-level theory and experiment is fortuitous, as any result which does not account for the quantum mechanical nature of the nuclei is inherently flawed. 

The low frequency region of the computed spectrum is not consistent with the results from experiment.  This result is consistent with previous work\cite{kurapothula_hydrogen-bonding_2022}. Much better agreement is obtained when the computed vibrational densities of states are compared with results from inelastic neutron scattering \cite{smrcok_combined_2010, white_structure_2013} (see SI). We therefore conclude that using path integral techniques to accurately model the NQEs is essential when performing these comparisons wsith experiment that determine the level of theory that is required to describe the electronic degrees of freedom accurately. 

\section{Conclusion}
\label{sec:conclusion}
In this work, we have fitted neural network potentials for kaolinite using data extracted from single point calculations that were performed using the revPBE + D3 and revPBE + vdW functionals.  We find that revPBE + vdW does an excellent job of reproducing the experimentally observed static properties of the material.  However, revPBE + D3 is better at reproducing the IR spectrum.  

We also compared the static and dynamic properties that are obtained when a fully-quantum-treatment of nuclei is employed with the results that are obtained from classical simulations. We find that NQEs do not have a substantial impact on the static properties. However, there is a substantial change in the dynamic properties when NQEs are included.

Overall, our work further illustrates the rich physical insight that can be obtained by comparing simulation and experiment. By employing functionals that use different levels of theory researchers can extract meaningful insight into the physics that matter for a particular system.  Furthermore, the advances in machine learning that have taken place over the last 15 years now mean that it is possible to perform these investigations on chemically-complex systems such as kaolinite.
Future work will focus on gauging the influence of long-ranged electrostatics on complex layered systems. These effects will be particularly important at the surface of a material, and to incorporate them will require the use of specialised NNPs~\cite{gao_self-consistent_2022, ko_general-purpose_2021, ko_fourth-generation_2021, yao_tensormol-01_2018}.

\section*{Supplementary Material}

The supplementary material contains further information on the HDNNP model training as well as further static and dynamical results for kaolinite.

\section*{Acknowledgements}
S.S. gratefully acknowledges the support of the Northern Ireland Department for the Economy (NI-DfE). We also appreciate the provision of any and all computational resources used throughout the creation of this work. We thank both the UK Materials and Molecular Modelling Hub for computational resources, which is partially funded by EPSRC (EP/P020194/1) and EP/T022213/1), and the Northern Ireland High Performance Computing (NI-HPC) service funded by EPSRC (EP/T022175). The authors thank Rhys Bunting and Ethan Crawford for helpful comments on the manuscript.

\clearpage

\onecolumngrid

\begin{center}
{\Large \textbf{Supplementary Information}}
\end{center}

\section{Github Repository}
Exemplary input files are provided for a number of the calculations and simulations performed throughout this work. These are available within our  \href{https://github.com/sshepherd637/ClayNN}{github repository}. Efforts have been taken to make these files as `general' as possible, however we recommend anyone wishing to use these models `out of the box' to first familiarise themselves with the I-PI\cite{kapil_i-pi_2019}, LAMMPS\cite{thompson_lammps_2022} and N2P2\cite{singraber_library-based_2019, singraber_parallel_2019} codes.

\section{Overview of ClayNN Model}

We first discuss the atomic descriptors that are used to translate the chemical structure of the kaolinite system to a machine learnable input. We then follow this with a brief description of the architecture and training/expansion procedure of the potentials.

\subsection{Symmetry Functions}
The atomic structure of the kaolinite system is translated to a more `machine-learnable' form of input through the use of atom-centered symmetry functions (ACSFs). Three varieties of these functions were used to provide a suitable environmental descriptor for all atomic configurations. These functions are given within the literature\cite{behler_generalized_2007, behler_atom-centered_2011} as $G^2$, $G^3$ and $G^9$ type symmetry functions respectively. The form of the cutoff function and the symmetry functions are shown here as Eqns \ref{equation:cutoff}, \ref{equation:G2}, \ref{equation:G3} and \ref{equation:G9} respectively.

\begin{equation}
\label{equation:cutoff}
    f_c(r) = \tanh^3(1 - \frac{r}{r_c})
\end{equation}

\begin{equation}
\label{equation:G2}
    G^2_i = \sum_{j\neq i} e^{-\eta(r_{ij}-r_s)^2}f_c(r_{ij})
\end{equation}

\begin{equation}
\label{equation:G3}
    G^3_i = 2^{1-\zeta}\sum_{\substack{j\neq i\\j<k}}(1+\lambda \cos \theta_{ijk})^{\zeta}e^{-\eta[(r_{ij}-r_s)^2 + (r_{ik}-r_s)^2 + (r_{jk}-r_s)^2]}f_c(r_{ij})f_c(r_{ik})f_c(r_{jk})
\end{equation}

\begin{equation}
\label{equation:G9}
    G^9_i = 2^{1-\zeta}\sum_{\substack{j\neq i\\j<k}}(1+\lambda \cos \theta_{ijk})^{\zeta}e^{-\eta[(r_{ij}-r_s)^2 + (r_{ik}-r_s)^2]}f_c(r_{ij})f_c(r_{ik})
\end{equation}

The complete set of symmetry functions we generated is provided within our github repository. We use cutoffs of 8 and 12 Bohr when creating the symmetry function set, yielding a `fine' description of atomic environments at short ranges whilst maintaining distinction over longer distances.

These functions were produced using an iterative python script (which we provide within our github). In order to adequately lower the number of functions used to a more representative set, we utilised the pruning feature within of N2P2. This procedure calculated all symmetry function values encountered within the training data and removed any functions which didn't sufficiently contribute to the distinction of atomic environments. The metric used to determine this is the range of values spanned by the symmetry function when considering all the training data. This process results in a final symmetry function set of 511 functions, a further breakdown of these functions is given in Table. \ref{tab:1} 

\begin{table}[h!]
    \centering
    \begin{tabular}{|c|ccc|c|}
        \hline
        & \multicolumn{3}{c|}{Function} &  \\
        \hline
        Element & $G^2$ & $G^3$ & $G^9$ & Total   \\
        \hline
        H & 29 & 11 & 116 & 156 \\
        O & 39 & 14 & 160 & 213 \\
        Al & 10 & 0 & 52 & 62   \\
        Si & 15 & 4 & 61 & 80   \\
        \hline
    \end{tabular}
    \caption{Final Symmetry Functions sets assigned to each element broken down by function type.}
    \label{tab:1}
\end{table}

\subsection{Architecture and Training}

We use a typical architecture for the elemental neural networks by employing a single input layer with the number of nodes determined by the number of symmetry functions for that element. This is then followed by two hidden layers, each with 25 nodes, and a final output layer consisting of a single node.

We provide the reference free atomic energies, system energies, and atomic forces to the neural network potential in atomic units. These can be seen within the respective input files in the github repository. We further sanitize our data by adjusting the energies provided within the training data. This is achieved through the subtraction of the mean system energy of the dataset from the system energy of every configuration.

The weights are updated through the use of the Kalman filter\cite{blank_adaptive_1994} implemented within N2P2. The training procedure is allowed to run for 25 epochs. After these epochs have completed, we select the most accurate epochs weights, taking care to ensure that the potential has not been overfit to the training data, and use these as the weights for our potential. 

The expansion and improvement of the potentials is achieved through a procedure named \textit{query by committee}, a process first employed by Schran et al\cite{schran_automated_2020}. We trained ten other potentials using the same training data but differing randomization seeds. This procedure enables us to distinguish between inaccuracies within the training data and regions of the potential energy surface that are inadequately sampled through analysis of the standard deviation of the committee predictions. This procedure was repeated until adequately good coverage over all required phase spaces was achieved. This procedure was used initially to create a NVT `stable' potential and then repeated for the NST ensemble.

\subsection{Validation of ClayNN Potentials}
\begin{figure}[h!]
    \centering
    \includegraphics[width=\textwidth]{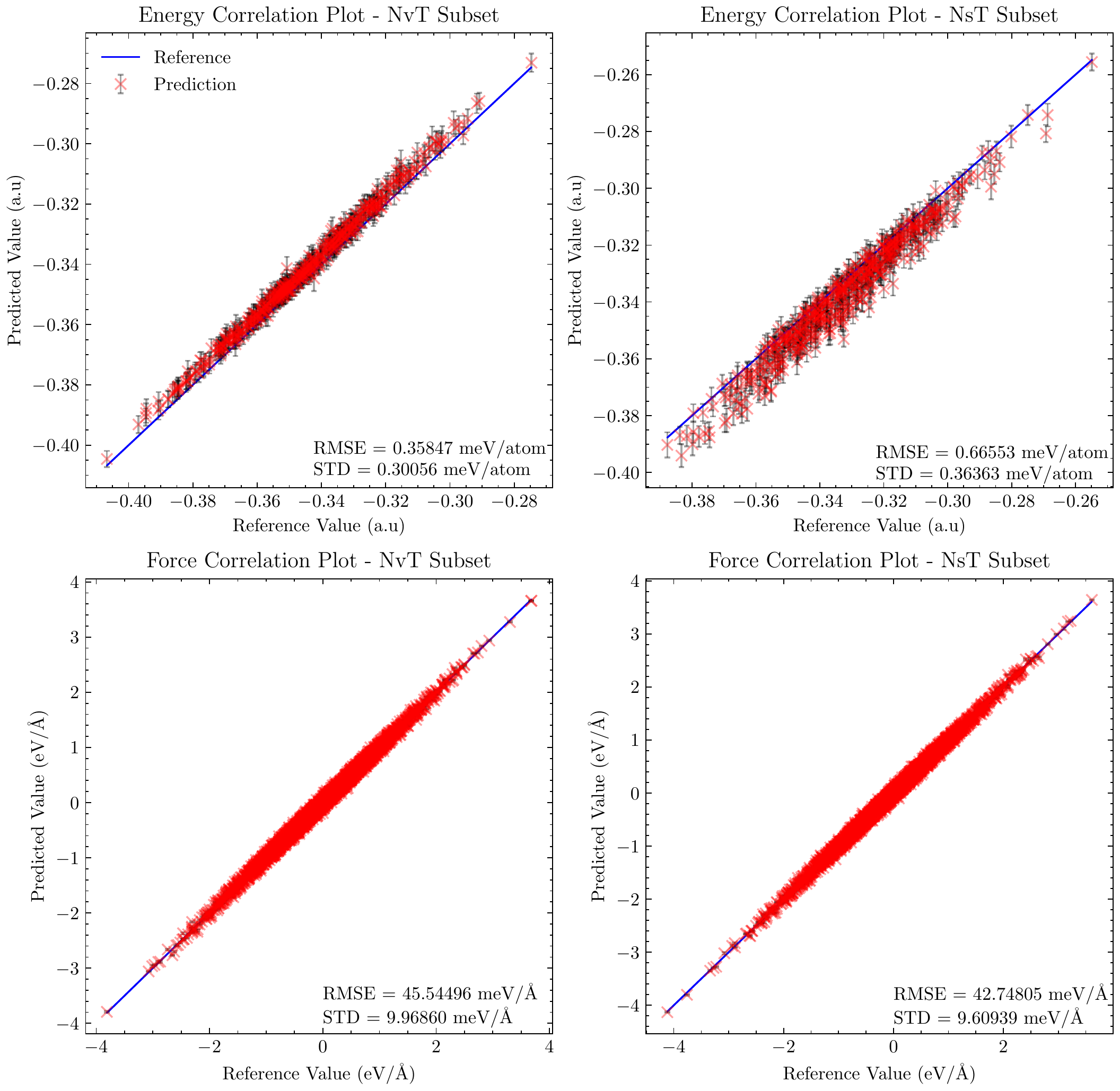}
    \caption{Committee validation plots for the revPBE + D3 potential, ClayNN-D3 with red points and black lines indicating the mean committee prediction and standard deviation of the committee prediction respectively, with the blue lines representing the reference value. Upper plots show energy validation for both NVT (left) and NST (right) while lower plots show force validation for both NVT (left) and NST (right).}
    \label{SI:Figs-revPBED3-Valid}
\end{figure}

\begin{figure}[h!]
    \centering
    \includegraphics[width=\textwidth]{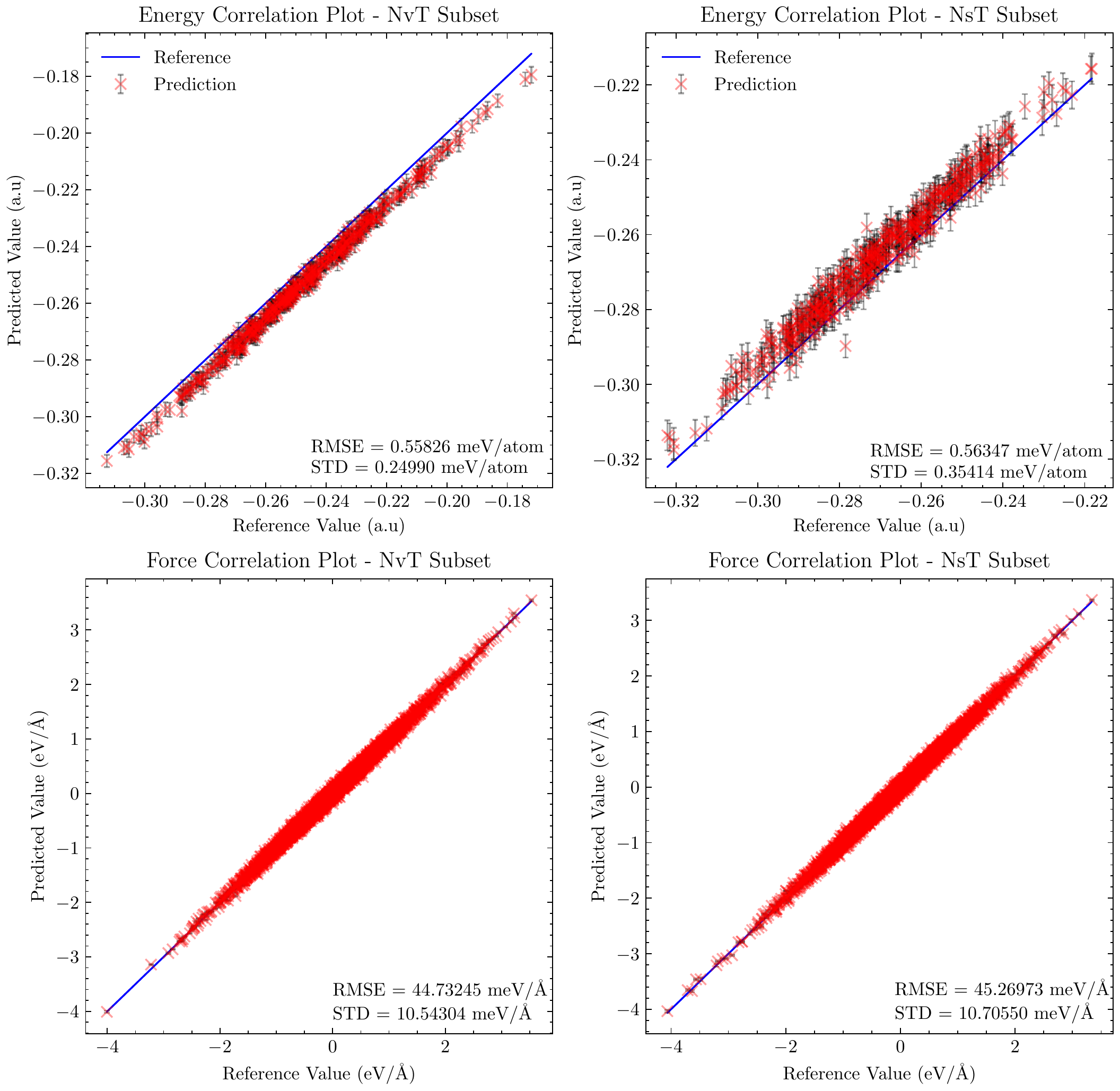}
    \caption{Committee validation plots for the revPBE + vdW potential, ClayNN-vdW with red points and black lines indicating the mean committee prediction and standard deviation of the committee prediction respectively, with the blue lines representing the reference value. Upper plots show energy validation for both NVT (left) and NST (right) while lower plots show force validation for both NVT (left) and NST (right).}
    \label{SI:Figs-revPBEvdW-Valid}
\end{figure}

We validate the models used throughout by the process of query by committee.  We computed the RMSE and standard deviation of the predictions of the committee of potentials. To this end, we randomly sample and recompute 500 frames from both the ClayNN-D3 and ClayNN-vdW driven classical NVT and NST simulations run at 300K. This provides us with a measure of both static and dynamic accuracy of the potentials. These are presented in Fig. \ref{SI:Figs-revPBED3-Valid} and Fig. \ref{SI:Figs-revPBEvdW-Valid} for ClayNN-D3 and ClayNN-vdW respectively. We split these into frames from both the NVT and NST simulations to provide clearer distinction between the two ensembles and the regions of phase space which they explore. A similar procedure for the quantum mechanical simulations was not completed. This recalculation with the baseline DFT functional would incur the same computational expense as running \textit{ab initio} path integral molecular dynamics and ultimately defeat the purpose of using MLIPs.

As can be seen from both Fig. \ref{SI:Figs-revPBED3-Valid} and Fig. \ref{SI:Figs-revPBEvdW-Valid}, the decrease of accuracy moving from NVT to NST is low. The RMSE increases whilst remaining relatively low for both the predicted energy and forces of the kaolinite system studied using both potentials. This is reasonably sensible, as the change of ensemble from NVT, which ensured a constant cell volume, to the NST ensemble allowed for a much larger region of chemical phase space to be explored. This results in slightly larger errors and standard deviation values across both potentials.

We note the relatively large error bars associated with the energetic predictions for both ClayNN-D3 and ClayNN-vdW with NST predictions. We computed the errors as the standard deviation of the committee potentials at 0.36 meV/atom (54.64\%) and 0.35 meV/atom (62.85\%) respectively. We attribute this to the disparity of available system energies within the reference data. The amount of data available equates to $N$ and $3n(N)$ points respectively where $N$ is the number of frames and $n$ the number of atoms. Ultimately, while the energy predictions are a useful indicator of phase space coverage, they provide a limited understanding into the performance of the potential when driving molecular dynamics.  Regardless of this fact, we present the energy predictions of the potentials as a general `sign of wellness' for both NVT and NST simulations at the conditions studied.

In contrast to the energy predictions, the force predictions of both potentials remains very accurate throughout both ensembles. We computed a maximum error of 45.54 meV/\r{A} and 45.27 meV/\r{A} corresponding to the ClayNN-D3 NVT and ClayNN-vdW NST predictions respectively. In addition to this, we report extremely low standard deviation values for both potentials for the sampled trajectory at 9.97 meV/\r{A} (21.89\%)and 10.7 meV/\r{A} (23.64\%) respectively. This corresponds to well covered regions of the phase space under the conditions we study within this work, and allows us to confidently use these potentials as surrogates to DFT level calculations. 

Overall, we observed that throughout all simulations, the potential was extremely accurate when predicting on structures with cell parameters similar to those that were prevalent within the training data. As such, the energy and force errors are extremely low throughout these simulations and we observe very few (<5) symmetry function extrapolation warnings during these simulations. When we consider NST simulations, it is of little surprise that deformations away from the cell parameters that dominate the training data incur an increase in the errors associated with the predicted energy and forces of the system. In order to correct this, a thorough coverage of the phase space accessible to the system when simulating using NST at 300 K would need to be accounted for and shown to the potential. We deem this as unnecessary, as ultimately, we see no statistically significant increase in the errors of the predicted atomic forces when simulated using the NST ensemble. We therefore conclude that both ClayNN-D3 and ClayNN-vdW are suitable for both ensembles in their current form.

\section{Kaolinite Characterisation}

\subsection{Cell Parameters}
All reported cell parameters were obtained from geometry/cell optimization calculations using the reference DFT functionals or respective NNPs, or obtained as ensemble averages from NST molecular dynamics simulations. DFT cell optimization was completed using the 2 x 2 x 2 kaolinite supercell with CP2K while NNP optimization was completed using LAMMPS. We provide input files for both of these calculations within our github. 

In addition, we report that during simulations involving the ClayNN-D3 potential with and without NQEs, we observed definite warpings of the cell, with extreme deviations of cell parameter C, $\alpha$ and $\beta$ at prolonged simulation times (> 1 ns). We attribute the revPBE + D3 functional and it's description of the interlayer interactions, as this was not something we observed when using the ClayNN-vdW potential. When we included NQEs within these simulations, this warping occurred on a shorter timescale (> 250 ps), and as a result, we attribute this to the description of the intermolecular interactions between the kaolinite layers. We note that we did not include any portion of the trajectory after this phenomenon was observed within the statistical averages when reporting cell parameters or sampling frames for simulations involving dynamical properties.

\subsection{Radial Distribution Functions}
\begin{figure}[ht!]
    \centering
    \includegraphics[width=\textwidth,height=0.9\textheight]{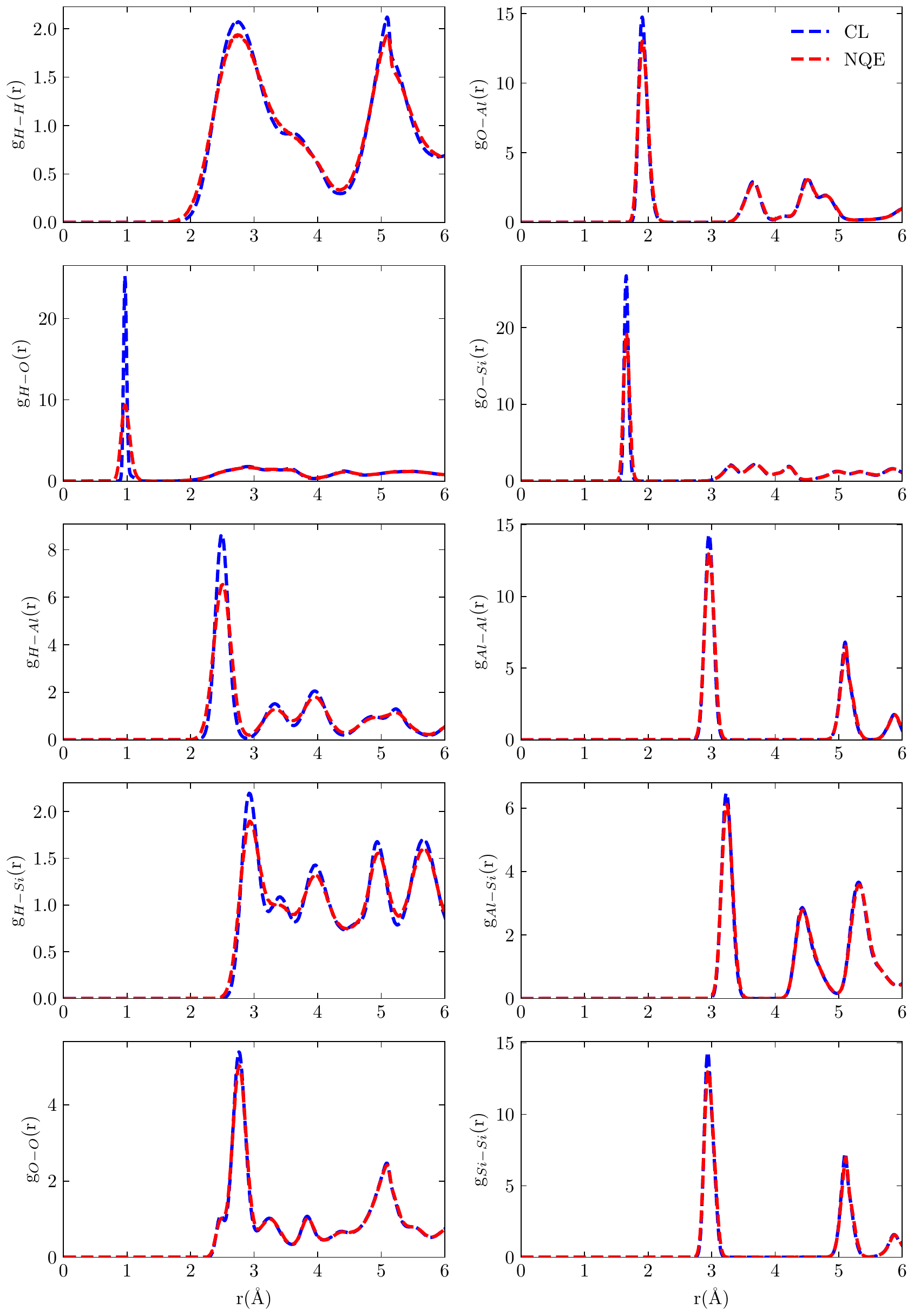}
    \caption{Radial distribution functions (RDFs) for all elemental pairs within the kaolinite system calculated from revPBE + D3 trained neural network driven molecular dynamics. Blue dashed lines show the RDF calculated from classical molecular dynamics while the red dashed lines show the RDF calculated when accounting for quantum effects using PIMD.}
    \label{SI:Figs-revPBED3-RDFS}
\end{figure}

\begin{figure}[ht!]
    \centering
    \includegraphics[width=\textwidth,height=0.9\textheight]{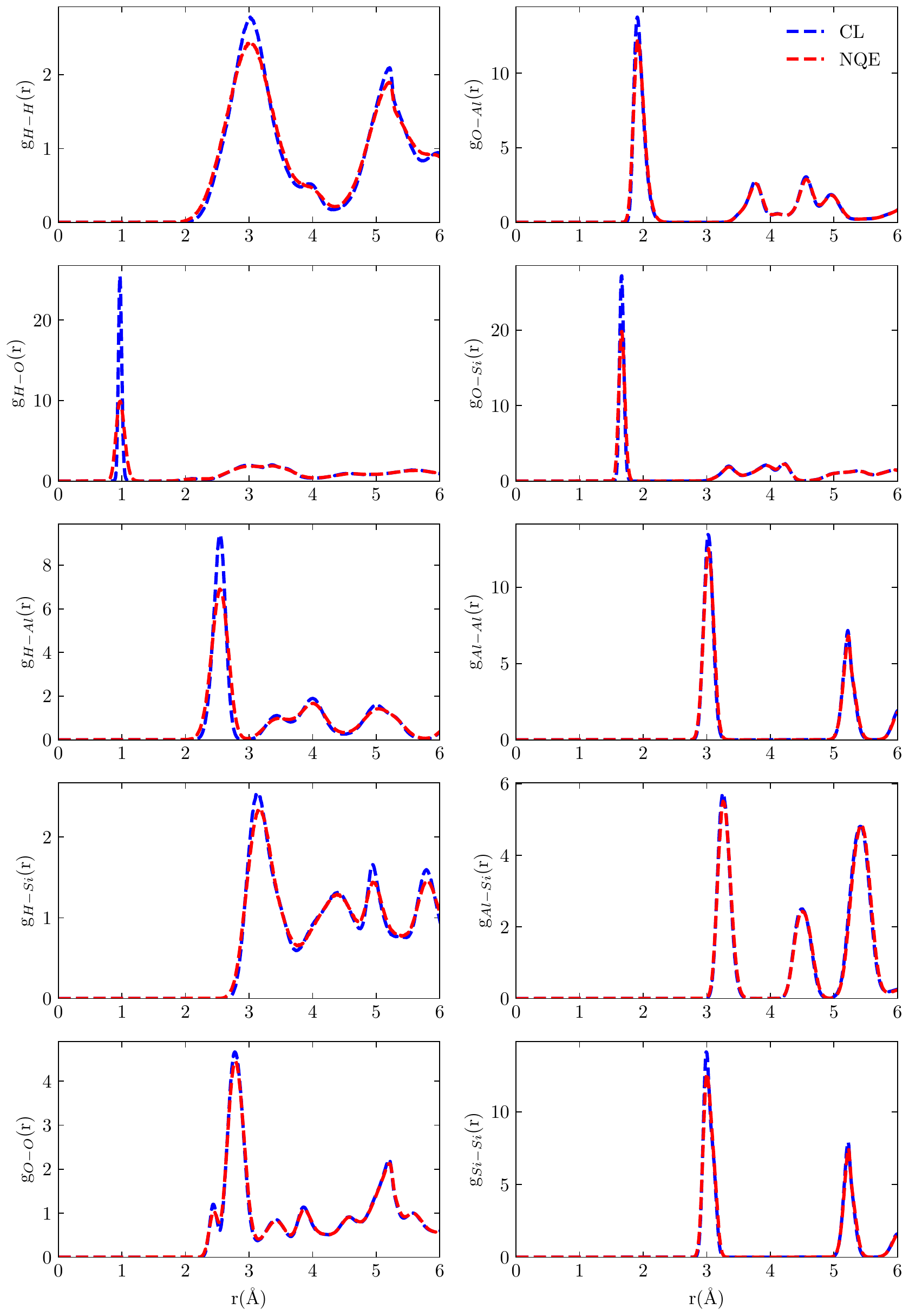}
    \caption{Radial distribution functions (RDFs) for all elemental pairs within the kaolinite system calculated from revPBE + vdW trained neural network driven molecular dynamics. Blue dashed lines show the RDF calculated from classical molecular dynamics while the red dashed lines show the RDF calculated when accounting for quantum effects using PIMD.}
    \label{SI:Figs-revPBEvdW-RDFS}
\end{figure}

\begin{figure}[ht!]
    \centering
    \includegraphics[width=\textwidth,height=0.9\textheight]{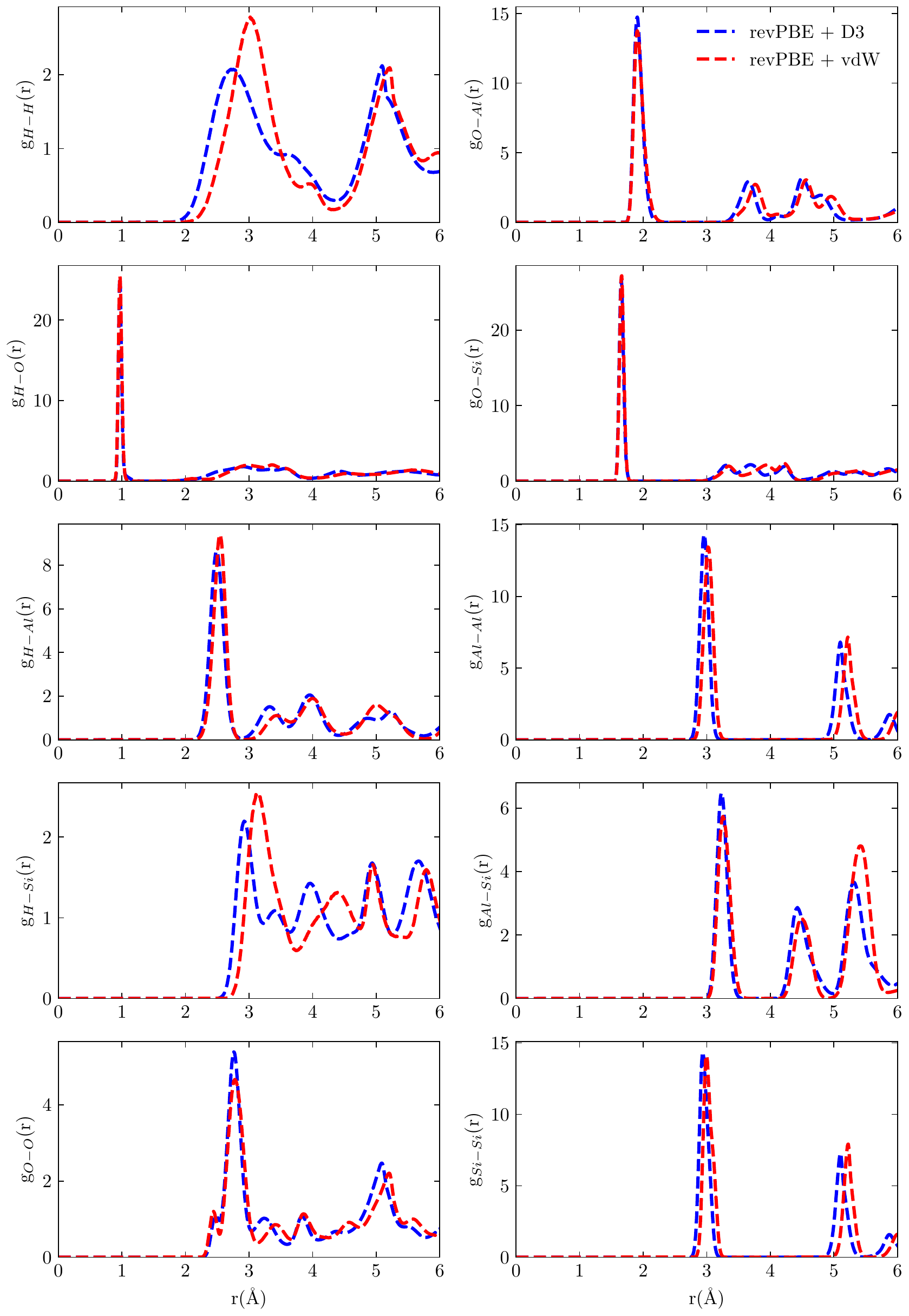}
    \caption{Radial distribution functions (RDFs) for all elemental pairs within the kaolinite system calculated using both the revPBE + D3 (blue) and revPBE + vdW (red) functionals. All functionals shown here are obtained from classical molecular dynamics, in order to isolate the effects of different dispersion corrections when simulating the kaolinite system.}
    \label{SI:Figs-D3vsvdW-RDFS}
\end{figure}

Figures \ref{SI:Figs-revPBED3-RDFS} and \ref{SI:Figs-revPBEvdW-RDFS} show the radial distribution functions for all elemental pairs within kaolinite computed using ClayNN-D3 and ClayNN-vdW respectively. The dashed blue lines represent the classically obtained results while the dashed red lines represent those calculated when we include NQEs. Figure \ref{SI:Figs-D3vsvdW-RDFS} shows the classically obtained RDFs for ClayNN-D3 and ClayNN-vdW as the dashed blue line and dashed red line respectively. The main results from both Figure \ref{SI:Figs-revPBED3-RDFS} and \ref{SI:Figs-revPBEvdW-RDFS} have been discussed within the main text. We plot Figure \ref{SI:Figs-D3vsvdW-RDFS} to more clearly compare the differences in modelling provided by the respective functionals. It can be easily seen that both functionals provide a similar description of most interactomic interactions within the system. The largest differences are seen in g$_{\text{H--H}}(r)$ and g$_{\text{H--Si}}(r)$, which show a substantial difference in the location of the first maxima for both functions of around 0.1 - 0.2 \r{A} and further differences in the more distant regions of both functions.

\subsection{Vibrational Density of States}
All reported phonon densities of states (PDOS) were obtained through the use of phonopy interfaced with CP2K. The workflow that we followed during this process is available on our github, along with a short configuration file we used to create the final plots. We attribute the differences between the NNP predicted phonon density of states and that computed using DFT as being due to the intrinsic error incurred through the learning procedure.

\begin{figure}[h!]
    \centering
    \includegraphics[width=\textwidth]{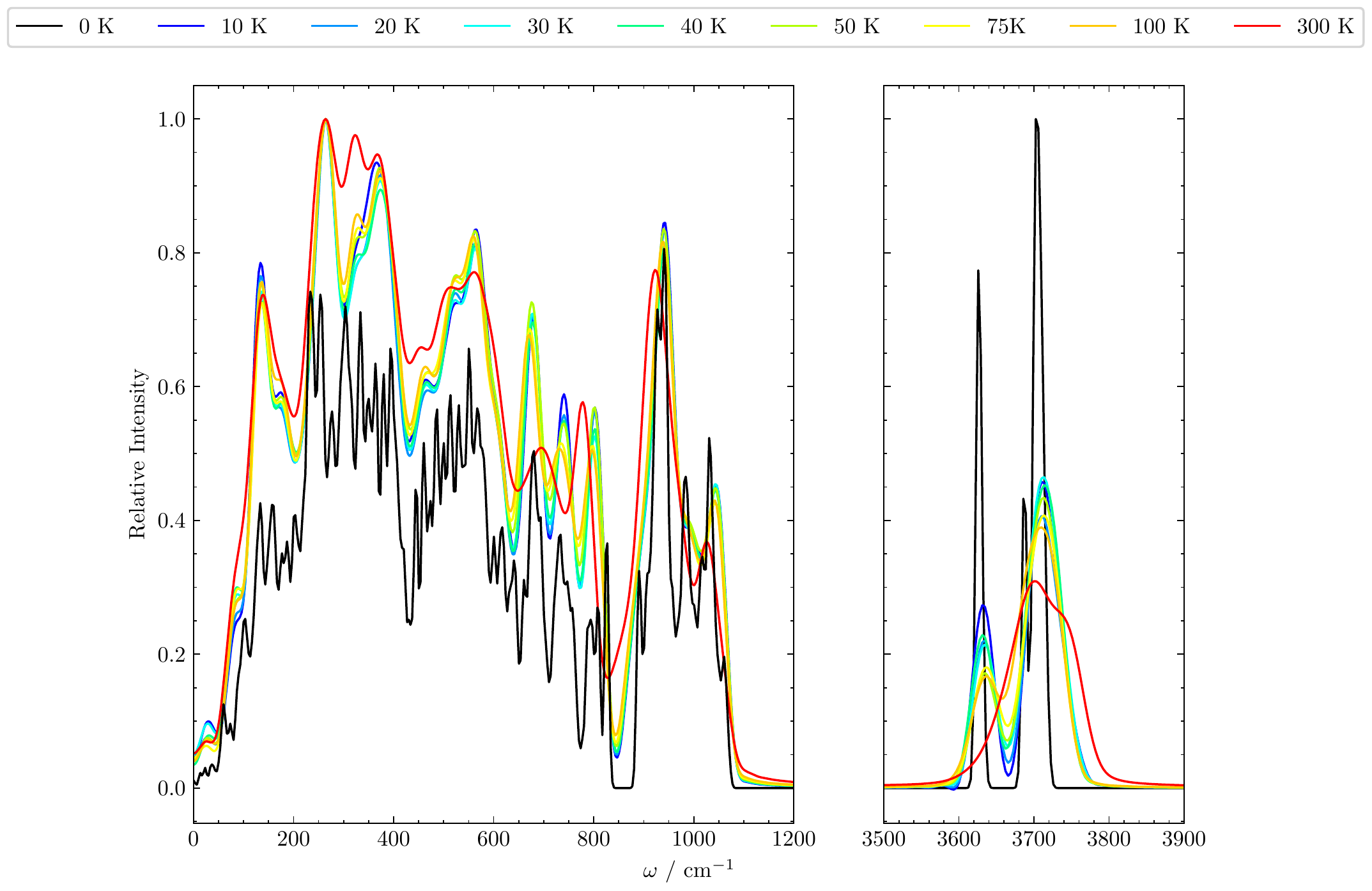}
    \caption{Computed VDOS for the various studied temperatures using ClayNN-D3 plotted against the computed PDOS. We present the low frequency ($<$ 1200 cm$^{-1}$ region and high frequency ($>$ 3500 cm$^{-1}$) on the left and right of the figure respectively.}
    \label{fig:TempEffectsSI}
\end{figure}

Figure \ref{fig:TempEffectsSI} shows the vibrational density of states (VDOS) computed using simulations driven by ClayNN-D3 at a range of temperatures. These temperatures range from 0 K (as in, the PDOS) to 300 K. A number of these were omitted from the figure presented within the main text for clarity. We observe significant degradation of the peak positions due to thermal effects upon an increase in temperature of 50 K. This is most clearly seen within the high frequency region of the VDOS (the right panel), where clear distinction of the two major peaks is sufficiently blurred at 50 K to leave it difficult to assign with confidence.  

\begin{figure}
    \centering
    \includegraphics{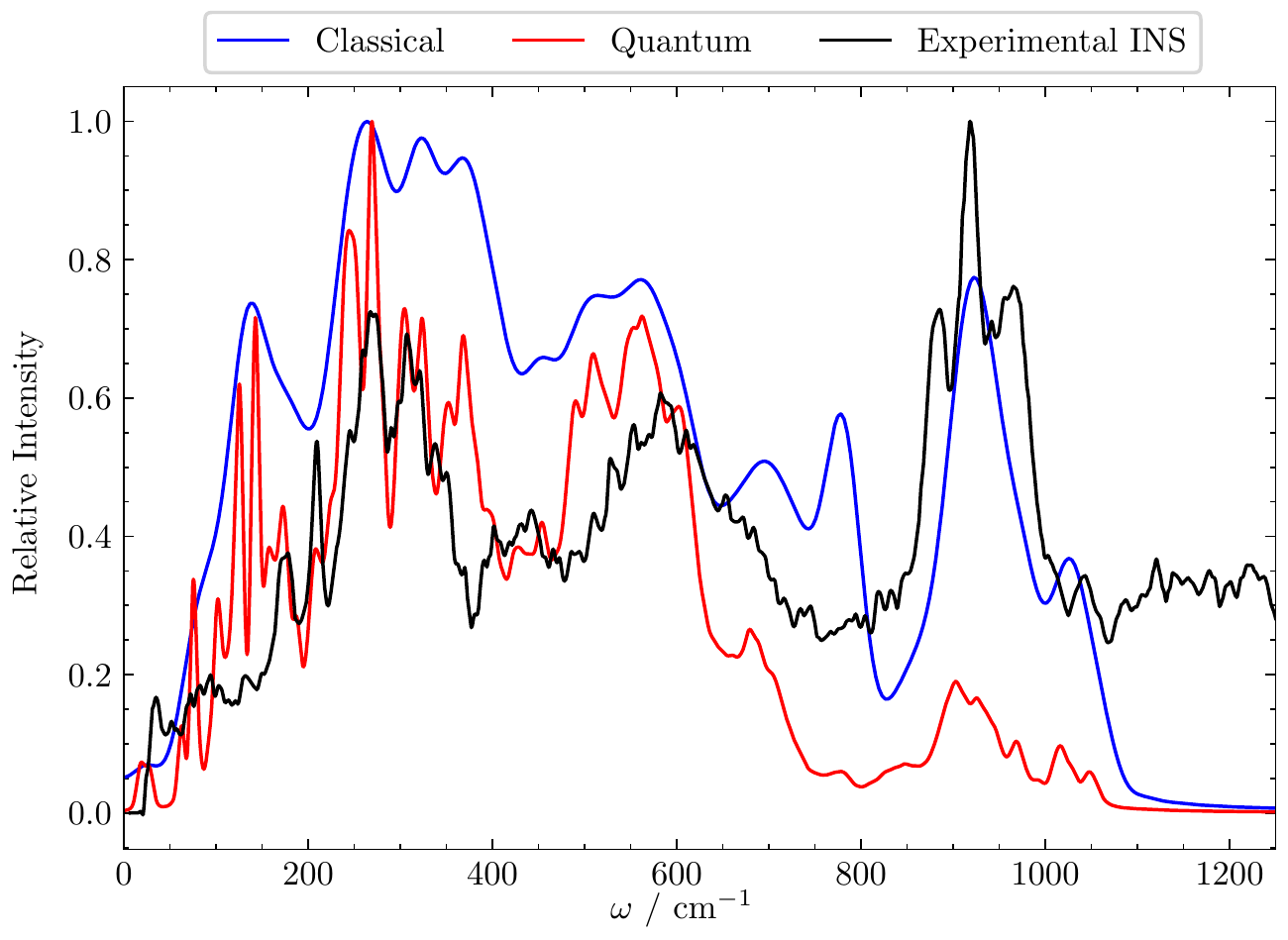}
    \caption{Vibrational density of states of kaolinite computed using ClayNN-D3 using both classical molecular dynamics (blue line) and PACMD (red line) pictured alongside the experimental inelastic neutron scattering spectrum reproduced from Ref. \citenum{smrcok_combined_2010}, with the permission of AIP Publishing (black line).}
    \label{fig:SI-INSSpectra}
\end{figure}

As mentioned within the main text, we present the computed VDOS (both classical and quantum mechanical) and the experimental inelastic neutron scattering spectrum\cite{smrcok_combined_2010}in figure \ref{fig:SI-INSSpectra}. We find a pleasantly surprising agreement between the computed spectra (both classical and quantum mechanical) and the experimental INS. 

We consider the classically computed VDOS (blue line) first. While suffering from a loss of peak precision, it is clear that the general shape of the INS is reproduced with reasonable accuracy. In particular, we note the existence of the peak around 925 cm$^{-1}$, occurring in the region assigned as the Al--O--H in-plane bending modes of the interlayer hydroxyl groups by Smr\v{c}ok \textit{et al}\cite{smrcok_combined_2010}. While distinction of individual contributing modes is not possible from our results, it is likely that all contributions due to differing strengths in hydrogen bonding exhibit themselves within the broadened peak in this region in the classical VDOS. This is further supported by the broadened peak within the O--H stretching region of the VDOS pictured within the main text, and provides yet another example of a phenomenon that can be attributed to the poor description of the O--H bond given by both studied dispersion corrected functionals. For modes below this region (< 800 cm$^{-1}$), the classical VDOS correctly reproduces the position of a number of peaks, arising from complex modes within the kaolinite system, but cannot account for the relative intensities of these modes easily. 

By contrast, the quantum mechanical VDOS shows a remarkable accuracy to the overall INS, but cannot reproduce the modes within the 850 - 1000 cm$^{-1}$ region with sufficient intensity. The agreement with experiment below 800 cm$^{-1}$ is encouraging, and suggests that the addition of NQEs affords an increase in the peak precision of these complex modes. When considering the intense modes within the 850 - 1000 cm$^{-1}$ region, we again find good agreement with the peak positions as computed by ClayNN-D3. In addition, we clearly show the existence of three peaks, corresponding to the experimentally observed three peaks within the INS spectra. This result is interesting and suggests that the addition of NQEs allows for the correct distinction of the individual contributing Al--O--H vibrational modes.

For a further discussion of the origin of these peaks, we direct the reader to the work of both Smr\v{c}ok and White\cite{smrcok_combined_2010, white_structure_2013}

\begin{figure}
    \centering
    \includegraphics{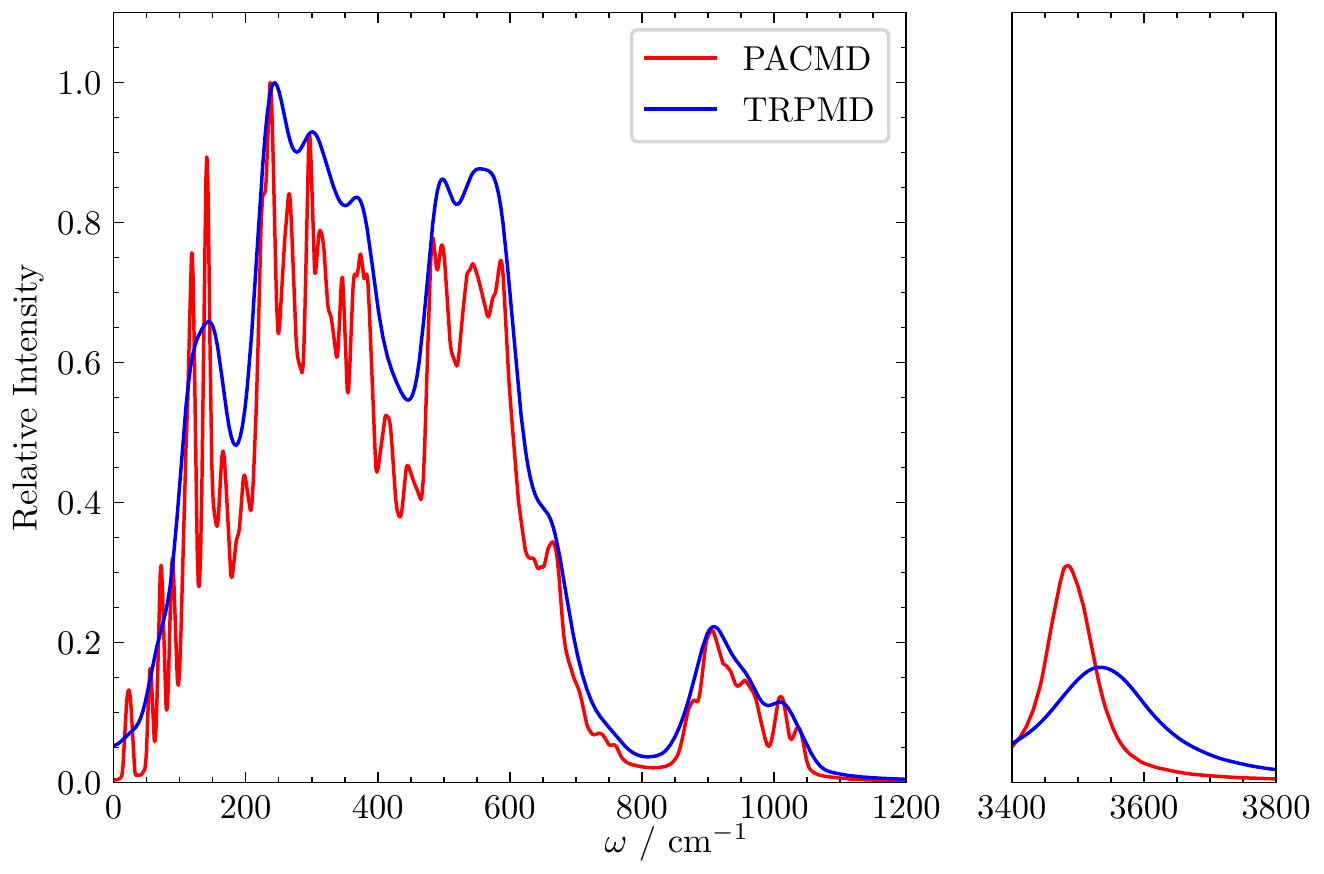}
    \caption{Vibrational density of states computed using ClayNN-vdW within both the TRPMD (blue line) and PACMD (red line) formalism of path integral molecular dynamics.}
    \label{fig:SI-revPBEvdW-PACMDvsTRPMD}
\end{figure}

When including NQEs in our study of the dynamical properties of kaolinite, we used both thermostatted ring polymer molecular dynamics\cite{habershon_ring-polymer_2013, rossi_how_2014} (TRPMD) and partially adiabatic centroid molecular dynamics\cite{hone_comparative_2006} (PACMD). Results obtained using PACMD are presented within the main text. The results obtained using TRPMD suffered from a well known `peak broadening'\cite{lbenson_which_2020} problem and were unsuitable for presentation. Figure \ref{fig:SI-revPBEvdW-PACMDvsTRPMD} shows the differences in vibrational spectrum obtained using PACMD and TRPMD as the red and blue lines respectively.

\begin{figure}
    \centering
    \includegraphics{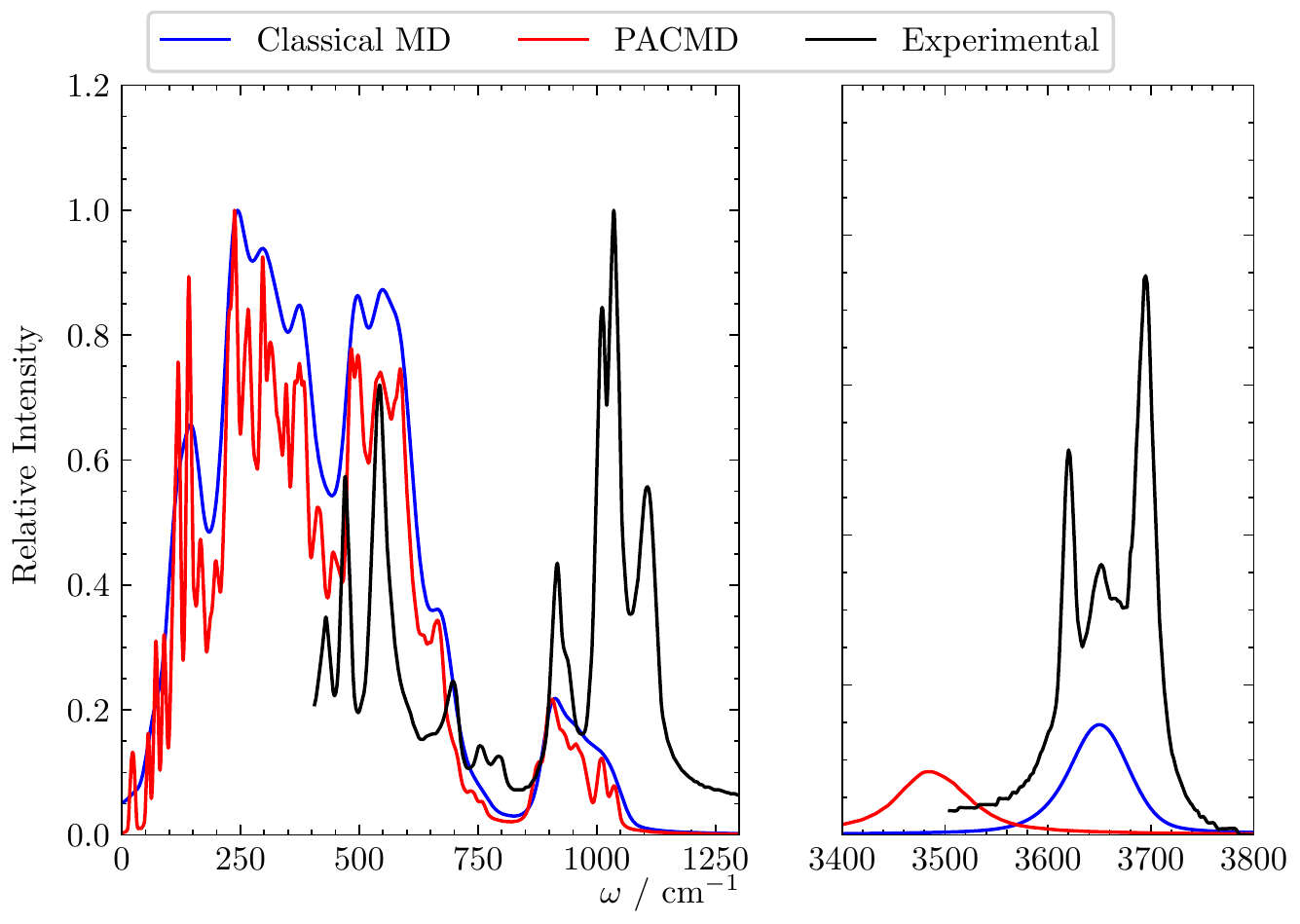}
    \caption{Vibrational density of states of kaolinite computed using ClayNN-vdW within both classical mechanics (blue line) and path integrals (red line). We include the experimentally obtained IR spectra from \citenum{madejova_ir_2017} to highlight the differences between the computationally obtained result and experiment when using this parameterization of \textit{ab initio} theory.}
    \label{fig:SI-revPBEvdW-VDOS}
\end{figure}

Due to our conclusion that ClayNN-D3, and by extension, revPBE + D3, provided a better description of the vibrational properties of the kaolinite systme, we did not present any results obtained using ClayNN-vdW. Figure \ref{fig:SI-revPBEvdW-VDOS} shows the VDOS computed classically and using PACMD in the blue and red lines respectively. These results were also obtained from simulations performed at 300 K. We note the relatively good agreement in peak position of the classical VDOS obtained using ClayNN-vdW to experimental results (pictured in black within the high frequency region). It is useful to reiterate that this result is entirely fortuitous, as the classically obtained results do not consider the quantum mechanical nature of the nuclei within the system. The agreement with experiment is therefore tainted as a result of this fact. Finally. figure \ref{fig:SI-revPBEvdW-PACMDvsTRPMD} shows the VDOS computed using PACMD and TRPMD as the red and blue lines respectively.

\section{Miscellaneous}
\subsection{Path Integral Bead Convergence}

\begin{figure}
    \centering
    \includegraphics{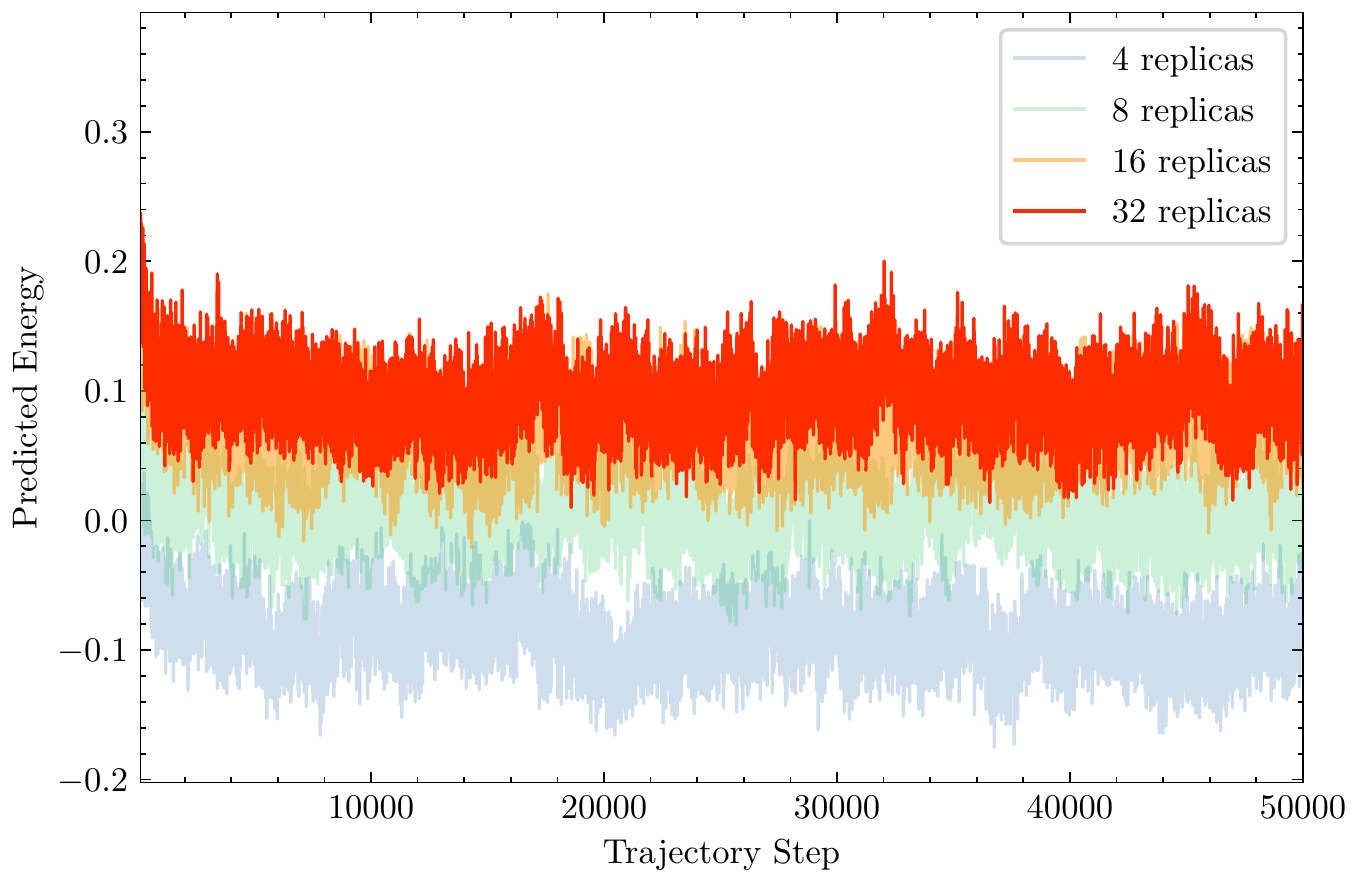}
    \caption{Total system energy of NVT simulations separated based on the number of replicas used within the simulation. We present these as 4 replica (blue lines), 8 replica (green lines), 16 replica (yellow lines) and 32 replica (red lines) simulations. The predicted energy is presented rather than the total energy, as the mean system energy is subtracted prior to training.}
    \label{fig:SI-BeadConvergence}
\end{figure}

To ensure we used a sufficient number of replicas when performing path integral molecular dynamics, we converged the number of beads with respect to the total system energy. We simulated the system in the NVT ensemble using 4, 8, 16 and 32 beads. The difference between 16 and 32 beads was calculated via the difference of the mean total energy, which we compute as 0.02 atomic units. As the total energy was computed as the referential energy within the ClayNN-D3 potential, we add this back to the values before computation of the percentage difference between the values. We present the system energy trajectories for these simulations within Fig \ref{fig:SI-BeadConvergence}.

\clearpage 

\vspace*{\fill}
\vspace*{\fill}\clearpage

\vspace*{\fill}
\vspace*{\fill}\clearpage

\clearpage
\vspace*{\fill}
\vspace*{\fill}\clearpage

\begin{figure}
    \centering
    \includegraphics{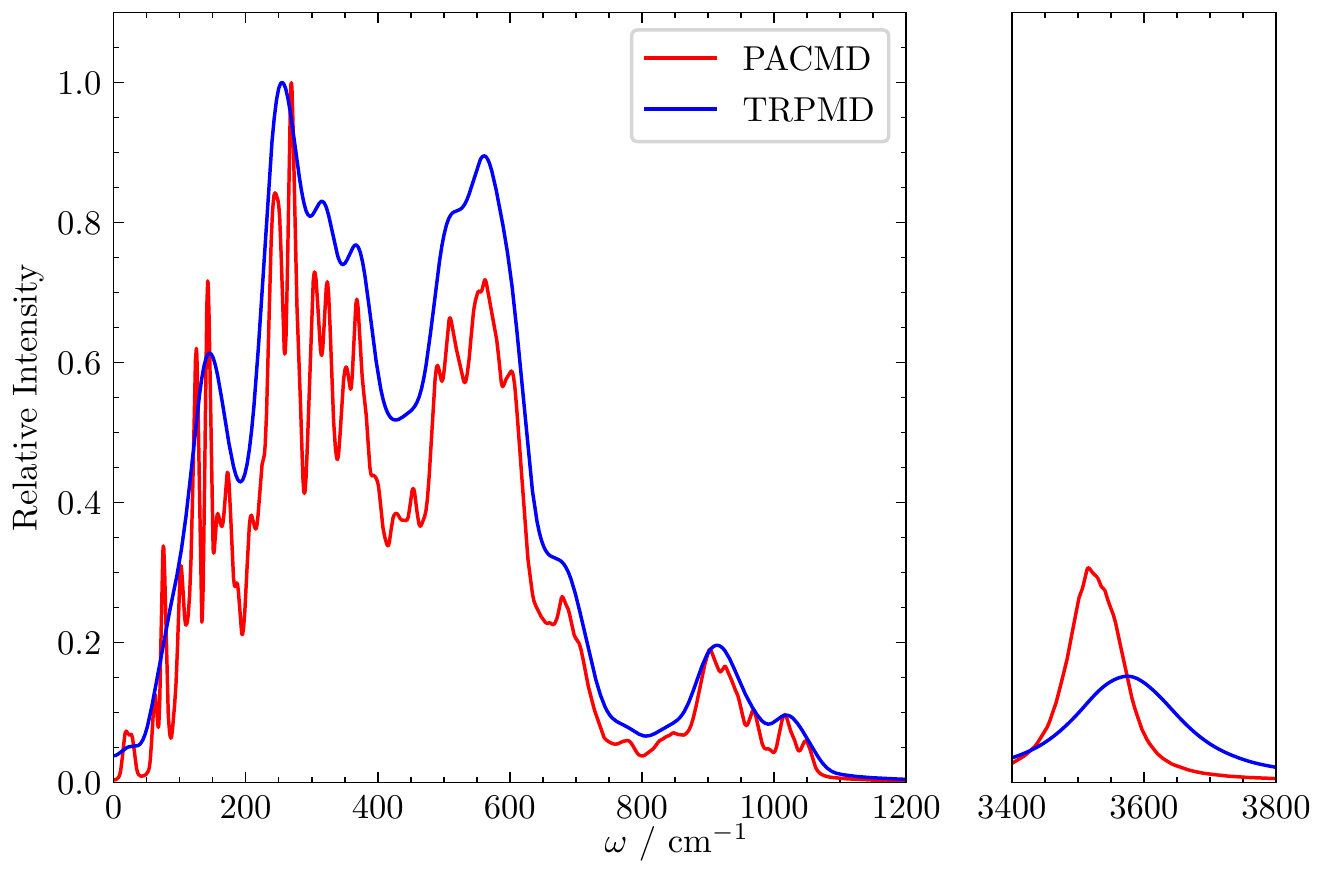}
    \caption{Vibrational density of states computing using ClayNN-D3 within both the TRPMD (blue line) and PACMD (red line) formalism of path integral molecular dynamics.}
    \label{fig:SI-revPBED3-VDOS}
\end{figure}

\begin{thebibliography}{106}%
\makeatletter
\providecommand \@ifxundefined [1]{%
 \@ifx{#1\undefined}
}%
\providecommand \@ifnum [1]{%
 \ifnum #1\expandafter \@firstoftwo
 \else \expandafter \@secondoftwo
 \fi
}%
\providecommand \@ifx [1]{%
 \ifx #1\expandafter \@firstoftwo
 \else \expandafter \@secondoftwo
 \fi
}%
\providecommand \natexlab [1]{#1}%
\providecommand \enquote  [1]{``#1''}%
\providecommand \bibnamefont  [1]{#1}%
\providecommand \bibfnamefont [1]{#1}%
\providecommand \citenamefont [1]{#1}%
\providecommand \href@noop [0]{\@secondoftwo}%
\providecommand \href [0]{\begingroup \@sanitize@url \@href}%
\providecommand \@href[1]{\@@startlink{#1}\@@href}%
\providecommand \@@href[1]{\endgroup#1\@@endlink}%
\providecommand \@sanitize@url [0]{\catcode `\\12\catcode `\$12\catcode
  `\&12\catcode `\#12\catcode `\^12\catcode `\_12\catcode `\%12\relax}%
\providecommand \@@startlink[1]{}%
\providecommand \@@endlink[0]{}%
\providecommand \url  [0]{\begingroup\@sanitize@url \@url }%
\providecommand \@url [1]{\endgroup\@href {#1}{\urlprefix }}%
\providecommand \urlprefix  [0]{URL }%
\providecommand \Eprint [0]{\href }%
\providecommand \doibase [0]{http://dx.doi.org/}%
\providecommand \selectlanguage [0]{\@gobble}%
\providecommand \bibinfo  [0]{\@secondoftwo}%
\providecommand \bibfield  [0]{\@secondoftwo}%
\providecommand \translation [1]{[#1]}%
\providecommand \BibitemOpen [0]{}%
\providecommand \bibitemStop [0]{}%
\providecommand \bibitemNoStop [0]{.\EOS\space}%
\providecommand \EOS [0]{\spacefactor3000\relax}%
\providecommand \BibitemShut  [1]{\csname bibitem#1\endcsname}%
\let\auto@bib@innerbib\@empty
\bibitem [{\citenamefont {Murray}(1991)}]{murray_overview_1991}%
  \BibitemOpen
  \bibfield  {author} {\bibinfo {author} {\bibfnamefont {H.~H.}\ \bibnamefont
  {Murray}},\ }\bibfield  {title} {\enquote {\bibinfo {title} {Overview —
  clay mineral applications},}\ }\href {\doibase 10.1016/0169-1317(91)90014-Z}
  {\bibfield  {journal} {\bibinfo  {journal} {Applied Clay Science}\ }\textbf
  {\bibinfo {volume} {5}},\ \bibinfo {pages} {379--395} (\bibinfo {year}
  {1991})}\BibitemShut {NoStop}%
\bibitem [{\citenamefont {Adams}\ and\ \citenamefont
  {McCabe}(2006)}]{adams_clay_2006}%
  \BibitemOpen
  \bibfield  {author} {\bibinfo {author} {\bibfnamefont {J.~M.}\ \bibnamefont
  {Adams}}\ and\ \bibinfo {author} {\bibfnamefont {R.~W.}\ \bibnamefont
  {McCabe}},\ }\bibfield  {title} {\enquote {\bibinfo {title} {Clay {Minerals}
  as {Catalysts}},}\ }in\ \href {\doibase 10.1016/S1572-4352(05)01017-2} {\emph
  {\bibinfo {booktitle} {Developments in {Clay} {Science}}}},\ \bibinfo
  {series} {Handbook of {Clay} {Science}}, Vol.~\bibinfo {volume} {1},\
  \bibinfo {editor} {edited by\ \bibinfo {editor} {\bibfnamefont
  {F.}~\bibnamefont {Bergaya}}, \bibinfo {editor} {\bibfnamefont {B.~K.~G.}\
  \bibnamefont {Theng}}, \ and\ \bibinfo {editor} {\bibfnamefont
  {G.}~\bibnamefont {Lagaly}}}\ (\bibinfo  {publisher} {Elsevier},\ \bibinfo
  {year} {2006})\ pp.\ \bibinfo {pages} {541--581}\BibitemShut {NoStop}%
\bibitem [{\citenamefont {Ferris}(2005)}]{ferris_mineral_2005}%
  \BibitemOpen
  \bibfield  {author} {\bibinfo {author} {\bibfnamefont {J.~P.}\ \bibnamefont
  {Ferris}},\ }\bibfield  {title} {\enquote {\bibinfo {title} {Mineral
  {Catalysis} and {Prebiotic} {Synthesis}: {Montmorillonite}-{Catalyzed}
  {Formation} of {RNA}},}\ }\href {\doibase 10.2113/gselements.1.3.145}
  {\bibfield  {journal} {\bibinfo  {journal} {Elements}\ }\textbf {\bibinfo
  {volume} {1}},\ \bibinfo {pages} {145--149} (\bibinfo {year}
  {2005})}\BibitemShut {NoStop}%
\bibitem [{\citenamefont {Johns}(1979)}]{johns_clay_1979}%
  \BibitemOpen
  \bibfield  {author} {\bibinfo {author} {\bibfnamefont {W.~D.}\ \bibnamefont
  {Johns}},\ }\bibfield  {title} {\enquote {\bibinfo {title} {Clay {Mineral}
  {Catalysis} and {Petroleum} {Generation}},}\ }\href@noop {} {\bibfield
  {journal} {\bibinfo  {journal} {Annual Reviews of Earth Planet Science}\ ,\
  \bibinfo {pages} {183--198}} (\bibinfo {year} {1979})}\BibitemShut {NoStop}%
\bibitem [{\citenamefont {Cecilia}\ and\ \citenamefont
  {Jiménez-Gómez}(2021)}]{cecilia_catalytic_2021}%
  \BibitemOpen
  \bibfield  {author} {\bibinfo {author} {\bibfnamefont {J.~A.}\ \bibnamefont
  {Cecilia}}\ and\ \bibinfo {author} {\bibfnamefont {C.~P.}\ \bibnamefont
  {Jiménez-Gómez}},\ }\bibfield  {title} {\enquote {\bibinfo {title}
  {Catalytic {Applications} of {Clay} {Minerals} and {Hydrotalcites}},}\ }\href
  {\doibase 10.3390/catal11010068} {\bibfield  {journal} {\bibinfo  {journal}
  {Catalysts}\ }\textbf {\bibinfo {volume} {11}},\ \bibinfo {pages} {68}
  (\bibinfo {year} {2021})}\BibitemShut {NoStop}%
\bibitem [{\citenamefont {Bhattacharyya}\ and\ \citenamefont
  {Gupta}(2008)}]{bhattacharyya_adsorption_2008}%
  \BibitemOpen
  \bibfield  {author} {\bibinfo {author} {\bibfnamefont {K.~G.}\ \bibnamefont
  {Bhattacharyya}}\ and\ \bibinfo {author} {\bibfnamefont {S.~S.}\ \bibnamefont
  {Gupta}},\ }\bibfield  {title} {\enquote {\bibinfo {title} {Adsorption of a
  few heavy metals on natural and modified kaolinite and montmorillonite: {A}
  review},}\ }\href {\doibase 10.1016/j.cis.2007.12.008} {\bibfield  {journal}
  {\bibinfo  {journal} {Advances in Colloid and Interface Science}\ }\textbf
  {\bibinfo {volume} {140}},\ \bibinfo {pages} {114--131} (\bibinfo {year}
  {2008})}\BibitemShut {NoStop}%
\bibitem [{\citenamefont {Struijk}, \citenamefont {Rocha},\ and\ \citenamefont
  {Detellier}(2017)}]{struijk_novel_2017}%
  \BibitemOpen
  \bibfield  {author} {\bibinfo {author} {\bibfnamefont {M.}~\bibnamefont
  {Struijk}}, \bibinfo {author} {\bibfnamefont {F.}~\bibnamefont {Rocha}}, \
  and\ \bibinfo {author} {\bibfnamefont {C.}~\bibnamefont {Detellier}},\
  }\bibfield  {title} {\enquote {\bibinfo {title} {Novel thio-kaolinite
  nanohybrid materials and their application as heavy metal adsorbents in
  wastewater},}\ }\href {\doibase 10.1016/j.clay.2017.09.024} {\bibfield
  {journal} {\bibinfo  {journal} {Applied Clay Science}\ }\textbf {\bibinfo
  {volume} {150}},\ \bibinfo {pages} {192--201} (\bibinfo {year}
  {2017})}\BibitemShut {NoStop}%
\bibitem [{\citenamefont {Crasto~de Lima}, \citenamefont {Miwa},\ and\
  \citenamefont {Miranda}(2017)}]{crasto_de_lima_retention_2017}%
  \BibitemOpen
  \bibfield  {author} {\bibinfo {author} {\bibfnamefont {F.~D.}\ \bibnamefont
  {Crasto~de Lima}}, \bibinfo {author} {\bibfnamefont {R.~H.}\ \bibnamefont
  {Miwa}}, \ and\ \bibinfo {author} {\bibfnamefont {C.~R.}\ \bibnamefont
  {Miranda}},\ }\bibfield  {title} {\enquote {\bibinfo {title} {Retention of
  contaminants {Cd} and {Hg} adsorbed and intercalated in aluminosilicate
  clays: {A} first principles study},}\ }\href {\doibase 10.1063/1.5009585}
  {\bibfield  {journal} {\bibinfo  {journal} {The Journal of Chemical Physics}\
  }\textbf {\bibinfo {volume} {147}},\ \bibinfo {pages} {174704} (\bibinfo
  {year} {2017})}\BibitemShut {NoStop}%
\bibitem [{\citenamefont {Zhang}\ \emph {et~al.}(2021)\citenamefont {Zhang},
  \citenamefont {Zhou}, \citenamefont {Zhuang},\ and\ \citenamefont
  {Zhao}}]{zhang_adsorption_2021}%
  \BibitemOpen
  \bibfield  {author} {\bibinfo {author} {\bibfnamefont {Z.}~\bibnamefont
  {Zhang}}, \bibinfo {author} {\bibfnamefont {Q.}~\bibnamefont {Zhou}},
  \bibinfo {author} {\bibfnamefont {L.}~\bibnamefont {Zhuang}}, \ and\ \bibinfo
  {author} {\bibfnamefont {Z.}~\bibnamefont {Zhao}},\ }\bibfield  {title}
  {\enquote {\bibinfo {title} {Adsorption of {Ca}({II}) and {K}({I}) on the
  kaolinite surface: a {DFT} study with an experimental verification},}\ }\href
  {\doibase 10.1080/00268976.2021.1896047} {\bibfield  {journal} {\bibinfo
  {journal} {Molecular Physics}\ }\textbf {\bibinfo {volume} {119}},\ \bibinfo
  {pages} {e1896047} (\bibinfo {year} {2021})}\BibitemShut {NoStop}%
\bibitem [{\citenamefont {Liu}(2007)}]{liu_polymer_2007}%
  \BibitemOpen
  \bibfield  {author} {\bibinfo {author} {\bibfnamefont {P.}~\bibnamefont
  {Liu}},\ }\bibfield  {title} {\enquote {\bibinfo {title} {Polymer modified
  clay minerals: {A} review},}\ }\href {\doibase 10.1016/j.clay.2007.01.004}
  {\bibfield  {journal} {\bibinfo  {journal} {Applied Clay Science}\ }\textbf
  {\bibinfo {volume} {38}},\ \bibinfo {pages} {64--76} (\bibinfo {year}
  {2007})}\BibitemShut {NoStop}%
\bibitem [{\citenamefont {Varma}(2002)}]{varma_clay_2002}%
  \BibitemOpen
  \bibfield  {author} {\bibinfo {author} {\bibfnamefont {R.~S.}\ \bibnamefont
  {Varma}},\ }\bibfield  {title} {\enquote {\bibinfo {title} {Clay and
  clay-supported reagents in organic synthesis},}\ }\href {\doibase
  10.1016/S0040-4020(01)01216-9} {\bibfield  {journal} {\bibinfo  {journal}
  {Tetrahedron}\ }\textbf {\bibinfo {volume} {58}},\ \bibinfo {pages}
  {1235--1255} (\bibinfo {year} {2002})}\BibitemShut {NoStop}%
\bibitem [{\citenamefont {Wilson}, \citenamefont {Wilson},\ and\ \citenamefont
  {Patey}(2014)}]{wilson_influence_2014}%
  \BibitemOpen
  \bibfield  {author} {\bibinfo {author} {\bibfnamefont {M.~J.}\ \bibnamefont
  {Wilson}}, \bibinfo {author} {\bibfnamefont {L.}~\bibnamefont {Wilson}}, \
  and\ \bibinfo {author} {\bibfnamefont {I.}~\bibnamefont {Patey}},\ }\bibfield
   {title} {\enquote {\bibinfo {title} {The influence of individual clay
  minerals on formation damage of reservoir sandstones: a critical review with
  some new insights},}\ }\href {\doibase 10.1180/claymin.2014.049.2.02}
  {\bibfield  {journal} {\bibinfo  {journal} {Clay Minerals}\ }\textbf
  {\bibinfo {volume} {49}},\ \bibinfo {pages} {147--164} (\bibinfo {year}
  {2014})}\BibitemShut {NoStop}%
\bibitem [{\citenamefont {Prempeh}\ \emph {et~al.}(2020)\citenamefont
  {Prempeh}, \citenamefont {Chequer}, \citenamefont {Badalyan},\ and\
  \citenamefont {Bedrikovetsky}}]{prempeh_effects_2020}%
  \BibitemOpen
  \bibfield  {author} {\bibinfo {author} {\bibfnamefont {K.}~\bibnamefont
  {Prempeh}}, \bibinfo {author} {\bibfnamefont {L.}~\bibnamefont {Chequer}},
  \bibinfo {author} {\bibfnamefont {A.}~\bibnamefont {Badalyan}}, \ and\
  \bibinfo {author} {\bibfnamefont {P.}~\bibnamefont {Bedrikovetsky}},\
  }\bibfield  {title} {\enquote {\bibinfo {title} {Effects of {Kaolinite} on
  {Fines} {Migration} and {Formation} {Damage}},}\ \ }(\bibinfo  {publisher}
  {OnePetro},\ \bibinfo {year} {2020})\BibitemShut {NoStop}%
\bibitem [{\citenamefont {Madejová}, \citenamefont {Gates},\ and\
  \citenamefont {Petit}(2017)}]{madejova_ir_2017}%
  \BibitemOpen
  \bibfield  {author} {\bibinfo {author} {\bibfnamefont {J.}~\bibnamefont
  {Madejová}}, \bibinfo {author} {\bibfnamefont {W.}~\bibnamefont {Gates}}, \
  and\ \bibinfo {author} {\bibfnamefont {S.}~\bibnamefont {Petit}},\ }\bibfield
   {title} {\enquote {\bibinfo {title} {{IR} {Spectra} of {Clay} {Minerals}},}\
  }in\ \href {\doibase 10.1016/B978-0-08-100355-8.00005-9} {\emph {\bibinfo
  {booktitle} {Developments in {Clay} {Science}}}},\ Vol.~\bibinfo {volume}
  {8}\ (\bibinfo  {publisher} {Elsevier},\ \bibinfo {year} {2017})\ pp.\
  \bibinfo {pages} {107--149}\BibitemShut {NoStop}%
\bibitem [{\citenamefont {Prost}\ \emph {et~al.}(1989)\citenamefont {Prost},
  \citenamefont {Huard}, \citenamefont {Driard},\ and\ \citenamefont
  {Leydecker}}]{prost_infared_1989}%
  \BibitemOpen
  \bibfield  {author} {\bibinfo {author} {\bibfnamefont {R.}~\bibnamefont
  {Prost}}, \bibinfo {author} {\bibfnamefont {E.}~\bibnamefont {Huard}},
  \bibinfo {author} {\bibfnamefont {J.}~\bibnamefont {Driard}}, \ and\ \bibinfo
  {author} {\bibfnamefont {J.~P.}\ \bibnamefont {Leydecker}},\ }\bibfield
  {title} {\enquote {\bibinfo {title} {Infared {Study} of {Structural} {OH} in
  {Kaolinite}. {Dickite}, {Nacrite} and {Poorly} {Crystalline} {Kaolintie} at 5
  to 600 {K}},}\ }\href@noop {} {\bibfield  {journal} {\bibinfo  {journal}
  {Clays and Clay Minerals}\ }\textbf {\bibinfo {volume} {37}},\ \bibinfo
  {pages} {464--468} (\bibinfo {year} {1989})}\BibitemShut {NoStop}%
\bibitem [{\citenamefont {Lanson}(1997)}]{lanson_decomposition_1997}%
  \BibitemOpen
  \bibfield  {author} {\bibinfo {author} {\bibfnamefont {B.}~\bibnamefont
  {Lanson}},\ }\href {\doibase 10.1346/CCMN.1997.0450202} {\bibfield  {journal}
  {\bibinfo  {journal} {Clays and Clay Minerals}\ }\textbf {\bibinfo {volume}
  {45}},\ \bibinfo {pages} {132--146} (\bibinfo {year} {1997})}\BibitemShut
  {NoStop}%
\bibitem [{\citenamefont {Pan}, \citenamefont {Yin},\ and\ \citenamefont
  {Iglauer}(2020)}]{pan_review_2020}%
  \BibitemOpen
  \bibfield  {author} {\bibinfo {author} {\bibfnamefont {B.}~\bibnamefont
  {Pan}}, \bibinfo {author} {\bibfnamefont {X.}~\bibnamefont {Yin}}, \ and\
  \bibinfo {author} {\bibfnamefont {S.}~\bibnamefont {Iglauer}},\ }\bibfield
  {title} {\enquote {\bibinfo {title} {A review on clay wettability: {From}
  experimental investigations to molecular dynamics simulations},}\ }\href
  {\doibase 10.1016/j.cis.2020.102266} {\bibfield  {journal} {\bibinfo
  {journal} {Advances in Colloid and Interface Science}\ }\textbf {\bibinfo
  {volume} {285}},\ \bibinfo {pages} {102266} (\bibinfo {year}
  {2020})}\BibitemShut {NoStop}%
\bibitem [{\citenamefont {Zhou}\ \emph {et~al.}(2018)\citenamefont {Zhou},
  \citenamefont {Liu}, \citenamefont {Bu}, \citenamefont {Deng}, \citenamefont
  {Liu}, \citenamefont {Yuan}, \citenamefont {Du},\ and\ \citenamefont
  {Song}}]{zhou_xrd-based_2018}%
  \BibitemOpen
  \bibfield  {author} {\bibinfo {author} {\bibfnamefont {X.}~\bibnamefont
  {Zhou}}, \bibinfo {author} {\bibfnamefont {D.}~\bibnamefont {Liu}}, \bibinfo
  {author} {\bibfnamefont {H.}~\bibnamefont {Bu}}, \bibinfo {author}
  {\bibfnamefont {L.}~\bibnamefont {Deng}}, \bibinfo {author} {\bibfnamefont
  {H.}~\bibnamefont {Liu}}, \bibinfo {author} {\bibfnamefont {P.}~\bibnamefont
  {Yuan}}, \bibinfo {author} {\bibfnamefont {P.}~\bibnamefont {Du}}, \ and\
  \bibinfo {author} {\bibfnamefont {H.}~\bibnamefont {Song}},\ }\bibfield
  {title} {\enquote {\bibinfo {title} {{XRD}-based quantitative analysis of
  clay minerals using reference intensity ratios, mineral intensity factors,
  {Rietveld}, and full pattern summation methods: {A} critical review},}\
  }\href {\doibase 10.1016/j.sesci.2017.12.002} {\bibfield  {journal} {\bibinfo
   {journal} {Solid Earth Sciences}\ }\textbf {\bibinfo {volume} {3}},\
  \bibinfo {pages} {16--29} (\bibinfo {year} {2018})}\BibitemShut {NoStop}%
\bibitem [{\citenamefont {Martins}\ \emph {et~al.}(2014)\citenamefont
  {Martins}, \citenamefont {Gates}, \citenamefont {Michot}, \citenamefont
  {Ferrage}, \citenamefont {Marry},\ and\ \citenamefont
  {Bordallo}}]{martins_neutron_2014}%
  \BibitemOpen
  \bibfield  {author} {\bibinfo {author} {\bibfnamefont {M.~L.}\ \bibnamefont
  {Martins}}, \bibinfo {author} {\bibfnamefont {W.~P.}\ \bibnamefont {Gates}},
  \bibinfo {author} {\bibfnamefont {L.}~\bibnamefont {Michot}}, \bibinfo
  {author} {\bibfnamefont {E.}~\bibnamefont {Ferrage}}, \bibinfo {author}
  {\bibfnamefont {V.}~\bibnamefont {Marry}}, \ and\ \bibinfo {author}
  {\bibfnamefont {H.~N.}\ \bibnamefont {Bordallo}},\ }\bibfield  {title}
  {\enquote {\bibinfo {title} {Neutron scattering, a powerful tool to study
  clay minerals},}\ }\href {\doibase 10.1016/j.clay.2014.05.004} {\bibfield
  {journal} {\bibinfo  {journal} {Applied Clay Science}\ }\bibinfo {series}
  {Clay {Science} \& {Technology} ({XV} {International} {Clay} {Conference})},\
  \textbf {\bibinfo {volume} {96}},\ \bibinfo {pages} {22--35} (\bibinfo {year}
  {2014})}\BibitemShut {NoStop}%
\bibitem [{\citenamefont {A. Harvey}\ \emph {et~al.}(2019)\citenamefont
  {A. Harvey}, \citenamefont {T. Johnston}, \citenamefont {J. Criscenti},\
  and\ \citenamefont {A. Greathouse}}]{aharvey_distinguishing_2019}%
  \BibitemOpen
  \bibfield  {author} {\bibinfo {author} {\bibfnamefont {J.}~\bibnamefont
  {A. Harvey}}, \bibinfo {author} {\bibfnamefont {C.}~\bibnamefont
  {T. Johnston}}, \bibinfo {author} {\bibfnamefont {L.}~\bibnamefont
  {J. Criscenti}}, \ and\ \bibinfo {author} {\bibfnamefont {J.}~\bibnamefont
  {A. Greathouse}},\ }\bibfield  {title} {\enquote {\bibinfo {title}
  {Distinguishing between bulk and edge hydroxyl vibrational properties of 2:1
  phyllosilicates via deuteration},}\ }\href {\doibase 10.1039/C9CC00164F}
  {\bibfield  {journal} {\bibinfo  {journal} {Chemical Communications}\
  }\textbf {\bibinfo {volume} {55}},\ \bibinfo {pages} {3453--3456} (\bibinfo
  {year} {2019})}\BibitemShut {NoStop}%
\bibitem [{\citenamefont {Lagaly}(1981)}]{lagaly_characterization_1981}%
  \BibitemOpen
  \bibfield  {author} {\bibinfo {author} {\bibfnamefont {G.}~\bibnamefont
  {Lagaly}},\ }\bibfield  {title} {\enquote {\bibinfo {title} {Characterization
  of clays by organic compounds},}\ }\href {\doibase
  10.1180/claymin.1981.016.1.01} {\bibfield  {journal} {\bibinfo  {journal}
  {Clay Minerals}\ }\textbf {\bibinfo {volume} {16}},\ \bibinfo {pages} {1--21}
  (\bibinfo {year} {1981})}\BibitemShut {NoStop}%
\bibitem [{\citenamefont {Awad}\ \emph {et~al.}(2017)\citenamefont {Awad},
  \citenamefont {López-Galindo}, \citenamefont {Setti}, \citenamefont
  {El-Rahmany},\ and\ \citenamefont {Iborra}}]{awad_kaolinite_2017}%
  \BibitemOpen
  \bibfield  {author} {\bibinfo {author} {\bibfnamefont {M.~E.}\ \bibnamefont
  {Awad}}, \bibinfo {author} {\bibfnamefont {A.}~\bibnamefont
  {López-Galindo}}, \bibinfo {author} {\bibfnamefont {M.}~\bibnamefont
  {Setti}}, \bibinfo {author} {\bibfnamefont {M.~M.}\ \bibnamefont
  {El-Rahmany}}, \ and\ \bibinfo {author} {\bibfnamefont {C.~V.}\ \bibnamefont
  {Iborra}},\ }\bibfield  {title} {\enquote {\bibinfo {title} {Kaolinite in
  pharmaceutics and biomedicine},}\ }\href {\doibase
  10.1016/j.ijpharm.2017.09.056} {\bibfield  {journal} {\bibinfo  {journal}
  {International Journal of Pharmaceutics}\ }\textbf {\bibinfo {volume}
  {533}},\ \bibinfo {pages} {34--48} (\bibinfo {year} {2017})}\BibitemShut
  {NoStop}%
\bibitem [{\citenamefont {Gianni}, \citenamefont {Avgoustakis},\ and\
  \citenamefont {Papoulis}(2020)}]{gianni_kaolinite_2020}%
  \BibitemOpen
  \bibfield  {author} {\bibinfo {author} {\bibfnamefont {E.}~\bibnamefont
  {Gianni}}, \bibinfo {author} {\bibfnamefont {K.}~\bibnamefont {Avgoustakis}},
  \ and\ \bibinfo {author} {\bibfnamefont {D.}~\bibnamefont {Papoulis}},\
  }\bibfield  {title} {\enquote {\bibinfo {title} {Kaolinite group minerals:
  {Applications} in cancer diagnosis and treatment},}\ }\href {\doibase
  10.1016/j.ejpb.2020.07.030} {\bibfield  {journal} {\bibinfo  {journal}
  {European Journal of Pharmaceutics and Biopharmaceutics}\ }\textbf {\bibinfo
  {volume} {154}},\ \bibinfo {pages} {359--376} (\bibinfo {year}
  {2020})}\BibitemShut {NoStop}%
\bibitem [{\citenamefont {Cao}, \citenamefont {Wang},\ and\ \citenamefont
  {Cheng}(2021)}]{cao_recent_2021}%
  \BibitemOpen
  \bibfield  {author} {\bibinfo {author} {\bibfnamefont {Z.}~\bibnamefont
  {Cao}}, \bibinfo {author} {\bibfnamefont {Q.}~\bibnamefont {Wang}}, \ and\
  \bibinfo {author} {\bibfnamefont {H.}~\bibnamefont {Cheng}},\ }\bibfield
  {title} {\enquote {\bibinfo {title} {Recent advances in kaolinite-based
  material for photocatalysts},}\ }\href {\doibase 10.1016/j.cclet.2021.01.009}
  {\bibfield  {journal} {\bibinfo  {journal} {Chinese Chemical Letters}\
  }\textbf {\bibinfo {volume} {32}},\ \bibinfo {pages} {2617--2628} (\bibinfo
  {year} {2021})}\BibitemShut {NoStop}%
\bibitem [{\citenamefont {Dedzo}\ and\ \citenamefont
  {Detellier}(2016)}]{dedzo_functional_2016}%
  \BibitemOpen
  \bibfield  {author} {\bibinfo {author} {\bibfnamefont {G.~K.}\ \bibnamefont
  {Dedzo}}\ and\ \bibinfo {author} {\bibfnamefont {C.}~\bibnamefont
  {Detellier}},\ }\bibfield  {title} {\enquote {\bibinfo {title} {Functional
  nanohybrid materials derived from kaolinite},}\ }\href {\doibase
  10.1016/j.clay.2016.01.010} {\bibfield  {journal} {\bibinfo  {journal}
  {Applied Clay Science}\ }\bibinfo {series} {{SI} {Euroclay} 2015 {Part}-1},\
  \textbf {\bibinfo {volume} {130}},\ \bibinfo {pages} {33--39} (\bibinfo
  {year} {2016})}\BibitemShut {NoStop}%
\bibitem [{\citenamefont {Bish}(1993)}]{bish_rietveld_1993}%
  \BibitemOpen
  \bibfield  {author} {\bibinfo {author} {\bibfnamefont {D.~L.}\ \bibnamefont
  {Bish}},\ }\bibfield  {title} {\enquote {\bibinfo {title} {Rietveld
  {Refinement} of the {Kaolinite} {Structure} at 1.5 {K}},}\ }\href {\doibase
  10.1346/CCMN.1993.0410613} {\bibfield  {journal} {\bibinfo  {journal} {Clays
  and Clay Minerals}\ }\textbf {\bibinfo {volume} {41}},\ \bibinfo {pages}
  {738--744} (\bibinfo {year} {1993})}\BibitemShut {NoStop}%
\bibitem [{\citenamefont {White}\ \emph {et~al.}(2009)\citenamefont {White},
  \citenamefont {Provis}, \citenamefont {Riley}, \citenamefont {Kearley},\ and\
  \citenamefont {van Deventer}}]{white_what_2009}%
  \BibitemOpen
  \bibfield  {author} {\bibinfo {author} {\bibfnamefont {C.~E.}\ \bibnamefont
  {White}}, \bibinfo {author} {\bibfnamefont {J.~L.}\ \bibnamefont {Provis}},
  \bibinfo {author} {\bibfnamefont {D.~P.}\ \bibnamefont {Riley}}, \bibinfo
  {author} {\bibfnamefont {G.~J.}\ \bibnamefont {Kearley}}, \ and\ \bibinfo
  {author} {\bibfnamefont {J.~S.~J.}\ \bibnamefont {van Deventer}},\ }\bibfield
   {title} {\enquote {\bibinfo {title} {What {Is} the {Structure} of
  {Kaolinite}? {Reconciling} {Theory} and {Experiment}},}\ }\href {\doibase
  10.1021/jp810448t} {\bibfield  {journal} {\bibinfo  {journal} {The Journal of
  Physical Chemistry B}\ }\textbf {\bibinfo {volume} {113}},\ \bibinfo {pages}
  {6756--6765} (\bibinfo {year} {2009})}\BibitemShut {NoStop}%
\bibitem [{\citenamefont {Yin}\ \emph {et~al.}(2012)\citenamefont {Yin},
  \citenamefont {Gupta}, \citenamefont {Du}, \citenamefont {Wang},\ and\
  \citenamefont {Miller}}]{yin_surface_2012}%
  \BibitemOpen
  \bibfield  {author} {\bibinfo {author} {\bibfnamefont {X.}~\bibnamefont
  {Yin}}, \bibinfo {author} {\bibfnamefont {V.}~\bibnamefont {Gupta}}, \bibinfo
  {author} {\bibfnamefont {H.}~\bibnamefont {Du}}, \bibinfo {author}
  {\bibfnamefont {X.}~\bibnamefont {Wang}}, \ and\ \bibinfo {author}
  {\bibfnamefont {J.~D.}\ \bibnamefont {Miller}},\ }\bibfield  {title}
  {\enquote {\bibinfo {title} {Surface charge and wetting characteristics of
  layered silicate minerals},}\ }\href {\doibase 10.1016/j.cis.2012.06.004}
  {\bibfield  {journal} {\bibinfo  {journal} {Advances in Colloid and Interface
  Science}\ }\bibinfo {series} {Interfaces, {Wettability}, {Surface} {Forces}
  and {Applications}: {Special} {Issue} in honour of the 65th {Birthday} of
  {John} {Ralston}},\ \textbf {\bibinfo {volume} {179-182}},\ \bibinfo {pages}
  {43--50} (\bibinfo {year} {2012})}\BibitemShut {NoStop}%
\bibitem [{\citenamefont {Volkova}\ \emph {et~al.}(2021)\citenamefont
  {Volkova}, \citenamefont {Narayanan~Nair}, \citenamefont {Engelbrecht},
  \citenamefont {Schwingenschlögl}, \citenamefont {Sun},\ and\ \citenamefont
  {Stenchikov}}]{volkova_molecular_2021}%
  \BibitemOpen
  \bibfield  {author} {\bibinfo {author} {\bibfnamefont {E.}~\bibnamefont
  {Volkova}}, \bibinfo {author} {\bibfnamefont {A.~K.}\ \bibnamefont
  {Narayanan~Nair}}, \bibinfo {author} {\bibfnamefont {J.}~\bibnamefont
  {Engelbrecht}}, \bibinfo {author} {\bibfnamefont {U.}~\bibnamefont
  {Schwingenschlögl}}, \bibinfo {author} {\bibfnamefont {S.}~\bibnamefont
  {Sun}}, \ and\ \bibinfo {author} {\bibfnamefont {G.}~\bibnamefont
  {Stenchikov}},\ }\bibfield  {title} {\enquote {\bibinfo {title} {Molecular
  {Dynamics} {Modeling} of {Kaolinite} {Particle} {Associations}},}\ }\href
  {\doibase 10.1021/acs.jpcc.1c06598} {\bibfield  {journal} {\bibinfo
  {journal} {The Journal of Physical Chemistry C}\ }\textbf {\bibinfo {volume}
  {125}},\ \bibinfo {pages} {24126--24136} (\bibinfo {year}
  {2021})}\BibitemShut {NoStop}%
\bibitem [{\citenamefont {Sosso}\ \emph {et~al.}(2016)\citenamefont {Sosso},
  \citenamefont {Tribello}, \citenamefont {Zen}, \citenamefont {Pedevilla},\
  and\ \citenamefont {Michaelides}}]{sosso_ice_2016}%
  \BibitemOpen
  \bibfield  {author} {\bibinfo {author} {\bibfnamefont {G.~C.}\ \bibnamefont
  {Sosso}}, \bibinfo {author} {\bibfnamefont {G.~A.}\ \bibnamefont {Tribello}},
  \bibinfo {author} {\bibfnamefont {A.}~\bibnamefont {Zen}}, \bibinfo {author}
  {\bibfnamefont {P.}~\bibnamefont {Pedevilla}}, \ and\ \bibinfo {author}
  {\bibfnamefont {A.}~\bibnamefont {Michaelides}},\ }\bibfield  {title}
  {\enquote {\bibinfo {title} {Ice formation on kaolinite: {Insights} from
  molecular dynamics simulations},}\ }\href {\doibase 10.1063/1.4968796}
  {\bibfield  {journal} {\bibinfo  {journal} {The Journal of Chemical Physics}\
  }\textbf {\bibinfo {volume} {145}},\ \bibinfo {pages} {211927} (\bibinfo
  {year} {2016})}\BibitemShut {NoStop}%
\bibitem [{\citenamefont {A. Prishchenko}\ \emph {et~al.}(2018)\citenamefont
  {A. Prishchenko}, \citenamefont {V. Zenkov}, \citenamefont {V. Mazurenko},
  \citenamefont {F. Fakhrullin}, \citenamefont {M. Lvov},\ and\ \citenamefont
  {G. Mazurenko}}]{aprishchenko_molecular_2018}%
  \BibitemOpen
  \bibfield  {author} {\bibinfo {author} {\bibfnamefont {D.}~\bibnamefont
  {A. Prishchenko}}, \bibinfo {author} {\bibfnamefont {E.}~\bibnamefont
  {V. Zenkov}}, \bibinfo {author} {\bibfnamefont {V.}~\bibnamefont
  {V. Mazurenko}}, \bibinfo {author} {\bibfnamefont {R.}~\bibnamefont
  {F. Fakhrullin}}, \bibinfo {author} {\bibfnamefont {Y.}~\bibnamefont
  {M. Lvov}}, \ and\ \bibinfo {author} {\bibfnamefont {V.}~\bibnamefont
  {G. Mazurenko}},\ }\bibfield  {title} {\enquote {\bibinfo {title} {Molecular
  dynamics of the halloysite nanotubes},}\ }\href {\doibase 10.1039/C7CP06575B}
  {\bibfield  {journal} {\bibinfo  {journal} {Physical Chemistry Chemical
  Physics}\ }\textbf {\bibinfo {volume} {20}},\ \bibinfo {pages} {5841--5849}
  (\bibinfo {year} {2018})}\BibitemShut {NoStop}%
\bibitem [{\citenamefont {Cygan}, \citenamefont {Liang},\ and\ \citenamefont
  {Kalinichev}(2004)}]{cygan_molecular_2004}%
  \BibitemOpen
  \bibfield  {author} {\bibinfo {author} {\bibfnamefont {R.~T.}\ \bibnamefont
  {Cygan}}, \bibinfo {author} {\bibfnamefont {J.-J.}\ \bibnamefont {Liang}}, \
  and\ \bibinfo {author} {\bibfnamefont {A.~G.}\ \bibnamefont {Kalinichev}},\
  }\bibfield  {title} {\enquote {\bibinfo {title} {Molecular {Models} of
  {Hydroxide}, {Oxyhydroxide}, and {Clay} {Phases} and the {Development} of a
  {General} {Force} {Field}},}\ }\href {\doibase 10.1021/jp0363287} {\bibfield
  {journal} {\bibinfo  {journal} {The Journal of Physical Chemistry B}\
  }\textbf {\bibinfo {volume} {108}},\ \bibinfo {pages} {1255--1266} (\bibinfo
  {year} {2004})}\BibitemShut {NoStop}%
\bibitem [{\citenamefont {Cygan}, \citenamefont {Greathouse},\ and\
  \citenamefont {Kalinichev}(2021)}]{cygan_advances_2021}%
  \BibitemOpen
  \bibfield  {author} {\bibinfo {author} {\bibfnamefont {R.~T.}\ \bibnamefont
  {Cygan}}, \bibinfo {author} {\bibfnamefont {J.~A.}\ \bibnamefont
  {Greathouse}}, \ and\ \bibinfo {author} {\bibfnamefont {A.~G.}\ \bibnamefont
  {Kalinichev}},\ }\bibfield  {title} {\enquote {\bibinfo {title} {Advances in
  {Clayff} {Molecular} {Simulation} of {Layered} and {Nanoporous} {Materials}
  and {Their} {Aqueous} {Interfaces}},}\ }\href {\doibase
  10.1021/acs.jpcc.1c04600} {\bibfield  {journal} {\bibinfo  {journal} {The
  Journal of Physical Chemistry C}\ }\textbf {\bibinfo {volume} {125}},\
  \bibinfo {pages} {17573--17589} (\bibinfo {year} {2021})}\BibitemShut
  {NoStop}%
\bibitem [{\citenamefont {Pouvreau}\ \emph {et~al.}(2017)\citenamefont
  {Pouvreau}, \citenamefont {Greathouse}, \citenamefont {Cygan},\ and\
  \citenamefont {Kalinichev}}]{pouvreau_structure_2017}%
  \BibitemOpen
  \bibfield  {author} {\bibinfo {author} {\bibfnamefont {M.}~\bibnamefont
  {Pouvreau}}, \bibinfo {author} {\bibfnamefont {J.~A.}\ \bibnamefont
  {Greathouse}}, \bibinfo {author} {\bibfnamefont {R.~T.}\ \bibnamefont
  {Cygan}}, \ and\ \bibinfo {author} {\bibfnamefont {A.~G.}\ \bibnamefont
  {Kalinichev}},\ }\bibfield  {title} {\enquote {\bibinfo {title} {Structure of
  {Hydrated} {Gibbsite} and {Brucite} {Edge} {Surfaces}: {DFT} {Results} and
  {Further} {Development} of the {ClayFF} {Classical} {Force} {Field} with
  {Metal}–{O}–{H} {Angle} {Bending} {Terms}},}\ }\href {\doibase
  10.1021/acs.jpcc.7b05362} {\bibfield  {journal} {\bibinfo  {journal} {The
  Journal of Physical Chemistry C}\ }\textbf {\bibinfo {volume} {121}},\
  \bibinfo {pages} {14757--14771} (\bibinfo {year} {2017})}\BibitemShut
  {NoStop}%
\bibitem [{\citenamefont {Greathouse}\ \emph {et~al.}(2009)\citenamefont
  {Greathouse}, \citenamefont {Durkin}, \citenamefont {Larentzos},\ and\
  \citenamefont {Cygan}}]{greathouse_implementation_2009}%
  \BibitemOpen
  \bibfield  {author} {\bibinfo {author} {\bibfnamefont {J.~A.}\ \bibnamefont
  {Greathouse}}, \bibinfo {author} {\bibfnamefont {J.~S.}\ \bibnamefont
  {Durkin}}, \bibinfo {author} {\bibfnamefont {J.~P.}\ \bibnamefont
  {Larentzos}}, \ and\ \bibinfo {author} {\bibfnamefont {R.~T.}\ \bibnamefont
  {Cygan}},\ }\bibfield  {title} {\enquote {\bibinfo {title} {Implementation of
  a {Morse} potential to model hydroxyl behavior in phyllosilicates},}\ }\href
  {\doibase 10.1063/1.3103886} {\bibfield  {journal} {\bibinfo  {journal} {The
  Journal of Chemical Physics}\ }\textbf {\bibinfo {volume} {130}},\ \bibinfo
  {pages} {134713} (\bibinfo {year} {2009})}\BibitemShut {NoStop}%
\bibitem [{\citenamefont {Liu}\ \emph {et~al.}(2019)\citenamefont {Liu},
  \citenamefont {Min}, \citenamefont {Chen}, \citenamefont {Lu},\ and\
  \citenamefont {Shen}}]{liu_adsorption_2019}%
  \BibitemOpen
  \bibfield  {author} {\bibinfo {author} {\bibfnamefont {L.}~\bibnamefont
  {Liu}}, \bibinfo {author} {\bibfnamefont {F.}~\bibnamefont {Min}}, \bibinfo
  {author} {\bibfnamefont {J.}~\bibnamefont {Chen}}, \bibinfo {author}
  {\bibfnamefont {F.}~\bibnamefont {Lu}}, \ and\ \bibinfo {author}
  {\bibfnamefont {L.}~\bibnamefont {Shen}},\ }\bibfield  {title} {\enquote
  {\bibinfo {title} {The adsorption of dodecylamine and oleic acid on kaolinite
  surfaces: {Insights} from {DFT} calculation and experimental
  investigation},}\ }\href {\doibase 10.1016/j.apsusc.2018.11.104} {\bibfield
  {journal} {\bibinfo  {journal} {Applied Surface Science}\ }\textbf {\bibinfo
  {volume} {470}},\ \bibinfo {pages} {27--35} (\bibinfo {year}
  {2019})}\BibitemShut {NoStop}%
\bibitem [{\citenamefont {Wang}\ \emph {et~al.}(2013)\citenamefont {Wang},
  \citenamefont {Qian}, \citenamefont {Song}, \citenamefont {Zhang},\ and\
  \citenamefont {Dong}}]{wang_dft_2013}%
  \BibitemOpen
  \bibfield  {author} {\bibinfo {author} {\bibfnamefont {X.}~\bibnamefont
  {Wang}}, \bibinfo {author} {\bibfnamefont {P.}~\bibnamefont {Qian}}, \bibinfo
  {author} {\bibfnamefont {K.}~\bibnamefont {Song}}, \bibinfo {author}
  {\bibfnamefont {C.}~\bibnamefont {Zhang}}, \ and\ \bibinfo {author}
  {\bibfnamefont {J.}~\bibnamefont {Dong}},\ }\bibfield  {title} {\enquote
  {\bibinfo {title} {The {DFT} study of adsorption of 2,4-dinitrotoluene on
  kaolinite surfaces},}\ }\href {\doibase 10.1016/j.comptc.2013.09.025}
  {\bibfield  {journal} {\bibinfo  {journal} {Computational and Theoretical
  Chemistry}\ }\textbf {\bibinfo {volume} {1025}},\ \bibinfo {pages} {16--23}
  (\bibinfo {year} {2013})}\BibitemShut {NoStop}%
\bibitem [{\citenamefont {Ren}\ \emph {et~al.}(2020)\citenamefont {Ren},
  \citenamefont {Min}, \citenamefont {Liu}, \citenamefont {Chen}, \citenamefont
  {Liu},\ and\ \citenamefont {Lv}}]{ren_adsorption_2020}%
  \BibitemOpen
  \bibfield  {author} {\bibinfo {author} {\bibfnamefont {B.}~\bibnamefont
  {Ren}}, \bibinfo {author} {\bibfnamefont {F.}~\bibnamefont {Min}}, \bibinfo
  {author} {\bibfnamefont {L.}~\bibnamefont {Liu}}, \bibinfo {author}
  {\bibfnamefont {J.}~\bibnamefont {Chen}}, \bibinfo {author} {\bibfnamefont
  {C.}~\bibnamefont {Liu}}, \ and\ \bibinfo {author} {\bibfnamefont
  {K.}~\bibnamefont {Lv}},\ }\bibfield  {title} {\enquote {\bibinfo {title}
  {Adsorption of different {PAM} structural units on kaolinite (0 0 1) surface:
  Density functional theory study},}\ }\href {\doibase
  10.1016/j.apsusc.2019.144324} {\bibfield  {journal} {\bibinfo  {journal}
  {Applied Surface Science}\ }\textbf {\bibinfo {volume} {504}},\ \bibinfo
  {pages} {144324} (\bibinfo {year} {2020})}\BibitemShut {NoStop}%
\bibitem [{\citenamefont {Chen}\ \emph
  {et~al.}(2019{\natexlab{a}})\citenamefont {Chen}, \citenamefont {Min},
  \citenamefont {Liu},\ and\ \citenamefont {Liu}}]{chen_mechanism_2019}%
  \BibitemOpen
  \bibfield  {author} {\bibinfo {author} {\bibfnamefont {J.}~\bibnamefont
  {Chen}}, \bibinfo {author} {\bibfnamefont {F.-f.}\ \bibnamefont {Min}},
  \bibinfo {author} {\bibfnamefont {L.-y.}\ \bibnamefont {Liu}}, \ and\
  \bibinfo {author} {\bibfnamefont {C.-f.}\ \bibnamefont {Liu}},\ }\bibfield
  {title} {\enquote {\bibinfo {title} {Mechanism research on surface hydration
  of kaolinite, insights from {DFT} and {MD} simulations},}\ }\href {\doibase
  10.1016/j.apsusc.2019.01.081} {\bibfield  {journal} {\bibinfo  {journal}
  {Applied Surface Science}\ }\textbf {\bibinfo {volume} {476}},\ \bibinfo
  {pages} {6--15} (\bibinfo {year} {2019}{\natexlab{a}})}\BibitemShut {NoStop}%
\bibitem [{\citenamefont {Chen}\ \emph
  {et~al.}(2019{\natexlab{b}})\citenamefont {Chen}, \citenamefont {Li},
  \citenamefont {Zhou}, \citenamefont {Xia},\ and\ \citenamefont
  {Yu}}]{chen_mechanism_2019-1}%
  \BibitemOpen
  \bibfield  {author} {\bibinfo {author} {\bibfnamefont {G.}~\bibnamefont
  {Chen}}, \bibinfo {author} {\bibfnamefont {X.}~\bibnamefont {Li}}, \bibinfo
  {author} {\bibfnamefont {L.}~\bibnamefont {Zhou}}, \bibinfo {author}
  {\bibfnamefont {S.}~\bibnamefont {Xia}}, \ and\ \bibinfo {author}
  {\bibfnamefont {L.}~\bibnamefont {Yu}},\ }\bibfield  {title} {\enquote
  {\bibinfo {title} {Mechanism insights into {Hg}({II}) adsorption on
  kaolinite(001) surface: {A} density functional study},}\ }\href {\doibase
  10.1016/j.apsusc.2019.05.227} {\bibfield  {journal} {\bibinfo  {journal}
  {Applied Surface Science}\ }\textbf {\bibinfo {volume} {488}},\ \bibinfo
  {pages} {494--502} (\bibinfo {year} {2019}{\natexlab{b}})}\BibitemShut
  {NoStop}%
\bibitem [{\citenamefont {Wang}\ \emph {et~al.}(2015)\citenamefont {Wang},
  \citenamefont {Huang}, \citenamefont {Pan}, \citenamefont {Wang},\ and\
  \citenamefont {Liu}}]{wang_theoretical_2015}%
  \BibitemOpen
  \bibfield  {author} {\bibinfo {author} {\bibfnamefont {X.}~\bibnamefont
  {Wang}}, \bibinfo {author} {\bibfnamefont {Y.}~\bibnamefont {Huang}},
  \bibinfo {author} {\bibfnamefont {Z.}~\bibnamefont {Pan}}, \bibinfo {author}
  {\bibfnamefont {Y.}~\bibnamefont {Wang}}, \ and\ \bibinfo {author}
  {\bibfnamefont {C.}~\bibnamefont {Liu}},\ }\bibfield  {title} {\enquote
  {\bibinfo {title} {Theoretical investigation of lead vapor adsorption on
  kaolinite surfaces with {DFT} calculations},}\ }\href {\doibase
  10.1016/j.jhazmat.2015.03.020} {\bibfield  {journal} {\bibinfo  {journal}
  {Journal of Hazardous Materials}\ }\textbf {\bibinfo {volume} {295}},\
  \bibinfo {pages} {43--54} (\bibinfo {year} {2015})}\BibitemShut {NoStop}%
\bibitem [{\citenamefont {Wang}\ \emph {et~al.}(2017)\citenamefont {Wang},
  \citenamefont {Kong}, \citenamefont {Zhang},\ and\ \citenamefont
  {Wang}}]{wang_adsorption_2017}%
  \BibitemOpen
  \bibfield  {author} {\bibinfo {author} {\bibfnamefont {Q.}~\bibnamefont
  {Wang}}, \bibinfo {author} {\bibfnamefont {X.-P.}\ \bibnamefont {Kong}},
  \bibinfo {author} {\bibfnamefont {B.-H.}\ \bibnamefont {Zhang}}, \ and\
  \bibinfo {author} {\bibfnamefont {J.}~\bibnamefont {Wang}},\ }\bibfield
  {title} {\enquote {\bibinfo {title} {Adsorption of {Zn}({II}) on the
  kaolinite(001) surfaces in aqueous environment: {A} combined {DFT} and
  molecular dynamics study},}\ }\href {\doibase 10.1016/j.apsusc.2017.04.062}
  {\bibfield  {journal} {\bibinfo  {journal} {Applied Surface Science}\
  }\textbf {\bibinfo {volume} {414}},\ \bibinfo {pages} {405--412} (\bibinfo
  {year} {2017})}\BibitemShut {NoStop}%
\bibitem [{\citenamefont {Zhao}\ and\ \citenamefont
  {He}(2014)}]{zhao_theoretical_2014}%
  \BibitemOpen
  \bibfield  {author} {\bibinfo {author} {\bibfnamefont {J.}~\bibnamefont
  {Zhao}}\ and\ \bibinfo {author} {\bibfnamefont {M.-C.}\ \bibnamefont {He}},\
  }\bibfield  {title} {\enquote {\bibinfo {title} {Theoretical study of heavy
  metal {Cd}, {Cu}, {Hg}, and {Ni}({II}) adsorption on the kaolinite(001)
  surface},}\ }\href {\doibase 10.1016/j.apsusc.2014.08.162} {\bibfield
  {journal} {\bibinfo  {journal} {Applied Surface Science}\ }\textbf {\bibinfo
  {volume} {317}},\ \bibinfo {pages} {718--723} (\bibinfo {year}
  {2014})}\BibitemShut {NoStop}%
\bibitem [{\citenamefont {He}, \citenamefont {Zhao},\ and\ \citenamefont
  {Wang}(2013)}]{he_adsorption_2013}%
  \BibitemOpen
  \bibfield  {author} {\bibinfo {author} {\bibfnamefont {M.-C.}\ \bibnamefont
  {He}}, \bibinfo {author} {\bibfnamefont {J.}~\bibnamefont {Zhao}}, \ and\
  \bibinfo {author} {\bibfnamefont {S.-X.}\ \bibnamefont {Wang}},\ }\bibfield
  {title} {\enquote {\bibinfo {title} {Adsorption and diffusion of {Pb}({II})
  on the kaolinite(001) surface: {A} density-functional theory study},}\ }\href
  {\doibase 10.1016/j.clay.2013.08.045} {\bibfield  {journal} {\bibinfo
  {journal} {Applied Clay Science}\ }\textbf {\bibinfo {volume} {85}},\
  \bibinfo {pages} {74--79} (\bibinfo {year} {2013})}\BibitemShut {NoStop}%
\bibitem [{\citenamefont {Schran}\ \emph {et~al.}(2021)\citenamefont {Schran},
  \citenamefont {Thiemann}, \citenamefont {Rowe}, \citenamefont {Müller},
  \citenamefont {Marsalek},\ and\ \citenamefont
  {Michaelides}}]{schran_machine_2021}%
  \BibitemOpen
  \bibfield  {author} {\bibinfo {author} {\bibfnamefont {C.}~\bibnamefont
  {Schran}}, \bibinfo {author} {\bibfnamefont {F.~L.}\ \bibnamefont
  {Thiemann}}, \bibinfo {author} {\bibfnamefont {P.}~\bibnamefont {Rowe}},
  \bibinfo {author} {\bibfnamefont {E.~A.}\ \bibnamefont {Müller}}, \bibinfo
  {author} {\bibfnamefont {O.}~\bibnamefont {Marsalek}}, \ and\ \bibinfo
  {author} {\bibfnamefont {A.}~\bibnamefont {Michaelides}},\ }\bibfield
  {title} {\enquote {\bibinfo {title} {Machine learning potentials for complex
  aqueous systems made simple},}\ }\href {\doibase 10.1073/pnas.2110077118}
  {\bibfield  {journal} {\bibinfo  {journal} {Proceedings of the National
  Academy of Sciences of the United States of America}\ }\textbf {\bibinfo
  {volume} {118}},\ \bibinfo {pages} {undefined--undefined} (\bibinfo {year}
  {2021})}\BibitemShut {NoStop}%
\bibitem [{\citenamefont {Schran}, \citenamefont {Brieuc},\ and\ \citenamefont
  {Marx}(2021)}]{schran_transferability_2021}%
  \BibitemOpen
  \bibfield  {author} {\bibinfo {author} {\bibfnamefont {C.}~\bibnamefont
  {Schran}}, \bibinfo {author} {\bibfnamefont {F.}~\bibnamefont {Brieuc}}, \
  and\ \bibinfo {author} {\bibfnamefont {D.}~\bibnamefont {Marx}},\ }\bibfield
  {title} {\enquote {\bibinfo {title} {Transferability of machine learning
  potentials: {Protonated} water neural network potential applied to the
  protonated water hexamer},}\ }\href {\doibase 10.1063/5.0035438} {\bibfield
  {journal} {\bibinfo  {journal} {The Journal of Chemical Physics}\ }\textbf
  {\bibinfo {volume} {154}},\ \bibinfo {pages} {051101} (\bibinfo {year}
  {2021})}\BibitemShut {NoStop}%
\bibitem [{\citenamefont {Deringer}, \citenamefont {Caro},\ and\ \citenamefont
  {Csányi}(2019)}]{deringer_machine_2019}%
  \BibitemOpen
  \bibfield  {author} {\bibinfo {author} {\bibfnamefont {V.}~\bibnamefont
  {Deringer}}, \bibinfo {author} {\bibfnamefont {M.}~\bibnamefont {Caro}}, \
  and\ \bibinfo {author} {\bibfnamefont {G.}~\bibnamefont {Csányi}},\
  }\bibfield  {title} {\enquote {\bibinfo {title} {Machine {Learning}
  {Interatomic} {Potentials} as {Emerging} {Tools} for {Materials}
  {Science}},}\ }\href {\doibase 10.1002/adma.201902765} {\bibfield  {journal}
  {\bibinfo  {journal} {Advanced Materials}\ }\textbf {\bibinfo {volume}
  {31}},\ \bibinfo {pages} {1902765} (\bibinfo {year} {2019})}\BibitemShut
  {NoStop}%
\bibitem [{\citenamefont {Bartók}\ \emph {et~al.}(2010)\citenamefont
  {Bartók}, \citenamefont {Payne}, \citenamefont {Kondor},\ and\ \citenamefont
  {Csányi}}]{bartok_gaussian_2010}%
  \BibitemOpen
  \bibfield  {author} {\bibinfo {author} {\bibfnamefont {A.~P.}\ \bibnamefont
  {Bartók}}, \bibinfo {author} {\bibfnamefont {M.~C.}\ \bibnamefont {Payne}},
  \bibinfo {author} {\bibfnamefont {R.}~\bibnamefont {Kondor}}, \ and\ \bibinfo
  {author} {\bibfnamefont {G.}~\bibnamefont {Csányi}},\ }\bibfield  {title}
  {\enquote {\bibinfo {title} {Gaussian {Approximation} {Potentials}: {The}
  {Accuracy} of {Quantum} {Mechanics}, without the {Electrons}},}\ }\href
  {\doibase 10.1103/PhysRevLett.104.136403} {\bibfield  {journal} {\bibinfo
  {journal} {Physical Review Letters}\ }\textbf {\bibinfo {volume} {104}},\
  \bibinfo {pages} {136403} (\bibinfo {year} {2010})}\BibitemShut {NoStop}%
\bibitem [{\citenamefont {Bartók}\ and\ \citenamefont
  {Csányi}(2015)}]{bartok_gaussian_2015}%
  \BibitemOpen
  \bibfield  {author} {\bibinfo {author} {\bibfnamefont {A.~P.}\ \bibnamefont
  {Bartók}}\ and\ \bibinfo {author} {\bibfnamefont {G.}~\bibnamefont
  {Csányi}},\ }\bibfield  {title} {\enquote {\bibinfo {title} {Gaussian
  approximation potentials: {A} brief tutorial introduction},}\ }\href
  {\doibase 10.1002/qua.24927} {\bibfield  {journal} {\bibinfo  {journal}
  {International Journal of Quantum Chemistry}\ }\textbf {\bibinfo {volume}
  {115}},\ \bibinfo {pages} {1051--1057} (\bibinfo {year} {2015})}\BibitemShut
  {NoStop}%
\bibitem [{\citenamefont {Chmiela}\ \emph {et~al.}(2018)\citenamefont
  {Chmiela}, \citenamefont {Sauceda}, \citenamefont {Müller},\ and\
  \citenamefont {Tkatchenko}}]{chmiela_towards_2018}%
  \BibitemOpen
  \bibfield  {author} {\bibinfo {author} {\bibfnamefont {S.}~\bibnamefont
  {Chmiela}}, \bibinfo {author} {\bibfnamefont {H.~E.}\ \bibnamefont
  {Sauceda}}, \bibinfo {author} {\bibfnamefont {K.-R.}\ \bibnamefont
  {Müller}}, \ and\ \bibinfo {author} {\bibfnamefont {A.}~\bibnamefont
  {Tkatchenko}},\ }\bibfield  {title} {\enquote {\bibinfo {title} {Towards
  exact molecular dynamics simulations with machine-learned force fields},}\
  }\href {\doibase 10.1038/s41467-018-06169-2} {\bibfield  {journal} {\bibinfo
  {journal} {Nature Communications}\ }\textbf {\bibinfo {volume} {9}},\
  \bibinfo {pages} {3887} (\bibinfo {year} {2018})}\BibitemShut {NoStop}%
\bibitem [{\citenamefont {Zhang}\ \emph {et~al.}(2020)\citenamefont {Zhang},
  \citenamefont {Ye}, \citenamefont {Zhang}, \citenamefont {Hu}, \citenamefont
  {Jiang},\ and\ \citenamefont {Jiang}}]{zhang_efficient_2020}%
  \BibitemOpen
  \bibfield  {author} {\bibinfo {author} {\bibfnamefont {Y.}~\bibnamefont
  {Zhang}}, \bibinfo {author} {\bibfnamefont {S.}~\bibnamefont {Ye}}, \bibinfo
  {author} {\bibfnamefont {J.}~\bibnamefont {Zhang}}, \bibinfo {author}
  {\bibfnamefont {C.}~\bibnamefont {Hu}}, \bibinfo {author} {\bibfnamefont
  {J.}~\bibnamefont {Jiang}}, \ and\ \bibinfo {author} {\bibfnamefont
  {B.}~\bibnamefont {Jiang}},\ }\bibfield  {title} {\enquote {\bibinfo {title}
  {Efficient and {Accurate} {Simulations} of {Vibrational} and {Electronic}
  {Spectra} with {Symmetry}-{Preserving} {Neural} {Network} {Models} for
  {Tensorial} {Properties}},}\ }\href {\doibase 10.1021/acs.jpcb.0c06926}
  {\bibfield  {journal} {\bibinfo  {journal} {The Journal of Physical Chemistry
  B}\ }\textbf {\bibinfo {volume} {124}},\ \bibinfo {pages} {7284--7290}
  (\bibinfo {year} {2020})}\BibitemShut {NoStop}%
\bibitem [{\citenamefont {Deringer}\ \emph {et~al.}(2021)\citenamefont
  {Deringer}, \citenamefont {Bartók}, \citenamefont {Bernstein}, \citenamefont
  {Wilkins}, \citenamefont {Ceriotti},\ and\ \citenamefont
  {Csányi}}]{deringer_gaussian_2021}%
  \BibitemOpen
  \bibfield  {author} {\bibinfo {author} {\bibfnamefont {V.~L.}\ \bibnamefont
  {Deringer}}, \bibinfo {author} {\bibfnamefont {A.~P.}\ \bibnamefont
  {Bartók}}, \bibinfo {author} {\bibfnamefont {N.}~\bibnamefont {Bernstein}},
  \bibinfo {author} {\bibfnamefont {D.~M.}\ \bibnamefont {Wilkins}}, \bibinfo
  {author} {\bibfnamefont {M.}~\bibnamefont {Ceriotti}}, \ and\ \bibinfo
  {author} {\bibfnamefont {G.}~\bibnamefont {Csányi}},\ }\bibfield  {title}
  {\enquote {\bibinfo {title} {Gaussian {Process} {Regression} for {Materials}
  and {Molecules}},}\ }\href {\doibase 10.1021/acs.chemrev.1c00022} {\bibfield
  {journal} {\bibinfo  {journal} {Chemical Reviews}\ }\textbf {\bibinfo
  {volume} {121}},\ \bibinfo {pages} {10073--10141} (\bibinfo {year}
  {2021})}\BibitemShut {NoStop}%
\bibitem [{\citenamefont {Musil}\ and\ \citenamefont
  {Ceriotti}(2019)}]{musil_machine_2019}%
  \BibitemOpen
  \bibfield  {author} {\bibinfo {author} {\bibfnamefont {F.}~\bibnamefont
  {Musil}}\ and\ \bibinfo {author} {\bibfnamefont {M.}~\bibnamefont
  {Ceriotti}},\ }\bibfield  {title} {\enquote {\bibinfo {title} {Machine
  {Learning} at the {Atomic} {Scale}},}\ }\href {\doibase
  10.2533/chimia.2019.972} {\bibfield  {journal} {\bibinfo  {journal} {CHIMIA}\
  }\textbf {\bibinfo {volume} {73}},\ \bibinfo {pages} {972--972} (\bibinfo
  {year} {2019})}\BibitemShut {NoStop}%
\bibitem [{\citenamefont {Kurapothula}, \citenamefont {Shepherd},\ and\
  \citenamefont {Wilkins}(2022)}]{kurapothula_hydrogen-bonding_2022}%
  \BibitemOpen
  \bibfield  {author} {\bibinfo {author} {\bibfnamefont {P.~K.~J.}\
  \bibnamefont {Kurapothula}}, \bibinfo {author} {\bibfnamefont
  {S.}~\bibnamefont {Shepherd}}, \ and\ \bibinfo {author} {\bibfnamefont
  {D.~M.}\ \bibnamefont {Wilkins}},\ }\bibfield  {title} {\enquote {\bibinfo
  {title} {Hydrogen-bonding and nuclear quantum effects in clays},}\ }\href
  {\doibase 10.1063/5.0083075} {\bibfield  {journal} {\bibinfo  {journal} {The
  Journal of Chemical Physics}\ }\textbf {\bibinfo {volume} {156}},\ \bibinfo
  {pages} {084702} (\bibinfo {year} {2022})}\BibitemShut {NoStop}%
\bibitem [{\citenamefont {Behler}(2014)}]{behler_representing_2014}%
  \BibitemOpen
  \bibfield  {author} {\bibinfo {author} {\bibfnamefont {J.}~\bibnamefont
  {Behler}},\ }\bibfield  {title} {\enquote {\bibinfo {title} {Representing
  potential energy surfaces by high-dimensional neural network potentials},}\
  }\href {\doibase 10.1088/0953-8984/26/18/183001} {\bibfield  {journal}
  {\bibinfo  {journal} {Journal of Physics: Condensed Matter}\ }\textbf
  {\bibinfo {volume} {26}},\ \bibinfo {pages} {183001} (\bibinfo {year}
  {2014})}\BibitemShut {NoStop}%
\bibitem [{\citenamefont {Behler}(2015)}]{behler_constructing_2015}%
  \BibitemOpen
  \bibfield  {author} {\bibinfo {author} {\bibfnamefont {J.}~\bibnamefont
  {Behler}},\ }\bibfield  {title} {\enquote {\bibinfo {title} {Constructing
  high-dimensional neural network potentials: {A} tutorial review},}\ }\href
  {\doibase 10.1002/qua.24890} {\bibfield  {journal} {\bibinfo  {journal}
  {International Journal of Quantum Chemistry}\ }\textbf {\bibinfo {volume}
  {115}},\ \bibinfo {pages} {1032--1050} (\bibinfo {year} {2015})}\BibitemShut
  {NoStop}%
\bibitem [{\citenamefont {Parrinello}\ and\ \citenamefont
  {Rahman}(1984)}]{parrinello_study_1984}%
  \BibitemOpen
  \bibfield  {author} {\bibinfo {author} {\bibfnamefont {M.}~\bibnamefont
  {Parrinello}}\ and\ \bibinfo {author} {\bibfnamefont {A.}~\bibnamefont
  {Rahman}},\ }\bibfield  {title} {\enquote {\bibinfo {title} {Study of an {F}
  center in molten {KCl}},}\ }\href {\doibase 10.1063/1.446740} {\bibfield
  {journal} {\bibinfo  {journal} {The Journal of Chemical Physics}\ }\textbf
  {\bibinfo {volume} {80}},\ \bibinfo {pages} {860--867} (\bibinfo {year}
  {1984})}\BibitemShut {NoStop}%
\bibitem [{\citenamefont {Bish}\ and\ \citenamefont
  {Von~Dreele}(1989)}]{bish_rietveld_1989}%
  \BibitemOpen
  \bibfield  {author} {\bibinfo {author} {\bibfnamefont {D.~L.}\ \bibnamefont
  {Bish}}\ and\ \bibinfo {author} {\bibfnamefont {R.~B.}\ \bibnamefont
  {Von~Dreele}},\ }\bibfield  {title} {\enquote {\bibinfo {title} {Rietveld
  {Refinement} of {Non}-{Hydrogen} {Atomic} {Positions} in {Kaolinite}},}\
  }\href {\doibase 10.1346/CCMN.1989.0370401} {\bibfield  {journal} {\bibinfo
  {journal} {Clays and Clay Minerals}\ }\textbf {\bibinfo {volume} {37}},\
  \bibinfo {pages} {289--296} (\bibinfo {year} {1989})}\BibitemShut {NoStop}%
\bibitem [{\citenamefont {Tunega}, \citenamefont {Bučko},\ and\ \citenamefont
  {Zaoui}(2012)}]{tunega_assessment_2012}%
  \BibitemOpen
  \bibfield  {author} {\bibinfo {author} {\bibfnamefont {D.}~\bibnamefont
  {Tunega}}, \bibinfo {author} {\bibfnamefont {T.}~\bibnamefont {Bučko}}, \
  and\ \bibinfo {author} {\bibfnamefont {A.}~\bibnamefont {Zaoui}},\ }\bibfield
   {title} {\enquote {\bibinfo {title} {Assessment of ten {DFT} methods in
  predicting structures of sheet silicates: {Importance} of dispersion
  corrections},}\ }\href {\doibase 10.1063/1.4752196} {\bibfield  {journal}
  {\bibinfo  {journal} {The Journal of Chemical Physics}\ }\textbf {\bibinfo
  {volume} {137}},\ \bibinfo {pages} {114105} (\bibinfo {year}
  {2012})}\BibitemShut {NoStop}%
\bibitem [{\citenamefont {Zen}\ \emph {et~al.}(2016)\citenamefont {Zen},
  \citenamefont {Roch}, \citenamefont {Cox}, \citenamefont {Hu}, \citenamefont
  {Sorella}, \citenamefont {Alfè},\ and\ \citenamefont
  {Michaelides}}]{zen_toward_2016}%
  \BibitemOpen
  \bibfield  {author} {\bibinfo {author} {\bibfnamefont {A.}~\bibnamefont
  {Zen}}, \bibinfo {author} {\bibfnamefont {L.~M.}\ \bibnamefont {Roch}},
  \bibinfo {author} {\bibfnamefont {S.~J.}\ \bibnamefont {Cox}}, \bibinfo
  {author} {\bibfnamefont {X.~L.}\ \bibnamefont {Hu}}, \bibinfo {author}
  {\bibfnamefont {S.}~\bibnamefont {Sorella}}, \bibinfo {author} {\bibfnamefont
  {D.}~\bibnamefont {Alfè}}, \ and\ \bibinfo {author} {\bibfnamefont
  {A.}~\bibnamefont {Michaelides}},\ }\bibfield  {title} {\enquote {\bibinfo
  {title} {Toward {Accurate} {Adsorption} {Energetics} on {Clay} {Surfaces}},}\
  }\href {\doibase 10.1021/acs.jpcc.6b09559} {\bibfield  {journal} {\bibinfo
  {journal} {The Journal of Physical Chemistry C}\ }\textbf {\bibinfo {volume}
  {120}},\ \bibinfo {pages} {26402--26413} (\bibinfo {year}
  {2016})}\BibitemShut {NoStop}%
\bibitem [{\citenamefont {Voora}, \citenamefont {Al-Saidi},\ and\ \citenamefont
  {Jordan}(2011)}]{voora_density_2011}%
  \BibitemOpen
  \bibfield  {author} {\bibinfo {author} {\bibfnamefont {V.~K.}\ \bibnamefont
  {Voora}}, \bibinfo {author} {\bibfnamefont {W.~A.}\ \bibnamefont {Al-Saidi}},
  \ and\ \bibinfo {author} {\bibfnamefont {K.~D.}\ \bibnamefont {Jordan}},\
  }\bibfield  {title} {\enquote {\bibinfo {title} {Density {Functional}
  {Theory} {Study} of {Pyrophyllite} and {M}-{Montmorillonites} ({M} = {Li},
  {Na}, {K}, {Mg}, and {Ca}): {Role} of {Dispersion} {Interactions}},}\ }\href
  {\doibase 10.1021/jp201277f} {\bibfield  {journal} {\bibinfo  {journal} {The
  Journal of Physical Chemistry A}\ }\textbf {\bibinfo {volume} {115}},\
  \bibinfo {pages} {9695--9703} (\bibinfo {year} {2011})}\BibitemShut {NoStop}%
\bibitem [{\citenamefont {Chatterjee}\ \emph {et~al.}(1999)\citenamefont
  {Chatterjee}, \citenamefont {Iwasaki}, \citenamefont {Ebina},\ and\
  \citenamefont {Miyamoto}}]{chatterjee_dft_1999}%
  \BibitemOpen
  \bibfield  {author} {\bibinfo {author} {\bibfnamefont {A.}~\bibnamefont
  {Chatterjee}}, \bibinfo {author} {\bibfnamefont {T.}~\bibnamefont {Iwasaki}},
  \bibinfo {author} {\bibfnamefont {T.}~\bibnamefont {Ebina}}, \ and\ \bibinfo
  {author} {\bibfnamefont {A.}~\bibnamefont {Miyamoto}},\ }\bibfield  {title}
  {\enquote {\bibinfo {title} {A {DFT} study on clay–cation–water
  interaction in montmorillonite and beidellite},}\ }\href {\doibase
  10.1016/S0927-0256(98)00083-4} {\bibfield  {journal} {\bibinfo  {journal}
  {Computational Materials Science}\ }\textbf {\bibinfo {volume} {14}},\
  \bibinfo {pages} {119--124} (\bibinfo {year} {1999})}\BibitemShut {NoStop}%
\bibitem [{\citenamefont {Wungu}\ \emph {et~al.}(2011)\citenamefont {Wungu},
  \citenamefont {Aspera}, \citenamefont {David}, \citenamefont {Dipojono},
  \citenamefont {Nakanishi},\ and\ \citenamefont
  {Kasai}}]{wungu_absorption_2011}%
  \BibitemOpen
  \bibfield  {author} {\bibinfo {author} {\bibfnamefont {T.~D.~K.}\
  \bibnamefont {Wungu}}, \bibinfo {author} {\bibfnamefont {S.~M.}\ \bibnamefont
  {Aspera}}, \bibinfo {author} {\bibfnamefont {M.~Y.}\ \bibnamefont {David}},
  \bibinfo {author} {\bibfnamefont {H.~K.}\ \bibnamefont {Dipojono}}, \bibinfo
  {author} {\bibfnamefont {H.}~\bibnamefont {Nakanishi}}, \ and\ \bibinfo
  {author} {\bibfnamefont {H.}~\bibnamefont {Kasai}},\ }\bibfield  {title}
  {\enquote {\bibinfo {title} {Absorption of {Lithium} in {Montmorillonite}:
  {A} {Density} {Functional} {Theory} ({DFT}) {Study}},}\ }\href {\doibase
  10.1166/jnn.2011.3913} {\bibfield  {journal} {\bibinfo  {journal} {Journal of
  Nanoscience and Nanotechnology}\ }\textbf {\bibinfo {volume} {11}},\ \bibinfo
  {pages} {2793--2801} (\bibinfo {year} {2011})}\BibitemShut {NoStop}%
\bibitem [{\citenamefont {Balan}\ \emph {et~al.}(2001)\citenamefont {Balan},
  \citenamefont {Saitta}, \citenamefont {Mauri},\ and\ \citenamefont
  {Calas}}]{balan_first-principles_2001}%
  \BibitemOpen
  \bibfield  {author} {\bibinfo {author} {\bibfnamefont {E.}~\bibnamefont
  {Balan}}, \bibinfo {author} {\bibfnamefont {A.~M.}\ \bibnamefont {Saitta}},
  \bibinfo {author} {\bibfnamefont {F.}~\bibnamefont {Mauri}}, \ and\ \bibinfo
  {author} {\bibfnamefont {G.}~\bibnamefont {Calas}},\ }\bibfield  {title}
  {\enquote {\bibinfo {title} {First-principles modeling of the infrared
  spectrum of kaolinite},}\ }\href {\doibase 10.2138/am-2001-11-1201}
  {\bibfield  {journal} {\bibinfo  {journal} {American Mineralogist}\ }\textbf
  {\bibinfo {volume} {86}},\ \bibinfo {pages} {1321--1330} (\bibinfo {year}
  {2001})}\BibitemShut {NoStop}%
\bibitem [{\citenamefont {Balan}\ \emph {et~al.}(2005)\citenamefont {Balan},
  \citenamefont {Lazzeri}, \citenamefont {Saitta}, \citenamefont {Allard},
  \citenamefont {Fuchs},\ and\ \citenamefont
  {Mauri}}]{balan_first-principles_2005}%
  \BibitemOpen
  \bibfield  {author} {\bibinfo {author} {\bibfnamefont {E.}~\bibnamefont
  {Balan}}, \bibinfo {author} {\bibfnamefont {M.}~\bibnamefont {Lazzeri}},
  \bibinfo {author} {\bibfnamefont {A.~M.}\ \bibnamefont {Saitta}}, \bibinfo
  {author} {\bibfnamefont {T.}~\bibnamefont {Allard}}, \bibinfo {author}
  {\bibfnamefont {Y.}~\bibnamefont {Fuchs}}, \ and\ \bibinfo {author}
  {\bibfnamefont {F.}~\bibnamefont {Mauri}},\ }\bibfield  {title} {\enquote
  {\bibinfo {title} {First-principles study of {OH}-stretching modes in
  kaolinite, dickite, and nacrite},}\ }\href {\doibase 10.2138/am.2005.1675}
  {\bibfield  {journal} {\bibinfo  {journal} {American Mineralogist}\ }\textbf
  {\bibinfo {volume} {90}},\ \bibinfo {pages} {50--60} (\bibinfo {year}
  {2005})}\BibitemShut {NoStop}%
\bibitem [{\citenamefont {Balan}\ \emph {et~al.}(2011)\citenamefont {Balan},
  \citenamefont {Fritsch}, \citenamefont {Allard}, \citenamefont {Morin},
  \citenamefont {Guillaumet}, \citenamefont {Delattre}, \citenamefont
  {Blanchard},\ and\ \citenamefont {Calas}}]{balan_spectroscopic_2011}%
  \BibitemOpen
  \bibfield  {author} {\bibinfo {author} {\bibfnamefont {E.}~\bibnamefont
  {Balan}}, \bibinfo {author} {\bibfnamefont {E.}~\bibnamefont {Fritsch}},
  \bibinfo {author} {\bibfnamefont {T.}~\bibnamefont {Allard}}, \bibinfo
  {author} {\bibfnamefont {G.}~\bibnamefont {Morin}}, \bibinfo {author}
  {\bibfnamefont {M.}~\bibnamefont {Guillaumet}}, \bibinfo {author}
  {\bibfnamefont {S.}~\bibnamefont {Delattre}}, \bibinfo {author}
  {\bibfnamefont {M.}~\bibnamefont {Blanchard}}, \ and\ \bibinfo {author}
  {\bibfnamefont {G.}~\bibnamefont {Calas}},\ }\bibfield  {title} {\enquote
  {\bibinfo {title} {Spectroscopic investigation and theoretical modeling of
  kaolinite-group minerals and other low-temperature phases {\textbar}
  {Elsevier} {Enhanced} {Reader}},}\ }\href {\doibase
  10.1016/j.crte.2010.10.006} {\bibfield  {journal} {\bibinfo  {journal}
  {Comptes Rendus Geoscience}\ }\textbf {\bibinfo {volume} {343}},\ \bibinfo
  {pages} {177--187} (\bibinfo {year} {2011})}\BibitemShut {NoStop}%
\bibitem [{\citenamefont {Elstner}\ \emph {et~al.}(1998)\citenamefont
  {Elstner}, \citenamefont {Porezag}, \citenamefont {Jungnickel}, \citenamefont
  {Elsner}, \citenamefont {Haugk}, \citenamefont {Frauenheim}, \citenamefont
  {Suhai},\ and\ \citenamefont
  {Seifert}}]{elstner_self-consistent-charge_1998}%
  \BibitemOpen
  \bibfield  {author} {\bibinfo {author} {\bibfnamefont {M.}~\bibnamefont
  {Elstner}}, \bibinfo {author} {\bibfnamefont {D.}~\bibnamefont {Porezag}},
  \bibinfo {author} {\bibfnamefont {G.}~\bibnamefont {Jungnickel}}, \bibinfo
  {author} {\bibfnamefont {J.}~\bibnamefont {Elsner}}, \bibinfo {author}
  {\bibfnamefont {M.}~\bibnamefont {Haugk}}, \bibinfo {author} {\bibfnamefont
  {T.}~\bibnamefont {Frauenheim}}, \bibinfo {author} {\bibfnamefont
  {S.}~\bibnamefont {Suhai}}, \ and\ \bibinfo {author} {\bibfnamefont
  {G.}~\bibnamefont {Seifert}},\ }\bibfield  {title} {\enquote {\bibinfo
  {title} {Self-consistent-charge density-functional tight-binding method for
  simulations of complex materials properties},}\ }\href {\doibase
  10.1103/PhysRevB.58.7260} {\bibfield  {journal} {\bibinfo  {journal}
  {Physical Review B}\ }\textbf {\bibinfo {volume} {58}},\ \bibinfo {pages}
  {7260--7268} (\bibinfo {year} {1998})}\BibitemShut {NoStop}%
\bibitem [{\citenamefont {Porezag}\ \emph {et~al.}(1995)\citenamefont
  {Porezag}, \citenamefont {Frauenheim}, \citenamefont {Köhler}, \citenamefont
  {Seifert},\ and\ \citenamefont {Kaschner}}]{porezag_construction_1995}%
  \BibitemOpen
  \bibfield  {author} {\bibinfo {author} {\bibfnamefont {D.}~\bibnamefont
  {Porezag}}, \bibinfo {author} {\bibfnamefont {T.}~\bibnamefont {Frauenheim}},
  \bibinfo {author} {\bibfnamefont {T.}~\bibnamefont {Köhler}}, \bibinfo
  {author} {\bibfnamefont {G.}~\bibnamefont {Seifert}}, \ and\ \bibinfo
  {author} {\bibfnamefont {R.}~\bibnamefont {Kaschner}},\ }\bibfield  {title}
  {\enquote {\bibinfo {title} {Construction of tight-binding-like potentials on
  the basis of density-functional theory: {Application} to carbon},}\ }\href
  {\doibase 10.1103/PhysRevB.51.12947} {\bibfield  {journal} {\bibinfo
  {journal} {Physical Review B}\ }\textbf {\bibinfo {volume} {51}},\ \bibinfo
  {pages} {12947--12957} (\bibinfo {year} {1995})}\BibitemShut {NoStop}%
\bibitem [{\citenamefont {Seifert}, \citenamefont {Porezag},\ and\
  \citenamefont {Frauenheim}(1996)}]{seifert_calculations_1996}%
  \BibitemOpen
  \bibfield  {author} {\bibinfo {author} {\bibfnamefont {G.}~\bibnamefont
  {Seifert}}, \bibinfo {author} {\bibfnamefont {D.}~\bibnamefont {Porezag}}, \
  and\ \bibinfo {author} {\bibfnamefont {T.}~\bibnamefont {Frauenheim}},\
  }\bibfield  {title} {\enquote {\bibinfo {title} {Calculations of molecules,
  clusters, and solids with a simplified {LCAO}-{DFT}-{LDA} scheme},}\ }\href
  {\doibase 10.1002/(SICI)1097-461X(1996)58:2<185::AID-QUA7>3.0.CO;2-U}
  {\bibfield  {journal} {\bibinfo  {journal} {International Journal of Quantum
  Chemistry}\ }\textbf {\bibinfo {volume} {58}},\ \bibinfo {pages} {185--192}
  (\bibinfo {year} {1996})}\BibitemShut {NoStop}%
\bibitem [{\citenamefont {Frenzel}\ \emph {et~al.}(2009)\citenamefont
  {Frenzel}, \citenamefont {Oliveira}, \citenamefont {Jardillier},
  \citenamefont {Heine},\ and\ \citenamefont
  {Seifert}}]{frenzel_semirelativistic_nodate}%
  \BibitemOpen
  \bibfield  {author} {\bibinfo {author} {\bibfnamefont {J.}~\bibnamefont
  {Frenzel}}, \bibinfo {author} {\bibfnamefont {A.~F.}\ \bibnamefont
  {Oliveira}}, \bibinfo {author} {\bibfnamefont {N.}~\bibnamefont
  {Jardillier}}, \bibinfo {author} {\bibfnamefont {T.}~\bibnamefont {Heine}}, \
  and\ \bibinfo {author} {\bibfnamefont {G.}~\bibnamefont {Seifert}},\
  }\bibfield  {title} {\enquote {\bibinfo {title} {Semi­relativistic,
  self­consistent charge {Slater}­{Koster} tables for density­functional
  based tight­binding ({DFTB}) for materials science simulations},}\
  }\href@noop {} {\bibfield  {journal} {\bibinfo  {journal} {Zeolites}\
  }\textbf {\bibinfo {volume} {2}},\ \bibinfo {pages} {7} (\bibinfo {year}
  {2004--2009})}\BibitemShut {NoStop}%
\bibitem [{\citenamefont {Kühne}\ \emph {et~al.}(2020)\citenamefont {Kühne},
  \citenamefont {Iannuzzi}, \citenamefont {Del~Ben}, \citenamefont {Rybkin},
  \citenamefont {Seewald}, \citenamefont {Stein}, \citenamefont {Laino},
  \citenamefont {Khaliullin}, \citenamefont {Schütt}, \citenamefont
  {Schiffmann}, \citenamefont {Golze}, \citenamefont {Wilhelm}, \citenamefont
  {Chulkov}, \citenamefont {Bani-Hashemian}, \citenamefont {Weber},
  \citenamefont {Borštnik}, \citenamefont {Taillefumier}, \citenamefont
  {Jakobovits}, \citenamefont {Lazzaro}, \citenamefont {Pabst}, \citenamefont
  {Müller}, \citenamefont {Schade}, \citenamefont {Guidon}, \citenamefont
  {Andermatt}, \citenamefont {Holmberg}, \citenamefont {Schenter},
  \citenamefont {Hehn}, \citenamefont {Bussy}, \citenamefont {Belleflamme},
  \citenamefont {Tabacchi}, \citenamefont {Glöß}, \citenamefont {Lass},
  \citenamefont {Bethune}, \citenamefont {Mundy}, \citenamefont {Plessl},
  \citenamefont {Watkins}, \citenamefont {VandeVondele}, \citenamefont
  {Krack},\ and\ \citenamefont {Hutter}}]{kuhne_cp2k_2020}%
  \BibitemOpen
  \bibfield  {author} {\bibinfo {author} {\bibfnamefont {T.~D.}\ \bibnamefont
  {Kühne}}, \bibinfo {author} {\bibfnamefont {M.}~\bibnamefont {Iannuzzi}},
  \bibinfo {author} {\bibfnamefont {M.}~\bibnamefont {Del~Ben}}, \bibinfo
  {author} {\bibfnamefont {V.~V.}\ \bibnamefont {Rybkin}}, \bibinfo {author}
  {\bibfnamefont {P.}~\bibnamefont {Seewald}}, \bibinfo {author} {\bibfnamefont
  {F.}~\bibnamefont {Stein}}, \bibinfo {author} {\bibfnamefont
  {T.}~\bibnamefont {Laino}}, \bibinfo {author} {\bibfnamefont {R.~Z.}\
  \bibnamefont {Khaliullin}}, \bibinfo {author} {\bibfnamefont
  {O.}~\bibnamefont {Schütt}}, \bibinfo {author} {\bibfnamefont
  {F.}~\bibnamefont {Schiffmann}}, \bibinfo {author} {\bibfnamefont
  {D.}~\bibnamefont {Golze}}, \bibinfo {author} {\bibfnamefont
  {J.}~\bibnamefont {Wilhelm}}, \bibinfo {author} {\bibfnamefont
  {S.}~\bibnamefont {Chulkov}}, \bibinfo {author} {\bibfnamefont {M.~H.}\
  \bibnamefont {Bani-Hashemian}}, \bibinfo {author} {\bibfnamefont
  {V.}~\bibnamefont {Weber}}, \bibinfo {author} {\bibfnamefont
  {U.}~\bibnamefont {Borštnik}}, \bibinfo {author} {\bibfnamefont
  {M.}~\bibnamefont {Taillefumier}}, \bibinfo {author} {\bibfnamefont {A.~S.}\
  \bibnamefont {Jakobovits}}, \bibinfo {author} {\bibfnamefont
  {A.}~\bibnamefont {Lazzaro}}, \bibinfo {author} {\bibfnamefont
  {H.}~\bibnamefont {Pabst}}, \bibinfo {author} {\bibfnamefont
  {T.}~\bibnamefont {Müller}}, \bibinfo {author} {\bibfnamefont
  {R.}~\bibnamefont {Schade}}, \bibinfo {author} {\bibfnamefont
  {M.}~\bibnamefont {Guidon}}, \bibinfo {author} {\bibfnamefont
  {S.}~\bibnamefont {Andermatt}}, \bibinfo {author} {\bibfnamefont
  {N.}~\bibnamefont {Holmberg}}, \bibinfo {author} {\bibfnamefont {G.~K.}\
  \bibnamefont {Schenter}}, \bibinfo {author} {\bibfnamefont {A.}~\bibnamefont
  {Hehn}}, \bibinfo {author} {\bibfnamefont {A.}~\bibnamefont {Bussy}},
  \bibinfo {author} {\bibfnamefont {F.}~\bibnamefont {Belleflamme}}, \bibinfo
  {author} {\bibfnamefont {G.}~\bibnamefont {Tabacchi}}, \bibinfo {author}
  {\bibfnamefont {A.}~\bibnamefont {Glöß}}, \bibinfo {author} {\bibfnamefont
  {M.}~\bibnamefont {Lass}}, \bibinfo {author} {\bibfnamefont {I.}~\bibnamefont
  {Bethune}}, \bibinfo {author} {\bibfnamefont {C.~J.}\ \bibnamefont {Mundy}},
  \bibinfo {author} {\bibfnamefont {C.}~\bibnamefont {Plessl}}, \bibinfo
  {author} {\bibfnamefont {M.}~\bibnamefont {Watkins}}, \bibinfo {author}
  {\bibfnamefont {J.}~\bibnamefont {VandeVondele}}, \bibinfo {author}
  {\bibfnamefont {M.}~\bibnamefont {Krack}}, \ and\ \bibinfo {author}
  {\bibfnamefont {J.}~\bibnamefont {Hutter}},\ }\bibfield  {title} {\enquote
  {\bibinfo {title} {{CP2K}: {An} electronic structure and molecular dynamics
  software package - {Quickstep}: {Efficient} and accurate electronic structure
  calculations},}\ }\href {\doibase 10.1063/5.0007045} {\bibfield  {journal}
  {\bibinfo  {journal} {The Journal of Chemical Physics}\ }\textbf {\bibinfo
  {volume} {152}},\ \bibinfo {pages} {194103} (\bibinfo {year}
  {2020})}\BibitemShut {NoStop}%
\bibitem [{\citenamefont {Imbalzano}\ \emph {et~al.}(2018)\citenamefont
  {Imbalzano}, \citenamefont {Anelli}, \citenamefont {Giofré}, \citenamefont
  {Klees}, \citenamefont {Behler},\ and\ \citenamefont
  {Ceriotti}}]{imbalzano_automatic_2018}%
  \BibitemOpen
  \bibfield  {author} {\bibinfo {author} {\bibfnamefont {G.}~\bibnamefont
  {Imbalzano}}, \bibinfo {author} {\bibfnamefont {A.}~\bibnamefont {Anelli}},
  \bibinfo {author} {\bibfnamefont {D.}~\bibnamefont {Giofré}}, \bibinfo
  {author} {\bibfnamefont {S.}~\bibnamefont {Klees}}, \bibinfo {author}
  {\bibfnamefont {J.}~\bibnamefont {Behler}}, \ and\ \bibinfo {author}
  {\bibfnamefont {M.}~\bibnamefont {Ceriotti}},\ }\bibfield  {title} {\enquote
  {\bibinfo {title} {Automatic selection of atomic fingerprints and reference
  configurations for machine-learning potentials},}\ }\href {\doibase
  10.1063/1.5024611} {\bibfield  {journal} {\bibinfo  {journal} {The Journal of
  Chemical Physics}\ }\textbf {\bibinfo {volume} {148}},\ \bibinfo {pages}
  {241730} (\bibinfo {year} {2018})}\BibitemShut {NoStop}%
\bibitem [{\citenamefont {Perdew}, \citenamefont {Burke},\ and\ \citenamefont
  {Ernzerhof}(1996)}]{perdew_generalized_1996}%
  \BibitemOpen
  \bibfield  {author} {\bibinfo {author} {\bibfnamefont {J.~P.}\ \bibnamefont
  {Perdew}}, \bibinfo {author} {\bibfnamefont {K.}~\bibnamefont {Burke}}, \
  and\ \bibinfo {author} {\bibfnamefont {M.}~\bibnamefont {Ernzerhof}},\
  }\bibfield  {title} {\enquote {\bibinfo {title} {Generalized {Gradient}
  {Approximation} {Made} {Simple}},}\ }\href {\doibase
  10.1103/PhysRevLett.77.3865} {\bibfield  {journal} {\bibinfo  {journal}
  {Physical Review Letters}\ }\textbf {\bibinfo {volume} {77}},\ \bibinfo
  {pages} {3865--3868} (\bibinfo {year} {1996})}\BibitemShut {NoStop}%
\bibitem [{\citenamefont {Zhang}\ and\ \citenamefont
  {Yang}(1998)}]{zhang_comment_1998}%
  \BibitemOpen
  \bibfield  {author} {\bibinfo {author} {\bibfnamefont {Y.}~\bibnamefont
  {Zhang}}\ and\ \bibinfo {author} {\bibfnamefont {W.}~\bibnamefont {Yang}},\
  }\bibfield  {title} {\enquote {\bibinfo {title} {Comment on ``{Generalized}
  {Gradient} {Approximation} {Made} {Simple}''},}\ }\href {\doibase
  10.1103/PhysRevLett.80.890} {\bibfield  {journal} {\bibinfo  {journal}
  {Physical Review Letters}\ }\textbf {\bibinfo {volume} {80}},\ \bibinfo
  {pages} {890--890} (\bibinfo {year} {1998})}\BibitemShut {NoStop}%
\bibitem [{\citenamefont {Grimme}\ \emph {et~al.}(2010)\citenamefont {Grimme},
  \citenamefont {Antony}, \citenamefont {Ehrlich},\ and\ \citenamefont
  {Krieg}}]{grimme_consistent_2010}%
  \BibitemOpen
  \bibfield  {author} {\bibinfo {author} {\bibfnamefont {S.}~\bibnamefont
  {Grimme}}, \bibinfo {author} {\bibfnamefont {J.}~\bibnamefont {Antony}},
  \bibinfo {author} {\bibfnamefont {S.}~\bibnamefont {Ehrlich}}, \ and\
  \bibinfo {author} {\bibfnamefont {H.}~\bibnamefont {Krieg}},\ }\bibfield
  {title} {\enquote {\bibinfo {title} {A consistent and accurate ab initio
  parametrization of density functional dispersion correction ({DFT}-{D}) for
  the 94 elements {H}-{Pu}},}\ }\href {\doibase 10.1063/1.3382344} {\bibfield
  {journal} {\bibinfo  {journal} {The Journal of Chemical Physics}\ }\textbf
  {\bibinfo {volume} {132}},\ \bibinfo {pages} {154104} (\bibinfo {year}
  {2010})}\BibitemShut {NoStop}%
\bibitem [{\citenamefont {Grimme}, \citenamefont {Ehrlich},\ and\ \citenamefont
  {Goerigk}(2011)}]{grimme_effect_2011}%
  \BibitemOpen
  \bibfield  {author} {\bibinfo {author} {\bibfnamefont {S.}~\bibnamefont
  {Grimme}}, \bibinfo {author} {\bibfnamefont {S.}~\bibnamefont {Ehrlich}}, \
  and\ \bibinfo {author} {\bibfnamefont {L.}~\bibnamefont {Goerigk}},\
  }\bibfield  {title} {\enquote {\bibinfo {title} {Effect of the damping
  function in dispersion corrected density functional theory},}\ }\href
  {\doibase 10.1002/jcc.21759} {\bibfield  {journal} {\bibinfo  {journal}
  {Journal of Computational Chemistry}\ }\textbf {\bibinfo {volume} {32}},\
  \bibinfo {pages} {1456--1465} (\bibinfo {year} {2011})}\BibitemShut {NoStop}%
\bibitem [{\citenamefont {Dion}\ \emph {et~al.}(2004)\citenamefont {Dion},
  \citenamefont {Rydberg}, \citenamefont {Schröder}, \citenamefont
  {Langreth},\ and\ \citenamefont {Lundqvist}}]{dion_van_2004}%
  \BibitemOpen
  \bibfield  {author} {\bibinfo {author} {\bibfnamefont {M.}~\bibnamefont
  {Dion}}, \bibinfo {author} {\bibfnamefont {H.}~\bibnamefont {Rydberg}},
  \bibinfo {author} {\bibfnamefont {E.}~\bibnamefont {Schröder}}, \bibinfo
  {author} {\bibfnamefont {D.~C.}\ \bibnamefont {Langreth}}, \ and\ \bibinfo
  {author} {\bibfnamefont {B.~I.}\ \bibnamefont {Lundqvist}},\ }\bibfield
  {title} {\enquote {\bibinfo {title} {Van der {Waals} {Density} {Functional}
  for {General} {Geometries}},}\ }\href {\doibase
  10.1103/PhysRevLett.92.246401} {\bibfield  {journal} {\bibinfo  {journal}
  {Physical Review Letters}\ }\textbf {\bibinfo {volume} {92}},\ \bibinfo
  {pages} {246401} (\bibinfo {year} {2004})}\BibitemShut {NoStop}%
\bibitem [{\citenamefont {Ohto}\ \emph {et~al.}(2019)\citenamefont {Ohto},
  \citenamefont {Dodia}, \citenamefont {Xu}, \citenamefont {Imoto},
  \citenamefont {Tang}, \citenamefont {Zysk}, \citenamefont {Kühne},
  \citenamefont {Shigeta}, \citenamefont {Bonn}, \citenamefont {Wu},\ and\
  \citenamefont {Nagata}}]{ohto_accessing_2019}%
  \BibitemOpen
  \bibfield  {author} {\bibinfo {author} {\bibfnamefont {T.}~\bibnamefont
  {Ohto}}, \bibinfo {author} {\bibfnamefont {M.}~\bibnamefont {Dodia}},
  \bibinfo {author} {\bibfnamefont {J.}~\bibnamefont {Xu}}, \bibinfo {author}
  {\bibfnamefont {S.}~\bibnamefont {Imoto}}, \bibinfo {author} {\bibfnamefont
  {F.}~\bibnamefont {Tang}}, \bibinfo {author} {\bibfnamefont {F.}~\bibnamefont
  {Zysk}}, \bibinfo {author} {\bibfnamefont {T.~D.}\ \bibnamefont {Kühne}},
  \bibinfo {author} {\bibfnamefont {Y.}~\bibnamefont {Shigeta}}, \bibinfo
  {author} {\bibfnamefont {M.}~\bibnamefont {Bonn}}, \bibinfo {author}
  {\bibfnamefont {X.}~\bibnamefont {Wu}}, \ and\ \bibinfo {author}
  {\bibfnamefont {Y.}~\bibnamefont {Nagata}},\ }\bibfield  {title} {\enquote
  {\bibinfo {title} {Accessing the {Accuracy} of {Density} {Functional}
  {Theory} through {Structure} and {Dynamics} of the {Water}–{Air}
  {Interface}},}\ }\href {\doibase 10.1021/acs.jpclett.9b01983} {\bibfield
  {journal} {\bibinfo  {journal} {J. Phys. Chem. Lett.}\ }\textbf {\bibinfo
  {volume} {10}},\ \bibinfo {pages} {4914--4919} (\bibinfo {year}
  {2019})}\BibitemShut {NoStop}%
\bibitem [{\citenamefont {Kobayashi}, \citenamefont {Yamaguchi},\ and\
  \citenamefont {Okumura}(2022)}]{kobayashi_machine_2022}%
  \BibitemOpen
  \bibfield  {author} {\bibinfo {author} {\bibfnamefont {K.}~\bibnamefont
  {Kobayashi}}, \bibinfo {author} {\bibfnamefont {A.}~\bibnamefont
  {Yamaguchi}}, \ and\ \bibinfo {author} {\bibfnamefont {M.}~\bibnamefont
  {Okumura}},\ }\bibfield  {title} {\enquote {\bibinfo {title} {Machine
  learning potentials of kaolinite based on the potential energy surfaces of
  {GGA} and meta-{GGA} density functional theory},}\ }\href {\doibase
  10.1016/j.clay.2022.106596} {\bibfield  {journal} {\bibinfo  {journal}
  {Applied Clay Science}\ }\textbf {\bibinfo {volume} {228}},\ \bibinfo {pages}
  {106596} (\bibinfo {year} {2022})}\BibitemShut {NoStop}%
\bibitem [{\citenamefont {Behler}(2021)}]{behler_four_2021}%
  \BibitemOpen
  \bibfield  {author} {\bibinfo {author} {\bibfnamefont {J.}~\bibnamefont
  {Behler}},\ }\bibfield  {title} {\enquote {\bibinfo {title} {Four
  {Generations} of {High}-{Dimensional} {Neural} {Network} {Potentials}},}\
  }\href {\doibase 10.1021/acs.chemrev.0c00868} {\bibfield  {journal} {\bibinfo
   {journal} {Chemical Reviews}\ }\textbf {\bibinfo {volume} {121}},\ \bibinfo
  {pages} {10037--10072} (\bibinfo {year} {2021})}\BibitemShut {NoStop}%
\bibitem [{\citenamefont {Singraber}, \citenamefont {Behler},\ and\
  \citenamefont {Dellago}(2019)}]{singraber_library-based_2019}%
  \BibitemOpen
  \bibfield  {author} {\bibinfo {author} {\bibfnamefont {A.}~\bibnamefont
  {Singraber}}, \bibinfo {author} {\bibfnamefont {J.}~\bibnamefont {Behler}}, \
  and\ \bibinfo {author} {\bibfnamefont {C.}~\bibnamefont {Dellago}},\
  }\bibfield  {title} {\enquote {\bibinfo {title} {Library-{Based} {LAMMPS}
  {Implementation} of {High}-{Dimensional} {Neural} {Network} {Potentials}},}\
  }\href {\doibase 10.1021/acs.jctc.8b00770} {\bibfield  {journal} {\bibinfo
  {journal} {Journal of Chemical Theory and Computation}\ }\textbf {\bibinfo
  {volume} {15}},\ \bibinfo {pages} {1827--1840} (\bibinfo {year}
  {2019})}\BibitemShut {NoStop}%
\bibitem [{\citenamefont {Singraber}\ \emph {et~al.}(2019)\citenamefont
  {Singraber}, \citenamefont {Morawietz}, \citenamefont {Behler},\ and\
  \citenamefont {Dellago}}]{singraber_parallel_2019}%
  \BibitemOpen
  \bibfield  {author} {\bibinfo {author} {\bibfnamefont {A.}~\bibnamefont
  {Singraber}}, \bibinfo {author} {\bibfnamefont {T.}~\bibnamefont
  {Morawietz}}, \bibinfo {author} {\bibfnamefont {J.}~\bibnamefont {Behler}}, \
  and\ \bibinfo {author} {\bibfnamefont {C.}~\bibnamefont {Dellago}},\
  }\bibfield  {title} {\enquote {\bibinfo {title} {Parallel {Multistream}
  {Training} of {High}-{Dimensional} {Neural} {Network} {Potentials}},}\ }\href
  {\doibase 10.1021/acs.jctc.8b01092} {\bibfield  {journal} {\bibinfo
  {journal} {Journal of Chemical Theory and Computation}\ }\textbf {\bibinfo
  {volume} {15}},\ \bibinfo {pages} {3075--3092} (\bibinfo {year}
  {2019})}\BibitemShut {NoStop}%
\bibitem [{\citenamefont {Huber}\ \emph {et~al.}(2020)\citenamefont {Huber},
  \citenamefont {Zoupanos}, \citenamefont {Uhrin}, \citenamefont {Talirz},
  \citenamefont {Kahle}, \citenamefont {Häuselmann}, \citenamefont {Gresch},
  \citenamefont {Müller}, \citenamefont {Yakutovich}, \citenamefont
  {Andersen}, \citenamefont {Ramirez}, \citenamefont {Adorf}, \citenamefont
  {Gargiulo}, \citenamefont {Kumbhar}, \citenamefont {Passaro}, \citenamefont
  {Johnston}, \citenamefont {Merkys}, \citenamefont {Cepellotti}, \citenamefont
  {Mounet}, \citenamefont {Marzari}, \citenamefont {Kozinsky},\ and\
  \citenamefont {Pizzi}}]{huber_aiida_2020}%
  \BibitemOpen
  \bibfield  {author} {\bibinfo {author} {\bibfnamefont {S.~P.}\ \bibnamefont
  {Huber}}, \bibinfo {author} {\bibfnamefont {S.}~\bibnamefont {Zoupanos}},
  \bibinfo {author} {\bibfnamefont {M.}~\bibnamefont {Uhrin}}, \bibinfo
  {author} {\bibfnamefont {L.}~\bibnamefont {Talirz}}, \bibinfo {author}
  {\bibfnamefont {L.}~\bibnamefont {Kahle}}, \bibinfo {author} {\bibfnamefont
  {R.}~\bibnamefont {Häuselmann}}, \bibinfo {author} {\bibfnamefont
  {D.}~\bibnamefont {Gresch}}, \bibinfo {author} {\bibfnamefont
  {T.}~\bibnamefont {Müller}}, \bibinfo {author} {\bibfnamefont {A.~V.}\
  \bibnamefont {Yakutovich}}, \bibinfo {author} {\bibfnamefont {C.~W.}\
  \bibnamefont {Andersen}}, \bibinfo {author} {\bibfnamefont {F.~F.}\
  \bibnamefont {Ramirez}}, \bibinfo {author} {\bibfnamefont {C.~S.}\
  \bibnamefont {Adorf}}, \bibinfo {author} {\bibfnamefont {F.}~\bibnamefont
  {Gargiulo}}, \bibinfo {author} {\bibfnamefont {S.}~\bibnamefont {Kumbhar}},
  \bibinfo {author} {\bibfnamefont {E.}~\bibnamefont {Passaro}}, \bibinfo
  {author} {\bibfnamefont {C.}~\bibnamefont {Johnston}}, \bibinfo {author}
  {\bibfnamefont {A.}~\bibnamefont {Merkys}}, \bibinfo {author} {\bibfnamefont
  {A.}~\bibnamefont {Cepellotti}}, \bibinfo {author} {\bibfnamefont
  {N.}~\bibnamefont {Mounet}}, \bibinfo {author} {\bibfnamefont
  {N.}~\bibnamefont {Marzari}}, \bibinfo {author} {\bibfnamefont
  {B.}~\bibnamefont {Kozinsky}}, \ and\ \bibinfo {author} {\bibfnamefont
  {G.}~\bibnamefont {Pizzi}},\ }\bibfield  {title} {\enquote {\bibinfo {title}
  {{AiiDA} 1.0, a scalable computational infrastructure for automated
  reproducible workflows and data provenance},}\ }\href {\doibase
  10.1038/s41597-020-00638-4} {\bibfield  {journal} {\bibinfo  {journal}
  {Scientific Data}\ }\textbf {\bibinfo {volume} {7}},\ \bibinfo {pages} {300}
  (\bibinfo {year} {2020})}\BibitemShut {NoStop}%
\bibitem [{\citenamefont {Uhrin}\ \emph {et~al.}(2021)\citenamefont {Uhrin},
  \citenamefont {Huber}, \citenamefont {Yu}, \citenamefont {Marzari},\ and\
  \citenamefont {Pizzi}}]{uhrin_workflows_2021}%
  \BibitemOpen
  \bibfield  {author} {\bibinfo {author} {\bibfnamefont {M.}~\bibnamefont
  {Uhrin}}, \bibinfo {author} {\bibfnamefont {S.~P.}\ \bibnamefont {Huber}},
  \bibinfo {author} {\bibfnamefont {J.}~\bibnamefont {Yu}}, \bibinfo {author}
  {\bibfnamefont {N.}~\bibnamefont {Marzari}}, \ and\ \bibinfo {author}
  {\bibfnamefont {G.}~\bibnamefont {Pizzi}},\ }\bibfield  {title} {\enquote
  {\bibinfo {title} {Workflows in {AiiDA}: {Engineering} a high-throughput,
  event-based engine for robust and modular computational workflows},}\ }\href
  {\doibase 10.1016/j.commatsci.2020.110086} {\bibfield  {journal} {\bibinfo
  {journal} {Computational Materials Science}\ }\textbf {\bibinfo {volume}
  {187}},\ \bibinfo {pages} {110086} (\bibinfo {year} {2021})}\BibitemShut
  {NoStop}%
\bibitem [{\citenamefont {Schran}, \citenamefont {Behler},\ and\ \citenamefont
  {Marx}(2020)}]{schran_automated_2020}%
  \BibitemOpen
  \bibfield  {author} {\bibinfo {author} {\bibfnamefont {C.}~\bibnamefont
  {Schran}}, \bibinfo {author} {\bibfnamefont {J.}~\bibnamefont {Behler}}, \
  and\ \bibinfo {author} {\bibfnamefont {D.}~\bibnamefont {Marx}},\ }\bibfield
  {title} {\enquote {\bibinfo {title} {Automated {Fitting} of {Neural}
  {Network} {Potentials} at {Coupled} {Cluster} {Accuracy}: {Protonated}
  {Water} {Clusters} as {Testing} {Ground}},}\ }\href {\doibase
  10.1021/acs.jctc.9b00805} {\bibfield  {journal} {\bibinfo  {journal} {Journal
  of Chemical Theory and Computation}\ }\textbf {\bibinfo {volume} {16}},\
  \bibinfo {pages} {88--99} (\bibinfo {year} {2020})}\BibitemShut {NoStop}%
\bibitem [{\citenamefont {Bussi}\ and\ \citenamefont
  {Parrinello}(2007)}]{bussi_accurate_2007}%
  \BibitemOpen
  \bibfield  {author} {\bibinfo {author} {\bibfnamefont {G.}~\bibnamefont
  {Bussi}}\ and\ \bibinfo {author} {\bibfnamefont {M.}~\bibnamefont
  {Parrinello}},\ }\bibfield  {title} {\enquote {\bibinfo {title} {Accurate
  sampling using {Langevin} dynamics},}\ }\href {\doibase
  10.1103/PhysRevE.75.056707} {\bibfield  {journal} {\bibinfo  {journal} {Phys.
  Rev. E}\ }\textbf {\bibinfo {volume} {75}},\ \bibinfo {pages} {056707}
  (\bibinfo {year} {2007})}\BibitemShut {NoStop}%
\bibitem [{\citenamefont {Ceriotti}\ \emph {et~al.}(2010)\citenamefont
  {Ceriotti}, \citenamefont {Parrinello}, \citenamefont {Markland},\ and\
  \citenamefont {Manolopoulos}}]{ceriotti_efficient_2010}%
  \BibitemOpen
  \bibfield  {author} {\bibinfo {author} {\bibfnamefont {M.}~\bibnamefont
  {Ceriotti}}, \bibinfo {author} {\bibfnamefont {M.}~\bibnamefont
  {Parrinello}}, \bibinfo {author} {\bibfnamefont {T.~E.}\ \bibnamefont
  {Markland}}, \ and\ \bibinfo {author} {\bibfnamefont {D.~E.}\ \bibnamefont
  {Manolopoulos}},\ }\bibfield  {title} {\enquote {\bibinfo {title} {Efficient
  stochastic thermostatting of path integral molecular dynamics},}\ }\href
  {\doibase 10.1063/1.3489925} {\bibfield  {journal} {\bibinfo  {journal} {J.
  Chem. Phys.}\ }\textbf {\bibinfo {volume} {133}},\ \bibinfo {pages} {124104}
  (\bibinfo {year} {2010})}\BibitemShut {NoStop}%
\bibitem [{\citenamefont {Habershon}\ \emph {et~al.}(2013)\citenamefont
  {Habershon}, \citenamefont {Manolopoulos}, \citenamefont {Markland},\ and\
  \citenamefont {Miller}}]{habershon_ring-polymer_2013}%
  \BibitemOpen
  \bibfield  {author} {\bibinfo {author} {\bibfnamefont {S.}~\bibnamefont
  {Habershon}}, \bibinfo {author} {\bibfnamefont {D.~E.}\ \bibnamefont
  {Manolopoulos}}, \bibinfo {author} {\bibfnamefont {T.~E.}\ \bibnamefont
  {Markland}}, \ and\ \bibinfo {author} {\bibfnamefont {T.~F.}\ \bibnamefont
  {Miller}},\ }\bibfield  {title} {\enquote {\bibinfo {title} {Ring-{Polymer}
  {Molecular} {Dynamics}: {Quantum} {Effects} in {Chemical} {Dynamics} from
  {Classical} {Trajectories} in an {Extended} {Phase} {Space}},}\ }\href
  {\doibase 10.1146/annurev-physchem-040412-110122} {\bibfield  {journal}
  {\bibinfo  {journal} {Annu. Rev. Phys. Chem.}\ }\textbf {\bibinfo {volume}
  {64}},\ \bibinfo {pages} {387--413} (\bibinfo {year} {2013})}\BibitemShut
  {NoStop}%
\bibitem [{\citenamefont {Rossi}, \citenamefont {Ceriotti},\ and\ \citenamefont
  {Manolopoulos}(2014)}]{rossi_how_2014}%
  \BibitemOpen
  \bibfield  {author} {\bibinfo {author} {\bibfnamefont {M.}~\bibnamefont
  {Rossi}}, \bibinfo {author} {\bibfnamefont {M.}~\bibnamefont {Ceriotti}}, \
  and\ \bibinfo {author} {\bibfnamefont {D.~E.}\ \bibnamefont {Manolopoulos}},\
  }\bibfield  {title} {\enquote {\bibinfo {title} {How to remove the spurious
  resonances from ring polymer molecular dynamics},}\ }\href {\doibase
  10.1063/1.4883861} {\bibfield  {journal} {\bibinfo  {journal} {J. Chem.
  Phys.}\ }\textbf {\bibinfo {volume} {140}},\ \bibinfo {pages} {234116}
  (\bibinfo {year} {2014})}\BibitemShut {NoStop}%
\bibitem [{\citenamefont {Hone}, \citenamefont {Rossky},\ and\ \citenamefont
  {Voth}(2006)}]{hone_comparative_2006}%
  \BibitemOpen
  \bibfield  {author} {\bibinfo {author} {\bibfnamefont {T.~D.}\ \bibnamefont
  {Hone}}, \bibinfo {author} {\bibfnamefont {P.~J.}\ \bibnamefont {Rossky}}, \
  and\ \bibinfo {author} {\bibfnamefont {G.~A.}\ \bibnamefont {Voth}},\
  }\bibfield  {title} {\enquote {\bibinfo {title} {A comparative study of
  imaginary time path integral based methods for quantum dynamics},}\ }\href
  {\doibase 10.1063/1.2186636} {\bibfield  {journal} {\bibinfo  {journal} {J.
  Chem. Phys.}\ }\textbf {\bibinfo {volume} {124}},\ \bibinfo {pages} {154103}
  (\bibinfo {year} {2006})}\BibitemShut {NoStop}%
\bibitem [{\citenamefont {Thompson}\ \emph {et~al.}(2022)\citenamefont
  {Thompson}, \citenamefont {Aktulga}, \citenamefont {Berger}, \citenamefont
  {Bolintineanu}, \citenamefont {Brown}, \citenamefont {Crozier}, \citenamefont
  {in~'t Veld}, \citenamefont {Kohlmeyer}, \citenamefont {Moore}, \citenamefont
  {Nguyen}, \citenamefont {Shan}, \citenamefont {Stevens}, \citenamefont
  {Tranchida}, \citenamefont {Trott},\ and\ \citenamefont
  {Plimpton}}]{thompson_lammps_2022}%
  \BibitemOpen
  \bibfield  {author} {\bibinfo {author} {\bibfnamefont {A.~P.}\ \bibnamefont
  {Thompson}}, \bibinfo {author} {\bibfnamefont {H.~M.}\ \bibnamefont
  {Aktulga}}, \bibinfo {author} {\bibfnamefont {R.}~\bibnamefont {Berger}},
  \bibinfo {author} {\bibfnamefont {D.~S.}\ \bibnamefont {Bolintineanu}},
  \bibinfo {author} {\bibfnamefont {W.~M.}\ \bibnamefont {Brown}}, \bibinfo
  {author} {\bibfnamefont {P.~S.}\ \bibnamefont {Crozier}}, \bibinfo {author}
  {\bibfnamefont {P.~J.}\ \bibnamefont {in~'t Veld}}, \bibinfo {author}
  {\bibfnamefont {A.}~\bibnamefont {Kohlmeyer}}, \bibinfo {author}
  {\bibfnamefont {S.~G.}\ \bibnamefont {Moore}}, \bibinfo {author}
  {\bibfnamefont {T.~D.}\ \bibnamefont {Nguyen}}, \bibinfo {author}
  {\bibfnamefont {R.}~\bibnamefont {Shan}}, \bibinfo {author} {\bibfnamefont
  {M.~J.}\ \bibnamefont {Stevens}}, \bibinfo {author} {\bibfnamefont
  {J.}~\bibnamefont {Tranchida}}, \bibinfo {author} {\bibfnamefont
  {C.}~\bibnamefont {Trott}}, \ and\ \bibinfo {author} {\bibfnamefont {S.~J.}\
  \bibnamefont {Plimpton}},\ }\bibfield  {title} {\enquote {\bibinfo {title}
  {{LAMMPS} - a flexible simulation tool for particle-based materials modeling
  at the atomic, meso, and continuum scales},}\ }\href {\doibase
  10.1016/j.cpc.2021.108171} {\bibfield  {journal} {\bibinfo  {journal}
  {Computer Physics Communications}\ }\textbf {\bibinfo {volume} {271}},\
  \bibinfo {pages} {108171} (\bibinfo {year} {2022})}\BibitemShut {NoStop}%
\bibitem [{\citenamefont {Kapil}\ \emph {et~al.}(2019)\citenamefont {Kapil},
  \citenamefont {Rossi}, \citenamefont {Marsalek}, \citenamefont {Petraglia},
  \citenamefont {Litman}, \citenamefont {Spura}, \citenamefont {Cheng},
  \citenamefont {Cuzzocrea}, \citenamefont {Meißner}, \citenamefont {Wilkins},
  \citenamefont {Helfrecht}, \citenamefont {Juda}, \citenamefont {Bienvenue},
  \citenamefont {Fang}, \citenamefont {Kessler}, \citenamefont {Poltavsky},
  \citenamefont {Vandenbrande}, \citenamefont {Wieme}, \citenamefont
  {Corminboeuf}, \citenamefont {Kühne}, \citenamefont {Manolopoulos},
  \citenamefont {Markland}, \citenamefont {Richardson}, \citenamefont
  {Tkatchenko}, \citenamefont {Tribello}, \citenamefont {Van~Speybroeck},\ and\
  \citenamefont {Ceriotti}}]{kapil_i-pi_2019}%
  \BibitemOpen
  \bibfield  {author} {\bibinfo {author} {\bibfnamefont {V.}~\bibnamefont
  {Kapil}}, \bibinfo {author} {\bibfnamefont {M.}~\bibnamefont {Rossi}},
  \bibinfo {author} {\bibfnamefont {O.}~\bibnamefont {Marsalek}}, \bibinfo
  {author} {\bibfnamefont {R.}~\bibnamefont {Petraglia}}, \bibinfo {author}
  {\bibfnamefont {Y.}~\bibnamefont {Litman}}, \bibinfo {author} {\bibfnamefont
  {T.}~\bibnamefont {Spura}}, \bibinfo {author} {\bibfnamefont
  {B.}~\bibnamefont {Cheng}}, \bibinfo {author} {\bibfnamefont
  {A.}~\bibnamefont {Cuzzocrea}}, \bibinfo {author} {\bibfnamefont {R.~H.}\
  \bibnamefont {Meißner}}, \bibinfo {author} {\bibfnamefont {D.~M.}\
  \bibnamefont {Wilkins}}, \bibinfo {author} {\bibfnamefont {B.~A.}\
  \bibnamefont {Helfrecht}}, \bibinfo {author} {\bibfnamefont {P.}~\bibnamefont
  {Juda}}, \bibinfo {author} {\bibfnamefont {S.~P.}\ \bibnamefont {Bienvenue}},
  \bibinfo {author} {\bibfnamefont {W.}~\bibnamefont {Fang}}, \bibinfo {author}
  {\bibfnamefont {J.}~\bibnamefont {Kessler}}, \bibinfo {author} {\bibfnamefont
  {I.}~\bibnamefont {Poltavsky}}, \bibinfo {author} {\bibfnamefont
  {S.}~\bibnamefont {Vandenbrande}}, \bibinfo {author} {\bibfnamefont
  {J.}~\bibnamefont {Wieme}}, \bibinfo {author} {\bibfnamefont
  {C.}~\bibnamefont {Corminboeuf}}, \bibinfo {author} {\bibfnamefont {T.~D.}\
  \bibnamefont {Kühne}}, \bibinfo {author} {\bibfnamefont {D.~E.}\
  \bibnamefont {Manolopoulos}}, \bibinfo {author} {\bibfnamefont {T.~E.}\
  \bibnamefont {Markland}}, \bibinfo {author} {\bibfnamefont {J.~O.}\
  \bibnamefont {Richardson}}, \bibinfo {author} {\bibfnamefont
  {A.}~\bibnamefont {Tkatchenko}}, \bibinfo {author} {\bibfnamefont {G.~A.}\
  \bibnamefont {Tribello}}, \bibinfo {author} {\bibfnamefont {V.}~\bibnamefont
  {Van~Speybroeck}}, \ and\ \bibinfo {author} {\bibfnamefont {M.}~\bibnamefont
  {Ceriotti}},\ }\bibfield  {title} {\enquote {\bibinfo {title} {i-{PI} 2.0:
  {A} universal force engine for advanced molecular simulations},}\ }\href
  {\doibase 10.1016/j.cpc.2018.09.020} {\bibfield  {journal} {\bibinfo
  {journal} {Computer Physics Communications}\ }\textbf {\bibinfo {volume}
  {236}},\ \bibinfo {pages} {214--223} (\bibinfo {year} {2019})}\BibitemShut
  {NoStop}%
\bibitem [{\citenamefont {Togo}\ and\ \citenamefont
  {Tanaka}(2015)}]{togo_first_2015}%
  \BibitemOpen
  \bibfield  {author} {\bibinfo {author} {\bibfnamefont {A.}~\bibnamefont
  {Togo}}\ and\ \bibinfo {author} {\bibfnamefont {I.}~\bibnamefont {Tanaka}},\
  }\bibfield  {title} {\enquote {\bibinfo {title} {First principles phonon
  calculations in materials science},}\ }\href {\doibase
  10.1016/j.scriptamat.2015.07.021} {\bibfield  {journal} {\bibinfo  {journal}
  {Scripta Materialia}\ }\textbf {\bibinfo {volume} {108}},\ \bibinfo {pages}
  {1--5} (\bibinfo {year} {2015})}\BibitemShut {NoStop}%
\bibitem [{\citenamefont {Farmer}(2000)}]{farmer_transverse_2000}%
  \BibitemOpen
  \bibfield  {author} {\bibinfo {author} {\bibfnamefont {V.~C.}\ \bibnamefont
  {Farmer}},\ }\bibfield  {title} {\enquote {\bibinfo {title} {Transverse and
  longitudinal crystal modes associated with {OH} stretching vibrations in
  single crystals of kaolinite and dickite},}\ }\href {\doibase
  10.1016/S1386-1425(99)00182-1} {\bibfield  {journal} {\bibinfo  {journal}
  {Spectrochimica Acta Part A: Molecular and Biomolecular Spectroscopy}\
  }\textbf {\bibinfo {volume} {56}},\ \bibinfo {pages} {927--930} (\bibinfo
  {year} {2000})}\BibitemShut {NoStop}%
\bibitem [{\citenamefont {Marsalek}\ and\ \citenamefont
  {Markland}(2017)}]{marsalek_quantum_2017}%
  \BibitemOpen
  \bibfield  {author} {\bibinfo {author} {\bibfnamefont {O.}~\bibnamefont
  {Marsalek}}\ and\ \bibinfo {author} {\bibfnamefont {T.~E.}\ \bibnamefont
  {Markland}},\ }\bibfield  {title} {\enquote {\bibinfo {title} {Quantum
  {Dynamics} and {Spectroscopy} of {Ab} {Initio} {Liquid} {Water}:
  {The} {Interplay} of {Nuclear} and {Electronic} {Quantum} {Effects}},}\
  }\href {\doibase 10.1021/acs.jpclett.7b00391} {\bibfield  {journal} {\bibinfo
   {journal} {The Journal of Physical Chemistry Letters}\ }\textbf {\bibinfo
  {volume} {8}},\ \bibinfo {pages} {1545--1551} (\bibinfo {year}
  {2017})}\BibitemShut {NoStop}%
\bibitem [{\citenamefont {T.~Johnston}\ \emph {et~al.}(2008)\citenamefont
  {T.~Johnston}, \citenamefont {Elzea~Kogel}, \citenamefont {L.~Bish},
  \citenamefont {Kogure},\ and\ \citenamefont
  {H.~Murray}}]{t_johnston_low-temperature_2008}%
  \BibitemOpen
  \bibfield  {author} {\bibinfo {author} {\bibfnamefont {C.}~\bibnamefont
  {T.~Johnston}}, \bibinfo {author} {\bibfnamefont {J.}~\bibnamefont
  {Elzea~Kogel}}, \bibinfo {author} {\bibfnamefont {D.}~\bibnamefont
  {L.~Bish}}, \bibinfo {author} {\bibfnamefont {T.}~\bibnamefont {Kogure}}, \
  and\ \bibinfo {author} {\bibfnamefont {H.}~\bibnamefont {H.~Murray}},\
  }\bibfield  {title} {\enquote {\bibinfo {title} {Low-temperature {Ftir}
  {Study} of {Kaolin}-{Group} {Minerals}},}\ }\href {\doibase
  10.1346/CCMN.2008.0560408} {\bibfield  {journal} {\bibinfo  {journal} {Clays
  Clay Miner.}\ }\textbf {\bibinfo {volume} {56}},\ \bibinfo {pages} {470--485}
  (\bibinfo {year} {2008})}\BibitemShut {NoStop}%
\bibitem [{\citenamefont {Smr\v{c}ok}\ \emph {et~al.}(2010)\citenamefont
  {Smr\v{c}ok}, \citenamefont {Tunega}, \citenamefont {Ramirez-Cuesta},\ and\
  \citenamefont {Scholtzov\'{a}}}]{smrcok_combined_2010}%
  \BibitemOpen
  \bibfield  {author} {\bibinfo {author} {\bibfnamefont {v.}~\bibnamefont
  {Smr\v{c}ok}}, \bibinfo {author} {\bibfnamefont {D.}~\bibnamefont {Tunega}},
  \bibinfo {author} {\bibfnamefont {A.~J.}\ \bibnamefont {Ramirez-Cuesta}}, \
  and\ \bibinfo {author} {\bibfnamefont {E.}~\bibnamefont {Scholtzov\'{a}}},\
  }\bibfield  {title} {\enquote {\bibinfo {title} {The combined inelastic
  neutron scattering and solid state {DFT} study of hydrogen atoms dynamics in
  a highly ordered kaolinite},}\ }\href {\doibase 10.1007/s00269-010-0358-3}
  {\bibfield  {journal} {\bibinfo  {journal} {Phys Chem Minerals}\ }\textbf
  {\bibinfo {volume} {37}},\ \bibinfo {pages} {571--579} (\bibinfo {year}
  {2010})}\BibitemShut {NoStop}%
\bibitem [{\citenamefont {White}\ \emph {et~al.}(2013)\citenamefont {White},
  \citenamefont {Kearley}, \citenamefont {Provis},\ and\ \citenamefont
  {Riley}}]{white_structure_2013}%
  \BibitemOpen
  \bibfield  {author} {\bibinfo {author} {\bibfnamefont {C.~E.}\ \bibnamefont
  {White}}, \bibinfo {author} {\bibfnamefont {G.~J.}\ \bibnamefont {Kearley}},
  \bibinfo {author} {\bibfnamefont {J.~L.}\ \bibnamefont {Provis}}, \ and\
  \bibinfo {author} {\bibfnamefont {D.~P.}\ \bibnamefont {Riley}},\ }\bibfield
  {title} {\enquote {\bibinfo {title} {Structure of kaolinite and influence of
  stacking faults: {Reconciling} theory and experiment using inelastic neutron
  scattering analysis},}\ }\href {\doibase 10.1063/1.4804306} {\bibfield
  {journal} {\bibinfo  {journal} {J. Chem. Phys.}\ }\textbf {\bibinfo {volume}
  {138}},\ \bibinfo {pages} {194501} (\bibinfo {year} {2013})},\ \bibinfo
  {note} {publisher: American Institute of Physics}\BibitemShut {NoStop}%
\bibitem [{\citenamefont {Gao}\ and\ \citenamefont
  {Remsing}(2022)}]{gao_self-consistent_2022}%
  \BibitemOpen
  \bibfield  {author} {\bibinfo {author} {\bibfnamefont {A.}~\bibnamefont
  {Gao}}\ and\ \bibinfo {author} {\bibfnamefont {R.~C.}\ \bibnamefont
  {Remsing}},\ }\bibfield  {title} {\enquote {\bibinfo {title} {Self-consistent
  determination of long-range electrostatics in neural network potentials},}\
  }\href {\doibase 10.1038/s41467-022-29243-2} {\bibfield  {journal} {\bibinfo
  {journal} {Nature Communications}\ }\textbf {\bibinfo {volume} {13}},\
  \bibinfo {pages} {1572} (\bibinfo {year} {2022})},\ \bibinfo {note} {number:
  1 Publisher: Nature Publishing Group}\BibitemShut {NoStop}%
\bibitem [{\citenamefont {Ko}\ \emph {et~al.}(2021{\natexlab{a}})\citenamefont
  {Ko}, \citenamefont {Finkler}, \citenamefont {Goedecker},\ and\ \citenamefont
  {Behler}}]{ko_general-purpose_2021}%
  \BibitemOpen
  \bibfield  {author} {\bibinfo {author} {\bibfnamefont {T.~W.}\ \bibnamefont
  {Ko}}, \bibinfo {author} {\bibfnamefont {J.~A.}\ \bibnamefont {Finkler}},
  \bibinfo {author} {\bibfnamefont {S.}~\bibnamefont {Goedecker}}, \ and\
  \bibinfo {author} {\bibfnamefont {J.}~\bibnamefont {Behler}},\ }\bibfield
  {title} {\enquote {\bibinfo {title} {General-{Purpose} {Machine} {Learning}
  {Potentials} {Capturing} {Nonlocal} {Charge} {Transfer}},}\ }\href {\doibase
  10.1021/acs.accounts.0c00689} {\bibfield  {journal} {\bibinfo  {journal}
  {Accounts of Chemical Research}\ }\textbf {\bibinfo {volume} {54}},\ \bibinfo
  {pages} {808--817} (\bibinfo {year} {2021}{\natexlab{a}})},\ \bibinfo {note}
  {publisher: American Chemical Society}\BibitemShut {NoStop}%
\bibitem [{\citenamefont {Ko}\ \emph {et~al.}(2021{\natexlab{b}})\citenamefont
  {Ko}, \citenamefont {Finkler}, \citenamefont {Goedecker},\ and\ \citenamefont
  {Behler}}]{ko_fourth-generation_2021}%
  \BibitemOpen
  \bibfield  {author} {\bibinfo {author} {\bibfnamefont {T.~W.}\ \bibnamefont
  {Ko}}, \bibinfo {author} {\bibfnamefont {J.~A.}\ \bibnamefont {Finkler}},
  \bibinfo {author} {\bibfnamefont {S.}~\bibnamefont {Goedecker}}, \ and\
  \bibinfo {author} {\bibfnamefont {J.}~\bibnamefont {Behler}},\ }\bibfield
  {title} {\enquote {\bibinfo {title} {A fourth-generation high-dimensional
  neural network potential with accurate electrostatics including non-local
  charge transfer},}\ }\href {\doibase 10.1038/s41467-020-20427-2} {\bibfield
  {journal} {\bibinfo  {journal} {Nature Communications}\ }\textbf {\bibinfo
  {volume} {12}},\ \bibinfo {pages} {398} (\bibinfo {year}
  {2021}{\natexlab{b}})},\ \bibinfo {note} {number: 1 Publisher: Nature
  Publishing Group}\BibitemShut {NoStop}%
\bibitem [{\citenamefont {Yao}\ \emph {et~al.}(2018)\citenamefont {Yao},
  \citenamefont {E. Herr}, \citenamefont {W. Toth}, \citenamefont
  {Mckintyre},\ and\ \citenamefont {Parkhill}}]{yao_tensormol-01_2018}%
  \BibitemOpen
  \bibfield  {author} {\bibinfo {author} {\bibfnamefont {K.}~\bibnamefont
  {Yao}}, \bibinfo {author} {\bibfnamefont {J.}~\bibnamefont {E. Herr}},
  \bibinfo {author} {\bibfnamefont {D.}~\bibnamefont {W. Toth}}, \bibinfo
  {author} {\bibfnamefont {R.}~\bibnamefont {Mckintyre}}, \ and\ \bibinfo
  {author} {\bibfnamefont {J.}~\bibnamefont {Parkhill}},\ }\bibfield  {title}
  {\enquote {\bibinfo {title} {The {TensorMol}-0.1 model chemistry: a neural
  network augmented with long-range physics},}\ }\href {\doibase
  10.1039/C7SC04934J} {\bibfield  {journal} {\bibinfo  {journal} {Chemical
  Science}\ }\textbf {\bibinfo {volume} {9}},\ \bibinfo {pages} {2261--2269}
  (\bibinfo {year} {2018})},\ \bibinfo {note} {publisher: Royal Society of
  Chemistry}\BibitemShut {NoStop}%
\bibitem [{\citenamefont {Behler}\ and\ \citenamefont
  {Parrinello}(2007)}]{behler_generalized_2007}%
  \BibitemOpen
  \bibfield  {author} {\bibinfo {author} {\bibfnamefont {J.}~\bibnamefont
  {Behler}}\ and\ \bibinfo {author} {\bibfnamefont {M.}~\bibnamefont
  {Parrinello}},\ }\bibfield  {title} {\enquote {\bibinfo {title} {Generalized
  {Neural}-{Network} {Representation} of {High}-{Dimensional}
  {Potential}-{Energy} {Surfaces}},}\ }\href {\doibase
  10.1103/PhysRevLett.98.146401} {\bibfield  {journal} {\bibinfo  {journal}
  {Phys. Rev. Lett.}\ }\textbf {\bibinfo {volume} {98}},\ \bibinfo {pages}
  {146401} (\bibinfo {year} {2007})}\BibitemShut {NoStop}%
\bibitem [{\citenamefont {Behler}(2011)}]{behler_atom-centered_2011}%
  \BibitemOpen
  \bibfield  {author} {\bibinfo {author} {\bibfnamefont {J.}~\bibnamefont
  {Behler}},\ }\bibfield  {title} {\enquote {\bibinfo {title} {Atom-centered
  symmetry functions for constructing high-dimensional neural network
  potentials},}\ }\href {\doibase 10.1063/1.3553717} {\bibfield  {journal}
  {\bibinfo  {journal} {J. Chem. Phys.}\ }\textbf {\bibinfo {volume} {134}},\
  \bibinfo {pages} {074106} (\bibinfo {year} {2011})}\BibitemShut {NoStop}%
\bibitem [{\citenamefont {Blank}\ and\ \citenamefont
  {Brown}(1994)}]{blank_adaptive_1994}%
  \BibitemOpen
  \bibfield  {author} {\bibinfo {author} {\bibfnamefont {T.~B.}\ \bibnamefont
  {Blank}}\ and\ \bibinfo {author} {\bibfnamefont {S.~D.}\ \bibnamefont
  {Brown}},\ }\bibfield  {title} {\enquote {\bibinfo {title} {Adaptive, global,
  extended {Kalman} filters for training feedforward neural networks},}\ }\href
  {\doibase 10.1002/cem.1180080605} {\bibfield  {journal} {\bibinfo  {journal}
  {J. Chemometrics}\ }\textbf {\bibinfo {volume} {8}},\ \bibinfo {pages}
  {391--407} (\bibinfo {year} {1994})}\BibitemShut {NoStop}%
\bibitem [{\citenamefont {L. Benson}, \citenamefont {Trenins},\ and\
  \citenamefont {C. Althorpe}(2020)}]{lbenson_which_2020}%
  \BibitemOpen
  \bibfield  {author} {\bibinfo {author} {\bibfnamefont {R.}~\bibnamefont
  {L. Benson}}, \bibinfo {author} {\bibfnamefont {G.}~\bibnamefont {Trenins}},
  \ and\ \bibinfo {author} {\bibfnamefont {S.}~\bibnamefont {C. Althorpe}},\
  }\bibfield  {title} {\enquote {\bibinfo {title} {Which quantum
  statistics–classical dynamics method is best for water?}}\ }\href {\doibase
  10.1039/C9FD00077A} {\bibfield  {journal} {\bibinfo  {journal} {Faraday
  Discussions}\ }\textbf {\bibinfo {volume} {221}},\ \bibinfo {pages}
  {350--366} (\bibinfo {year} {2020})}\BibitemShut {NoStop}%
\end{thebibliography}
\end{document}